\documentclass[10pt]{article}
\usepackage{float}
\usepackage[symbol]{footmisc} 
\usepackage{algorithm}
\usepackage{algpseudocode}
\usepackage[superscript,sort]{cite} 
\usepackage{array} 
\usepackage[pdftex,colorlinks=true,linkcolor=black,citecolor=black,urlcolor=black,bookmarks=true,breaklinks=true]{hyperref}
\usepackage[pdftex]{graphicx}
\usepackage[usenames,pdftex]{color}
\definecolor{Red}{rgb}{1.0,0.0,0.0}
\usepackage{amsmath,amssymb}
\usepackage{amsthm,amscd,amsxtra,amsfonts,amsmath,amssymb,multirow}
\usepackage{wrapfig}
\usepackage[footnotesize]{caption}
\usepackage[tiny,compact]{titlesec}
\usepackage{ctable}
\usepackage{helvet}

\usepackage{xcolor,colortbl}
\usepackage{subfigure}
\usepackage{tikz}
\usetikzlibrary{shapes,arrows}
\usepackage[T1]{fontenc}

\providecommand{\e}[1]{\ensuremath{\times 10^{#1}}} 
\titlespacing*{\section}{0pt}{*0}{*0}
\titlespacing*{\subsection}{0pt}{*0}{*0}
\titlespacing*{\subsubsection}{0pt}{*0}{*0}
\titlespacing{\paragraph}{0pt}{*0}{*1}

\definecolor{MyPurple}{rgb}{1,0,1}

\newcommand{\beq}[1]{\begin{equation} \label{#1}}
\newcommand{\eeq}{\end{equation}}
\newcommand{\barray}{\begin{array}{ll}}
\newcommand{\earray}{\end{array}}

\begin{document}
\pagenumbering{roman}

\clearpage \pagebreak \setcounter{page}{1}
\renewcommand{\thepage}{{\arabic{page}}}

\title{ Fast and Anisotropic Flexibility-Rigidity Index for Protein Flexibility and Fluctuation Analysis
}

\author{
Kristopher Opron$^1$,
Kelin Xia$^2$  and
Guo-Wei Wei$^{1,2,3}$ \footnote{ Address correspondences  to Guo-Wei Wei. E-mail:wei@math.msu.edu}\\
$^1$  Department of Biochemistry and Molecular Biology\\
Michigan State University, MI 48824, USA \\
$^2$ Department of Mathematics \\
Michigan State University, MI 48824, USA\\
$^3$ Department of Electrical and Computer Engineering \\
Michigan State University, MI 48824, USA \\
}

\date{\today}
\maketitle

\begin{abstract}
Protein structural fluctuation, typically measured by Debye-Waller factors, or B-factors, is a  manifestation of protein flexibility, which strongly correlates to protein function. The flexibility-rigidity  index (FRI) is a newly proposed method for the construction of atomic rigidity functions required in the theory of continuum elasticity with atomic rigidity (CEWAR), which is a  new multiscale formalism for describing  excessively large biomolecular systems. The FRI method analyzes protein rigidity and flexibility and is  capable of predicting protein B-factors without resorting to matrix diagonalization. A fundamental assumption used in the FRI is that protein structures are uniquely determined by various internal and external interactions, while the protein functions, such as stability and flexibility, are solely determined by the structure. As such, one can predict protein flexibility without resorting to the protein interaction Hamiltonian. Consequently, bypassing the matrix diagonalization, the original  FRI has a computational complexity of ${\cal O}(N^2)$. This work introduces a fast FRI (fFRI) algorithm for the flexibility analysis of large macromolecules. The proposed fFRI further reduces the computational complexity to ${\cal O}(N)$.   Additionally, we propose anisotropic FRI (aFRI) algorithms for the analysis of protein collective dynamics.  {The aFRI algorithms admit adaptive Hessian matrices, from a completely global $3N\times3N$ matrix  to completely local $3\times3$ matrices.  However, these local $3\times3$ matrices have built in much non-local correlation.  Eigenvectors obtained the proposed aFRI lagorithms are able to demonstrate collective motions.}  Moreover,  we investigate the performance of FRI by employing four families of radial basis correlation functions. Both parameter optimized and parameter-free FRI methods are explored. Furthermore,  we compare the accuracy and efficiency of FRI with some  {established} approaches to flexibility analysis, namely, normal mode analysis (NMA) and  Gaussian network model (GNM). The accuracy of the FRI method is tested using four sets of proteins, three sets of relatively small-, medium- and large-sized structures and an extended set of 365 proteins. A fifth set of proteins is used to compare the efficiency of the FRI, fFRI, aFRI and GNM methods. Intensive validation and comparison indicate that the FRI, particularly the fFRI, is orders of magnitude more efficient and about 10\% more accurate overall than some of the most popular methods in the field.  The proposed fFRI is able to predict  B-factors for $\alpha$-carbons of the HIV virus capsid (313,236 residues) in less than 30 seconds on a single processor using only one core.  { Finally, we demonstrate the application of FRI and aFRI  to protein domain analysis}. 

\end{abstract}
\maketitle

\pagebreak

\section{Introduction}\label{sec:Intro}

Proteins provide the structural basis for living organisms and are essential research subjects of the biological sciences. Although the original sequence-structure-function dogma  \cite{Anfinsen:1973} has been seriously challenged \cite{Schroder:2005,Chiti:2006}, the protein structure, in either folded or unfolded form, still determines its function. Therefore, the understanding of the structure of a protein holds the key to the prediction of the protein's function \cite{Onuchic:1997,White:1999}. Unfortunately, it remains a major challenge in the biological sciences to predict a protein's functions from its known structure.

There are a few essential factors, including protein geometry, electrostatics, and flexibility, that strongly correlate to protein function. There is no need to elaborate on the importance of protein geometry and electrostatics to protein function and dynamics. However, the impact of protein flexibility to protein function is often underestimated or even overlooked. In general, protein flexibility is the ability to deform from the equilibrium state under external forces, such as docking of ligands, docking with other proteins or random  bombardments of small molecules in liquid and/or gas phases and lattice phonons in a solid phase \cite{Frauenfelder:1991}. Under physiological conditions, proteins experience everlasting motion or structural fluctuation in a wide variety of spatiotemporal scales because of their  flexibility and  uninterrupted external forces. However, at absolute zero temperature, there is no protein motion or fluctuation. Therefore protein motion or fluctuation is just the molecule's response to the external stimuli, while protein flexibility is an intrinsic property of the structure. In fact, protein flexibility varies from protein to protein and is a signature of the protein.  In a given protein, flexibility can be  different from  atom to atom, from residue to residue  and from domain to domain. Typically, the performance of a theoretical model for  protein flexibility analysis can be validated via its prediction of protein structural fluctuation.

X-ray crystallography is one of the most important techniques for protein flexibility analysis. The atomic mean-square-fluctuations are reflected in X-ray diffraction or other diffraction data and can be estimated in terms of the Debye-Waller factor, also known as the B-factor or temperature factor. Typically, reported B-factors are not corrected for  the variations in atomic diffraction cross sections and chemical stability during the diffraction data collection, which perhaps contributes to the fact that all-atom models usually do not work as well as coarse-grained models in the B-factor prediction \cite{JKPark:2013}. Nuclear magnetic resonance (NMR) is another important method, which is particularly valuable for flexibility analysis under physiological conditions.  NMR is able to investigate  protein flexibility at a variety of timescales.

Apart from  experimental approaches, a number of theoretical methods  have been developed for flexibility analysis and B-factor prediction in the past. Protein collective motion and fluctuation can be elucidated by  molecular dynamics (MD), which  has considerably expanded our understanding of the conformational landscapes of proteins.  However, the excessively large number of degrees of freedom associated with the all-atom representation and long time integration becomes computationally inefficient with increasing size of the system and can obscure larger scale motions.  Alternative time-independent approaches, such as normal mode analysis (NMA)  \cite{Go:1983,Tasumi:1982,Brooks:1983,Levitt:1985},  {graph theory \cite{Jacobs:2001} and elastic network model (ENM) \cite{Tirion:1996} theories, including  Gaussian network model (GNM)   \cite{Flory:1976, Bahar:1997,Bahar:1998}  and anisotropic network model (ANM) \cite{Atilgan:2001}}, have been developed for protein flexibility analysis in the past few decades. These methods can be derived from their corresponding Newton's equations by using  the time-harmonic approximation \cite{JKPark:2013}. The low order eigenmodes computed from diagonalizing the Kirchhoff matrix or the  Hessian matrix can shed light on the   long-time behavior of the protein dynamics beyond the reach of MD simulations  \cite{Tasumi:1982,Brooks:1983,Levitt:1985,Tirion:1996,Bahar:1997}.  Coarse-grained based ENM and GNM approaches  have become popular recently due to their simplified potential and computational efficiency \cite{Bahar:1997,Bahar:1998,Atilgan:2001,Hinsen:1998,Tama:2001,LiGH:2002}.  It was shown that the GNM is about one order more efficient than most other approaches \cite{LWYang:2008}. Improvements to these approaches have been developed for many aspects, including crystal periodicity and cofactor corrections  \cite{Kundu:2002,Kondrashov:2007,Hisen:2008,GSong:2007}, and  density - cluster  rotational - translational blocking \cite{Demerdash:2012}.   These approaches have been  applied to the study of large proteins or protein complexes, such as, hemoglobin \cite{CXu:2003}, F1 ATPase \cite{WZheng:2003,QCui:2004}, chaperonin GroEL \cite{Keskin:2002,WZheng:2007}, viral capsids \cite{Rader:2005,Tama:2005}, and ribosome \cite{Tama:2003,YWang:2004}. Flexibility also plays an important role in stability \cite{Livesay:2004} and docking analysis \cite{Gerek:2010}. For further detail in their status and application,  the reader is referred to recent review papers \cite{JMa:2005,LWYang:2008,Skjaven:2009,QCui:2010}.

Recently, we have developed a new multiscale formalism called continuum elasticity with atomic rigidity (CEWAR) for the elastic analysis of excessively large macromolecules. In the CEWAR approach, a continuous atomic rigidity function is required to characterize the shear modulus in the stress tensor of elasticity equations. To this end, a simple method, called flexibility-rigidity index  (FRI), is introduced to evaluate macromolecular flexibility and rigidity    \cite{KLXia:2013d}.
We noted after the publication of our  earlier work \cite{KLXia:2013d} that the name of ``flexibility index'' was proposed independently by von der Lieth et al. \cite{vonderLieth:1996} and Jacobs et al. \cite{Jacobs:2001} for two different quantities to describe bond strengths. Both of these flexibility indices are distinct from our proposed FRI.  The FRI is a solely structural based algorithm that does not reconstruct any protein interaction Hamiltonian.  Only elementary arithmetics is needed in the FRI method for proteins. In particular,  the FRI prediction of protein B-factors does not   require a stringently minimized structure and  time consuming matrix diagonalization or matrix decomposition, nor does it involve any training procedure. Two types of monotonically decaying correlation functions, namely, exponential type and Lorentz type, have been utilized previously for the construction of  protein correlation matrix. Parameter ranges for the FRI have been extensively tested  and the performance of the FRI for protein B-factor prediction has been carefully validated with a set of 263 proteins  \cite{KLXia:2013d}. It is found that for residue based  B-factor prediction, the FRI can be made parameter free. However, it is not clear how the FRI compares to alternative approaches in the field, particularly the state of the art methods such as GNM and NMA.

One of the objectives of the present work is to introduce a  fast FRI (fFRI) algorithm by using appropriate data structures.  Computational efficiency is a central issue. The computational complexity of the proposed fFRI is of ${\cal O}(N)$, compared to that of ${\cal O}(N^2)$ for the original FRI algorithm and of ${\cal O}(N^3)$  for the GNM, where $N$ is the number of atoms.   We use a cell lists  approach \cite{Allen:1987} to reduce the computational complexity.   Another objective is to introduce anisotropic FRI (aFRI) algorithms for the motion analysis of biomolecules. Unlike  ANM \cite{Atilgan:2001,JKPark:2013}, which is completely global and has $3N\times3N$ elements in its Hessian matrix, the proposed aFRI algorithms  {have adaptive  Hessian matrices, which vary from completely global to completely local.}   Despite of the localization, there are collective motions in three sets of  eigenvectors. The other objective of the present work is to further analyze the performance of the FRI methods for protein B-factor prediction. To this end, we examine the accuracy of FRI algorithms associated with four families of correlation functions  and carry out a comparative study of the FRI and fFRI vs. other cutting edge approaches, namely, NMA and GNM. Our investigation concerns three issues, i.e.,  accuracy, reliability, and  efficiency in the protein B-factor prediction.   { Finally, we also demonstrate the applications of  the FRI and aFRI to protein domain analysis.}

The rest of this paper is organized as follows. Section \ref{sec:methods} is devoted to  methods and algorithms.
 To establish notation and facilitate further discussion, the FRI approach is briefly discussed. We present  a new simplified version of the FRI that is  relevant to the B-factor prediction and visualization analysis studied in this work.  {Anisotropic FRI algorithms are proposed via two different ways. } The fFRI algorithm is developed by using  appropriate data structures to avoid the evaluation of insignificant correlation matrix elements, which leads to a sparse fFRI matrix. In Section \ref{sec:Numerical}, we first analyze the behavior of a few FRI correlation functions.  Additionally, we examine the parameter dependence of some FRI correlation functions.  Moreover,  the performance of fFRI is investigated. Further, we provide a comprehensive comparison of our FRI and other established methods.  We adopted three protein sets corresponding to relatively small-, medium-, and large-sized structures, proposed in the literature  \cite{JKPark:2013}. We also utilize an extended set of 365 proteins to further evaluate the performance of various methods.  Furthermore, the computational complexities of FRI, fFRI and GNM are compared over a set of 44 proteins.
Finally,  we show that the FRI offers a distinguished  visualization of biomolecular structure and interaction.  In Section \ref{Sec:Domains}, we demonstrate the usefulness of  { the FRI and aFRI by carrying out an in-depth study of  protein domains}.  This paper ends with concluding remarks.

\section{Theory and algorithm}\label{sec:methods}
 {
In the continuum elasticity with atomic rigidity (CEWAR) model, the dynamics of a biomolecular system under the given force ${\bf f}$ is governed by  the equation of motion  \cite{KLXia:2013d}
\begin{eqnarray}\label{eqn:elasticdynamics}
{\rho{\bf \ddot{\bf w}}
=
  \left[ \nabla \lambda \nabla\cdot {\bf w} + \nabla \mu \cdot [\nabla{\bf w} + (\nabla {\bf w})^T] + (\lambda +\mu )\nabla \nabla\cdot {\bf w}
  + \mu \nabla^2 {\bf w} \right]
+{\bf f},}
\end{eqnarray}
where $\rho$ is the density of the macromolecule, ${\bf w}$ is the displacement,   and $\lambda =\lambda ({\bf r})$ and $\mu =\mu ({\bf r})$ are respectively bulk modulus  and shear modulus. The FRI algorithm was proposed to evaluate   the shear modulus,  i.e., rigidity. }

   {
This section describes the theory and algorithm underpinning  the FRI method. We first briefly review the FRI theory to establish notation. Then, two anisotropic FRI algorithms are introduced for the analysis of the anisotropic motions of biomolecules. Finally,  
A fast FRI algorithm is proposed to reduce the computational complexity of the original FRI.}

\subsection{Flexibility-rigidity index}\label{sec:Flexibility}

We consider proteins as examples to illustrate our FRI algorithm, although other biomolecules, such as DNA and RNA, can be similarly treated with a minor modification of our algorithm.  We are particularly interested in a coarse-grained representation. However, methods for a full atom description can be formulated as well.

We seek a structure based algorithm to convert protein geometry into protein topology. To this end, we consider a protein with $N$  C$_\alpha$ atoms. Their locations are represented by $\{ {\bf r}_{j}| {\bf r}_{j}\in \mathbb{R}^{3}, j=1,2,\cdots,N\}$. We denote $  \|{\bf r}_i-{\bf r}_j\|$ the Euclidean space distance between  $i$th  C$_\alpha$ atom  and the $j$th C$_\alpha$ atom. The distance geometry of protein C$_\alpha$ atoms   is utilized to establish the topology connectivity by using monotonically decreasing radial basis functions,
\begin{eqnarray}\label{eq:couple_matrix0}
{C}_{ij} =  \Phi( \|{\bf r}_i - {\bf r}_j \|;\eta_{ij}),
\end{eqnarray}
where    $\eta_{ij}$ is a  characteristic distance between particles,   and
 $\Phi( \|{\bf r}_i - {\bf r}_j \|;\eta_{ij}) $ is a   correlation function, which is, in general,  a  real-valued monotonically decreasing function. As a correlation function, it satisfies
\begin{eqnarray}\label{eq:couple_matrix1-1}
\Phi( \|{\bf r}_i - {\bf r}_i \|;\eta_{ii})&=&1 \\
  \Phi( \|{\bf r}_i - {\bf r}_j \|;\eta_{ij})&=&0 \quad {\rm as }\quad  \|{\bf r}_i - {\bf r}_j \| \rightarrow\infty.
\end{eqnarray}
 Delta sequences of the positive type discussed in an earlier work \cite{GWei:2000} are all good choices.  For example,  { one can use generalized exponential  functions
\begin{eqnarray}\label{eq:couple_matrix1}
\Phi(\|{\bf r}_i - {\bf r}_j \|;\eta_{ij}) =    e^{-\left(\|{\bf r}_i - {\bf r}_j \|/\eta_{ij}\right)^\kappa},    \quad \kappa >0
\end{eqnarray}
and  generalized Lorentz functions
\begin{eqnarray}\label{eq:couple_matrix2}
 \Phi(\|{\bf r}_i - {\bf r}_j \|;\eta_{ij}) = \frac{1}{1+ \left( \|{\bf r}_i - {\bf r}_j \|/\eta_{ij}\right)^{\upsilon}},  \quad  \upsilon >0.
 \end{eqnarray}
}
 Essentially,  the correlation between any two particles should decay according to their distance. Therefore, many other alternatives can be used and some of them are investigated in Section \ref{sec:Numerical}.

The  correlation map or cross correlation is an important quantity for the GNM. We can define a similar correlation map by setting
${\bf C}=\{C_{ij}\}, i, j =1,2,\cdots,N$. The correlation map measures the connectivity of C$_\alpha$s in the protein. The similarity and difference of the present correlation map and that of the GNM are studied in   Section \ref{sec:Numerical}.

We define an atomic  rigidity index  $\mu_i$  as the summation of   topological connectivity
\begin{eqnarray}\label{eq:rigidity1}
 \mu_i = \sum_{j=1}^N w_{ij} \Phi( \|{\bf r}_i - {\bf  r}_j \|;\eta_{ij} ), \quad \forall i =1,2,\cdots,N,
\end{eqnarray}
 where $w_{ij}$ is a weight function related to the atomic type,
The atomic rigidity index  $\mu_i$  manifests the rigidity or stiffness at the $i$th atom. In a general sense,  the atomic rigidity index reflects the total interaction strength, including both bonded and non-bonded contributions. It is quite straightforward  to  define the averaged molecular rigidity index  as a summation of atomic rigidity indices
\begin{eqnarray}\label{eq:Averagerigidity}
 \bar{\mu}_{\rm MRI} = \frac{1}{N}\sum_{i=1}^N {\mu}_{i}.
\end{eqnarray}
 The averaged molecular rigidity index can be used to predict   molecular thermal stability, bulk modulus,  density (compactness),  boiling points of isomers, the ratio of surface area over volume,  surface tension,   etc. A detailed investigation of these aspects
 is beyond the scope of the present work.

We are now ready to define a position dependent shear modulus
\begin{eqnarray}\label{eq:rigidity3}
 \mu({\bf r}) & = & \sum_{j=1}^N w_{j}({\bf r}) \Phi( \|{\bf r} - {\bf  r}_j \|;\eta_{ij} ), \quad  {\bf r}\in \Omega_E,
 \end{eqnarray}
where $w_{j}({\bf r})$ is a weight function,  ${\bf r}$ is in the proximity of ${\bf r}_i$  and $\Omega_E$ is the macromolecular domain.   In order to   determine $w_{j}({\bf r})$,  we define an average rigidity (or averaged   rigidity index function) by
\begin{eqnarray}\label{eq:rigidity4}
 \bar{\mu} & = & \frac{1}{V } \int    \mu({\bf r}) d{\bf r},
\end{eqnarray}
where $V$ is the volume of the macromolecule. If $w_{j}({\bf r})$ is a constant, its value can be uniquely determined by  a comparison of $\bar{\mu}$ with experimental  shear modulus \cite{Sept:2010} for a given macromolecule and correlation function.

We also define   an atomic flexibility index as
\begin{eqnarray}\label{eq:flexibility1}
f_i= \frac{1}{\mu_i}, \quad \forall i =1,2,\cdots,N.
\end{eqnarray}
Since the flexibility at each atom is proportional to its temperature fluctuation, we can express B-factors  as
\begin{eqnarray}\label{eq:regression}
 B_i^t = a f_i + b, \quad \forall i =1,2,\cdots,N
\end{eqnarray}
where $ \{B_i^t\}$ are  theoretically predicted B-factors,  and $a$ and $b$ are two  constants to be determined by a simple linear regression.

We can also define the averaged molecular flexibility index (MFI) as a summation of atomic flexibility indices
\begin{eqnarray}\label{eq:Averageflexibility}
 \bar{f}_{\rm MFI} = \frac{1}{N}\sum_{i=1}^N {f}_{i}.
\end{eqnarray}
MFI should correlate with molecular stability and energy.

For the purpose of visualization, we define a continuous atomic flexibility function as
 \begin{eqnarray}\label{eq:flexibility3}
 F({\bf r}) & = &  \sum_{j=1}^N   B_i^t\Psi( \|{\bf r} - {\bf r}_j \|), \quad  {\bf r}\in \Omega_E.
 \end{eqnarray}
where $\Psi( \|{\bf r} - {\bf  r}_j \|)$ is a general interpolation function for scattered data. Wavelets, spline functions, and  modified Shepard's method \cite{Renka:1988,Thacker:2010} can be employed for the interpolation. One can map $f({\bf r})$ to the molecular surface to visualize the protein flexibility  \cite{KLXia:2013d}.  Alternatively, one can compute the continuous atomic flexibility function by
\begin{eqnarray}\label{eq:flexibility4}
F({\bf r}) & = & \frac{1}{\sum_{j=1}^N w_{j}({\bf r}) \Phi( \|{\bf r} - {\bf  r}_j \|;\eta_{ij} )}, \quad  {\bf r}\in \Omega_E.
 \end{eqnarray}

\subsection{Anisotropic flexibility-rigidity index}\label{sec:AnisoFlexibility}

In this section, we propose a new anisotropic model based on our FRI theory. In existing anisotropic methods, the Hessian matrix is always global, i.e., the matrix contains all the $3N\times 3N$ elements for $N$ particles in molecule. In our aFRI model, the Hessian matrix is inherently local and adaptive. Its size may vary from $3\times 3$ for a completely local aFRI to $3N\times 3N$ for a complete global aFRI, depending on the need of a physical problem.

Let us partition all the $N$ particles in a molecule into a total of $M$ clusters $\{c_1, c_2,\cdots,c_k,\cdots, c_M \}$. Cluster $c_k$ has $N_k$ particles or atoms so  that $N=\sum_{k=1}^M N_{k}$. A cluster may be of physical interest, i.e., an alpha helix, a domain, or a binding site of a protein.    One of two extreme cases is that there is only one particle in each cluster. We therefore have $N$ cluster. The other case is that    there is only one cluster, i.e., the whole molecule. The essential idea is to develop a Hessian matrix for each cluster individually without the information about other cluster properties  (However, information for nearby particles outside the cluster is still required). For example, if  we are interested in the thermal fluctuation of a particular cluster $c_k$ with  $N_{k}$ particles or atoms, we can find $3N_k$ eigenvectors for the cluster.  Let us keep in mind that each position vector in $\mathbb{R}^3$ has three components, i.e., ${\bf r}=(x,y,z)$. We denote 
\begin{eqnarray}\label{eq:Anisorigidity1}
 \Phi^{ij}_{uv}  = \frac{\partial}{\partial u_i} \frac{\partial}{\partial v_j} \Phi( \|{\bf r}_i - {\bf  r}_j \|; \eta_{ij} ), \quad  u,v= x, y, z; i,j =1,2,\cdots,N. 
\end{eqnarray}
Note that  for each given $ij$, we define $\Phi^{ij}=\left( \Phi^{ij}_{uv} \right)$ as a local anisotropic matrix 
\begin{equation}
\Phi^{ij}=\left(
\begin{array}{ccc}
\Phi^{ij}_{xx} & \Phi^{ij}_{xy}& \Phi^{ij}_{xz}\\
\Phi^{ij}_{yx} & \Phi^{ij}_{yy}& \Phi^{ij}_{yz}\\
\Phi^{ij}_{zx} & \Phi^{ij}_{zy}& \Phi^{ij}_{zz}  
\end{array}
\right).
\end{equation}

Since rigidity and flexibility can be both anisotropic, it is nature to propose two different aFRI algorithms based on rigidity Hessian  matrix and flexibility Hessian  matrix, respectively.

\subsubsection{Anisotropic rigidity}
 
The anisotropic rigidity is defined by  a  rigidity Hessian matrix for an arbitrary cluster $c_k$.  Let us denote $ \left(\mu_{uv}^{ij}(c_k)\right)$ a  rigidity Hessian  matrix for cluster $c_k$. Its  elements are chosen as 
\begin{eqnarray}\label{eq:Anisorigidity2}
\mu^{ij}_{uv}(c_k) =& - w_{ij}\Phi^{ij}_{uv},                &\quad   i,j \in c_k; i\neq j;  u,v= x, y, z \\ \label{eq:Anisorigidity3}
\mu^{ii}_{uv}(c_k)=&  \sum_{j=1}^N w_{ij} \Phi^{ij}_{uv},  &\quad   i \in c_k;  u,v= x, y, z \\ \label{eq:Anisorigidity4}
\mu^{ij}_{uv}(c_k)=&  0,                                   &\quad   i,j \notin  c_k; u,v= x, y, z. 
\end{eqnarray}
Hessian  matrix  $ \left(\mu_{uv}^{ij}(c_k)\right)$ is of  $3N_k\times 3N_k$ dimensions. Note that the diagonal part, $\mu^{ii}_{uv}(c_k)$, has built in information from all the particles in the system, even if the cluster is completely localized, i.e.,  $N_k=1,~ \forall k$.

An immediate test of the anisotropic rigidity is to check if it works for the B-factor prediction. To this end, we  collect the diagonal terms of the rigidity Hessian  matrix 
\begin{eqnarray}\label{eq:Anisorigidity33}
\mu^{i}_{\rm diag} &=& {\rm Tr}\left(\mu_{uv}^{i}\right)\\
                    &=&      \sum_{j=1}^N w_{ij} \left[\Phi^{ij}_{xx}+\Phi^{ij}_{yy}+   \Phi^{ij}_{zz}\right].  
\end{eqnarray}
We then define a set of anisotropic rigidity  (AR) based flexibility indices by
\begin{eqnarray}\label{eq:Anisorigidity34}
f_{i}^{\rm AR} =  \frac{1}{\mu^{i}_{\rm diag}}.
\end{eqnarray}
B-factors can be predicted with a set of $\{f_{i}^{\rm AR}\}$ by using the linear regression in Eq. (\ref{eq:regression}).

\subsubsection{Anisotropic flexibility}
 
To analyze biomolecular anisotropic motions in parallel to ANM, we need to examine their anisotropic flexibility. To this end, we further define a flexibility Hessian  matrix ${\bf F}(c_k)$ for cluster $c_k$ as  
\begin{eqnarray}\label{eq:Anisoflexibility}
{\bf F}^{ij}(c_k)     =&  - \frac{1}{w_{ij}} (\Phi^{ij})^{-1},                &\quad   i,j \in c_k; i\neq j;  u,v= x, y, z \\ \label{eq:Anisoflexibilityy3}
{\bf F}^{ii}(c_k)=&   \sum_{j=1}^N \frac{1}{w_{ij}} (\Phi^{ij})^{-1},  &\quad   i \in c_k;  u,v= x, y, z \\ \label{eq:Anisoflexibility4}
{\bf F}^{ij}(c_k)=&  0,                                     &\quad   i,j \notin  c_k; u,v= x, y, z. 
\end{eqnarray}
where $(\Phi^{ij})^{-1}$ denote the unscaled inverse of matrix $\Phi^{ij}$ such that $\Phi^{ij}(\Phi^{ij})^{-1}=| \Phi^{ij}|$. Similar to the anisotropic rigidity, the diagonal part ${\bf F}^{ii}(c_k)$ has built in information from all particles in the system. Therefore, even if the partition of clusters is completely localized (i.e., $N$ clusters), certain correlation among atomic motions is retained. By diagonalizing ${\bf F}(c_k)$, we obtain $3N_k$ eigenvectors   for the $N_k$ particles in cluster $c_k$.   Since the selection of $c_k$ is arbitrary, eigenvectors of all other clusters can be attained using the same procedure.

To obtain the B-factor prediction  from this anisotropic flexibility, we   define a set of anisotropic flexibility (AF) based flexibility indices by
\begin{eqnarray}\label{eq:Anisoflexibility2}
f_i^{\rm AF} &=&{\rm Tr} \left({\bf F}(c_k)\right)^{ii},   \\
                &=&  \left({\bf F}(c_k)\right)^{ii}_{xx}+ \left({\bf F}(c_k)\right)^{ii}_{yy}+ \left({\bf F}(c_k)\right)^{ii}_{zz}.
\end{eqnarray}
Then Eq. (\ref{eq:regression}) is employed to obtain B-factor predictions.

In this work, we only consider the coarse-grained model in which each residue is represented by its C$_\alpha$. To further simply the model, the differences between residues are ignored. The parameter $w_{ij}$ is assumed to be 1 and $\eta_{ij}$ is set to a constant $\eta$.

\subsection{Fast FRI algorithm}\label{sec:SpeedUp}

As discussed in our earlier work \cite{KLXia:2013d}, the original FRI algorithm has the computational complexity of ${\cal O}(N^2)$, mainly due to the construction of the correlation matrix. In the present work, we propose a fast  FRI (fFRI) algorithm, which computes only the significant elements of the correlation matrix and at the same time maintains the accuracy of our method. As a result, the computational complexity  of our  fFRI algorithm is of ${\cal O}(N)$.

The essential idea is to partition the residues in a protein into cubic boxes according to their spatial locations. For each residue in a given box, we only compute its correlation matrix elements with all residues within the given box and with all residues in the adjacent 26 boxes. The accuracy and efficiency of this approach are determined by the box dimension. We select a box size of $R$ such that
\begin{eqnarray}\label{eq:trancationSize}
\Phi(R;\eta) \leq  \varepsilon
\end{eqnarray}
where $\varepsilon>0 $ is a given truncation error.  Therefore, for   generalized exponential  functions
 (\ref{eq:couple_matrix1}), we have
\begin{eqnarray}\label{eq:trancationSize1}
R \geq    \eta \left(\ln \frac{1}{\varepsilon  }\right)^{\frac{1}{\kappa}}.
\end{eqnarray}
If we set $\varepsilon=10^{-2}$,  we have $R\approx 4.6 \eta$ for $\kappa=1$ and
$R\approx 2.15 \eta$ for $\kappa=2$. Note that different $\kappa$ values have different  optimal $\eta$ values. The higher the $\kappa$ value is, the
larger the optimal $\eta$ is.

Similarly, for  generalized Lorentz functions  (\ref{eq:couple_matrix2}), we choose the box size
\begin{eqnarray}\label{eq:trancationSize2}
R \geq    \eta \left(\frac{1-\varepsilon}{\varepsilon  }\right)^{\frac{1}{\upsilon}}.
\end{eqnarray}
Again, if we set $\varepsilon=10^{-2}$,  we have $R\approx 10 \eta$ for $\upsilon=2$ and $R\approx 4.6 \eta$ for $\upsilon=3$.

An optimal $R$ should balance the accuracy and efficiency.  In Section \ref{sec:fFRIparameters}, it is found that the selection of $R=12$\AA~ is  near optimal for both  exponential  and Lorentz functions. In Algorithm \ref{Alg}, we present a pseudocode to illustrate  the  truncation algorithm  of the fFRI.

\begin{algorithm}
\caption{fFRI algorithm}\label{AFRI}
\begin{algorithmic}
 \State \textbf{Input: } $ atoms(N)$  \Comment XYZ coordinates from PDB file
\vspace{3 mm}
 \State $ mincoor \gets minval(atoms)$ \Comment Compute dimensions of bounding box
 \State $ maxcoor \gets maxval(atoms)$
 \State $R \gets boxsize$ \Comment Set size of grid
 \State $Nbox \gets ceiling((maxcoor-mincoor)/R)$ \Comment Compute number of boxes in each direction
\vspace{3 mm}
  \For{ $ii\gets 1,Natoms$ }
			\State $i,j,k \gets ceiling((atoms(ii)-mincoor/R)) $ \Comment Count the number of atoms in each box
			\State $Natoms(i,j,k) \gets Natoms(i,j,k) + 1 $
  \EndFor
	\vspace{3 mm}
	 \For{ $k\gets 1,Nbox[3]$ }
	\For{ $j\gets 1,Nbox[2]$ }
	\For{ $i\gets 1,Nbox[1]$ }
 \State $allocate(box(i,j,k))$   \Comment Allocate space for each box
   \EndFor
	  \EndFor
		 \EndFor
\vspace{3 mm}
  \For{ $ {ii}\gets 1,Natoms$ }   \Comment  Copy coordinates to appropriate box based on 3D coordinates
			\State $i,j,k \gets ceiling((atoms(ii)-mincoor)/R) $
			\State $box(i,j,k) \gets atoms(ii) $
  \EndFor
	\vspace{3 mm}
	  \For{ $k\gets 1,Nbox[3]$ }   \Comment  Iterate over boxes
			  \For{ $j\gets 1,Nbox[2]$ }
					  \For{ $i\gets 1,Nbox[1]$ }
							\vspace{3 mm}
							
							  \For{ ${n_a}\gets 1,Natoms(i,j,k)$ }   \Comment  Iterate over atoms in current box
														\vspace{3 mm}
						  \For{ $n\gets k-1,k+1$ }   \Comment  Iterate over adjacent boxes
			  \For{ $m\gets j-1,j+1$ }
					  \For{ $l\gets i-1,i+1$ }
							\vspace{3 mm}
											  \For{ $ {n_b}\gets 1,Natoms(l,m,n) $ }   \Comment Iterate over atoms in adjacent boxes
			\State $ dist \gets distance(box(i,j,k)(n_a),box(l,m,n)(n_b)) $
			\State $FRI(n_a) \gets kernel(dist) $
			
				\vspace{3 mm}
				\EndFor
				
				\vspace{3 mm}
  \EndFor
	  \EndFor
		  \EndFor
			\vspace{3 mm}
			  \EndFor
							\vspace{3 mm}
  \EndFor
	  \EndFor
		  \EndFor
 \end{algorithmic}
\label{Alg}
  \end{algorithm}

\section{Numerical experiments}\label{sec:Numerical}

In this section, we validate the   FRI approach for protein B-factor prediction by a comparison of its performance with that of two established methods, namely   NMA    and GNM. We consider the accuracy, reliability  and efficiency of these methods. It is well known that the computational complexity of  matrix diagonalization is asymptotically close to ${\cal O}(N^3)$, while that  of correlation map   construction and two-parameter linear regression given in Eq. (\ref{eq:regression}) is asymptotically ${\cal O}(N^2)$.  The computational complexity of  the proposed fFRI algorithm is further reduced to ${\cal O}(N)$. Therefore, there is a dramatic reduction in the computational complexity. We demonstrate that the FRI method outperforms other methods in computational efficiency and is potentially useful for the flexibility analysis of excessively large macromolecules.

To test FRI against GNM and NMA, five sets of structures are utilized. Among them, three sets were used by Park,  Jernigan and  Wu  in their comparative study \cite{JKPark:2013}. These include  relatively small-, medium- and large-sized sets of structures. A fourth set of 44 structures  was created to test the efficiency of each algorithm. This set was created by randomly selecting protein-only structures from the Protein Data Bank database with varying size. The number of residues for proteins in this set range from 125 to 313,236 residues. The final set, called a superset, is a combination of sets including the three sets used by Park el al. \cite{JKPark:2013}, the first 40 structures of the efficiency set and a set of 263 high-resolution structures used in earlier tests of the FRI method. \cite{KLXia:2013d} The total number of structures in the superset is 365 after the removal of duplicate structures.

To quantitatively  assess the performance of the proposed FRI model for the B-factor  prediction, we  consider the correlation coefficient
\begin{eqnarray}\label{correlation}
   C_c=\frac{\Sigma^N_{i=1}\left(B^e_i-\bar{B}^e \right)\left( B^t_i-\bar{B}^t \right)}
   { \left[\Sigma^N_{i=1}(B^e_i- \bar{B}^e)^2\Sigma^N_{i=1}(B^t_i-\bar{B}^t)^2\right]^{1/2}},
\end{eqnarray}
where $\{B^t_i,  i=1,2,\cdots,N\}$ are a set of predicted B-factors by using the proposed method and $\{B^e_i, i=1,2,\cdots, N\}$ are a set of experimental B-factors read from the PDB file. Here $\bar{B}^t$ and $\bar{B}^e$ the statistical averages of theoretical and experimental B-factors, respectively.

\subsection{Analysis of FRI correlation functions }\label{sec:kernels}

\begin{table}[htbp]
  \centering
	\renewcommand{\arraystretch}{1.3}
		\caption{ Comparison of average correlation coefficients computed from various correlation functions.
		Each function was tested across a range of parameters and the best score was saved for each structure and used to calculate the average over a set of 365 structures. }
    \begin{tabular}{cccc}
    \toprule
    Correlation Function & Parameter Range & Average Correlation Coefficient \\
    \midrule
		$\begin{array} {lcl} e^{-(r/\eta)^\kappa}  \end{array}$ & $\begin{array} {lcl}  1.0 \leq \eta \leq 10.0 & 0.5 \leq \kappa \leq 10.0 \end{array}$ & 0.676  \\
    $\begin{array} {lcl} \frac{1}{1+(r/\eta)^\upsilon}  \end{array}$ & $\begin{array} {lcl} 1.0 \leq \eta \leq 10.0 & 0.5 \leq \upsilon \leq 10.0 \end{array}$ & 0.673  \\
		$\begin{array} {lcl} \frac{1}{1+(r/\eta)^\upsilon} e^{-(r/\eta)^\kappa}  \end{array}$ & $\begin{array} {lcl}  1.0 \leq \eta, \upsilon, \kappa \leq 10.0 \end{array}$ & 0.670  \\
		$\begin{array} {lcl} \frac{1}{\sqrt{1+ (r/\eta)^\upsilon}} \end{array}$ & $\begin{array} {lcl}  1.0 \leq \eta, \upsilon \leq 10.0 \end{array}$ &  0.577  \\
    \bottomrule
    \end{tabular}
  \label{othereq}%
\end{table}%

In order to further explore the FRI method, we test four types of correlation functions. Apart from the Lorentz and exponential functions, two alternative functions are employed in our study.  All correlation functions equal to unit at the origin and are monotonically decreasing with respect to an increasing distance ($r$). 
 Each correlation function  is tested with a range of parameter values for each of   365 structures  as listed in Table \ref{othereq}. The performance of the new correlation functions comes close to that of the exponential and Lorentz functions with the product of these two having the highest average correlation coefficient among alternative functions.

\subsection{Analysis of fFRI algorithms  }\label{sec:fFRIparameters}

\begin{figure}[ht!]
\begin{center}
\begin{tabular}{cc}
		\includegraphics[width=0.4\textwidth]{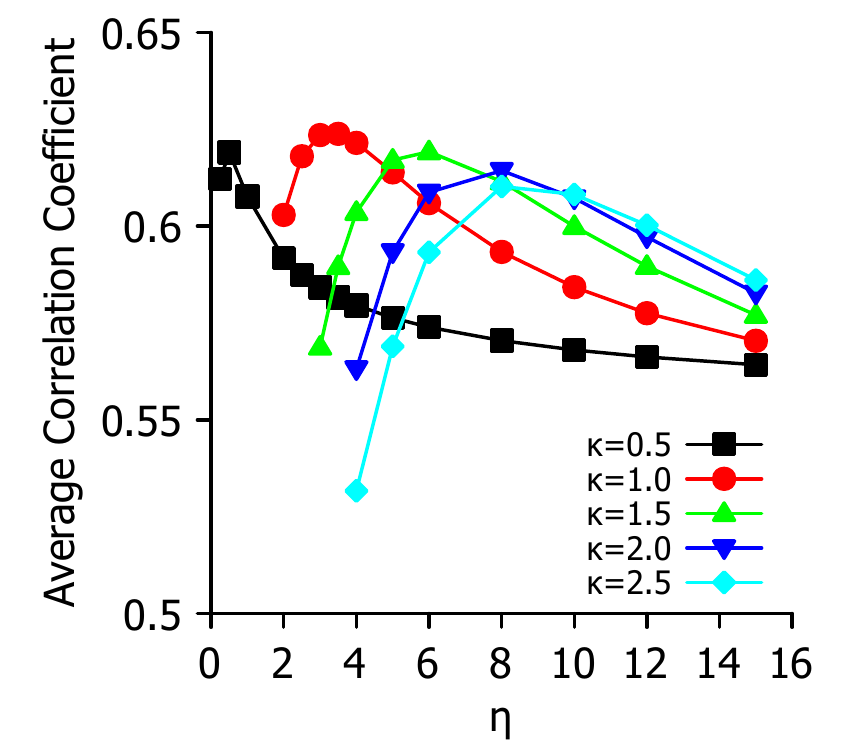}
		\includegraphics[width=0.4\textwidth]{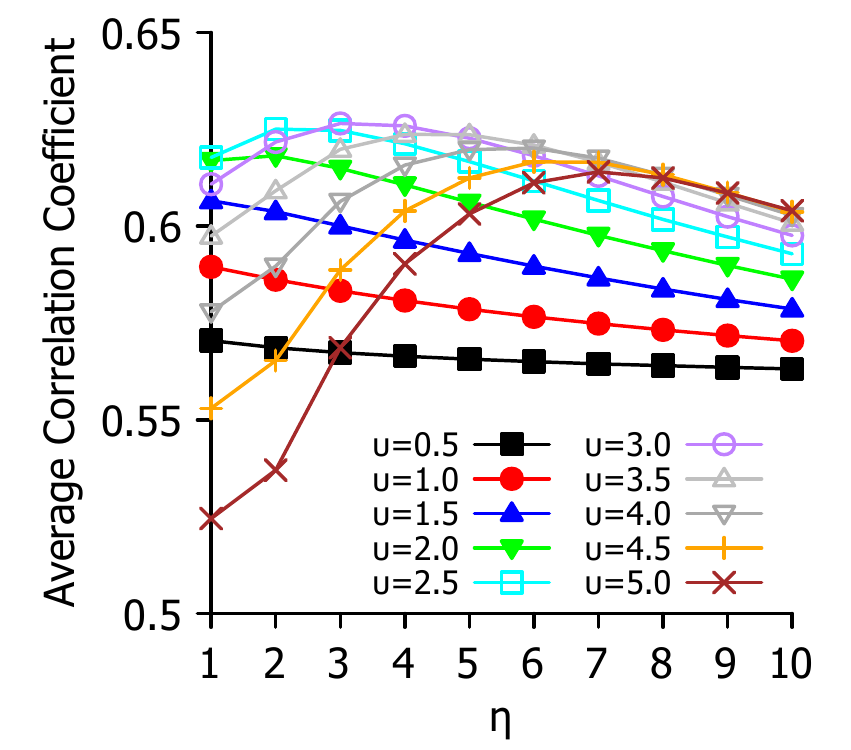}
\end{tabular}
\end{center}
\caption{ Parameter  testing for exponential (Left chart) and Lorentz (Right chart) functions. Average correlation coefficient of B-factor predctions of 365 proteins is plot against choice of $\eta$ for a range of values for $\kappa$ or $\upsilon$.  }
\label{funcs2}
\end{figure}

To analyze the best parameter for  Lorentz and exponential functions, we  study their behavior in Fig. \ref{funcs2}, where each function is tested over a range of parameters. For exponential  type of functions, $\kappa=1 $ and $\eta=3$\AA~ give rise to a near optimal parameter-free FRI. Similarly, for Lorentz type of functions, $\upsilon=3$, and $\eta=3$\AA~ offer near optimal results. It is seen from Fig. \ref{funcs2} that
 exponential functions are quite sensitive to $\eta$ values, while Lorentz functions are relatively robust with respect to $\eta$.  This study provides a basis for the selection of parameter free FRI (pfFRI) schemes.

\begin{figure}[ht!]
\begin{center}
\begin{tabular}{cc}
		\includegraphics[width=0.4\textwidth]{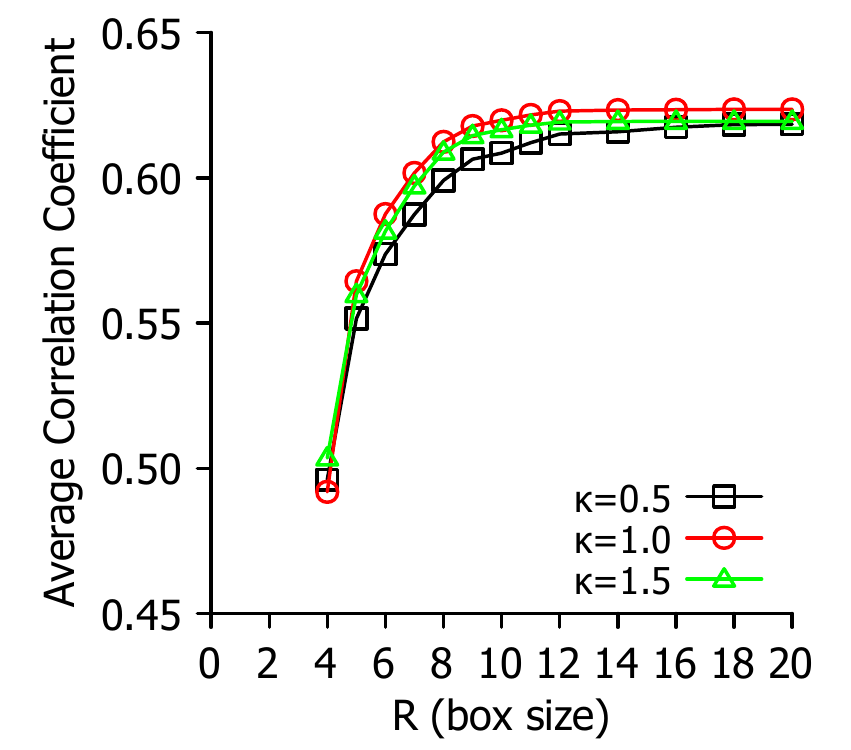}
		\includegraphics[width=0.4\textwidth]{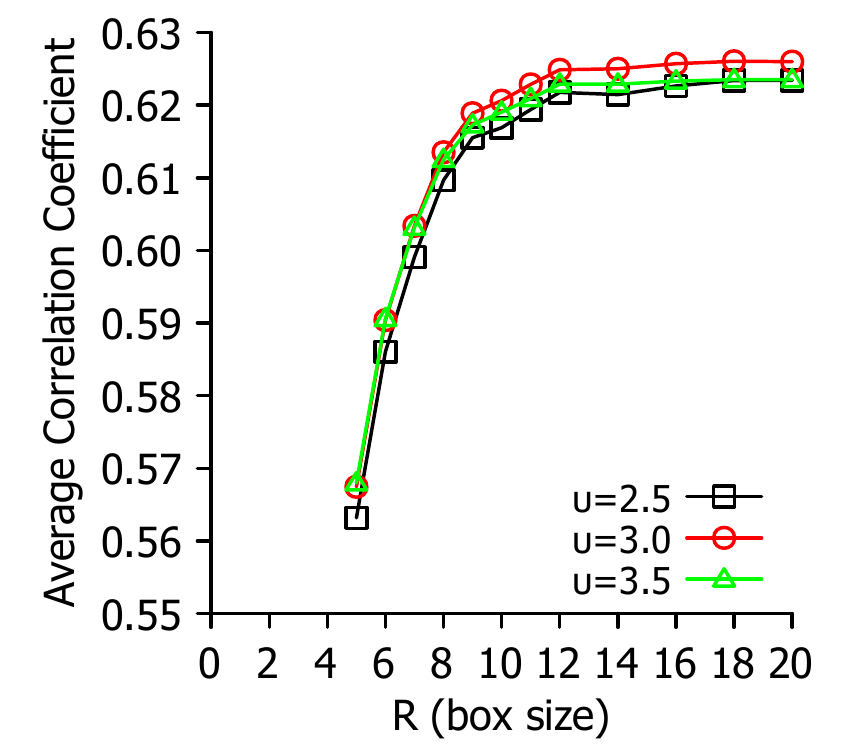}
\end{tabular}
\end{center}
\caption{ The impact of box size to the average correlation coefficient   for a set of 365 proteins. The fFRI is examined over a range of values for parameters ({$\kappa$ and $\upsilon$}) to illustrate the relationship between accuracy and choice of box size $R$.  }
\label{boxsize}
\end{figure}

It is interesting to analyze the performance of the proposed fFRI in terms of accuracy and efficiency.  To this end, we first explore the impact of box size to the correlation coefficients of a few fFRI schemes in Fig. \ref{boxsize}.  For each given $\kappa$ and $\upsilon$, the best $\eta$ found in Fig.  \ref{funcs2} is employed. It is seen from  Fig. \ref{boxsize} that  both exponential and Lorentz types of functions are able to achieve their near optimal performance at $R=12$\AA~.  Therefore, we recommend  $R=12$\AA, $\eta=3$\AA~ and $\kappa=1$ for the exponent type of fFRI method. Similarly,  $R=12$\AA, $\eta=3$\AA~  and  $\upsilon=3$ are  near optimal for Lorentz type of fFRI methods.

\subsection{Comparison of B-factor predictions  }\label{sec:B-factor}

\subsubsection{ FRI  vs GNM and NMA}\label{sec:B-factor1}

\begin{figure}[ht!]
\begin{center}
\begin{tabular}{cc}
\includegraphics[width=0.4\textwidth]{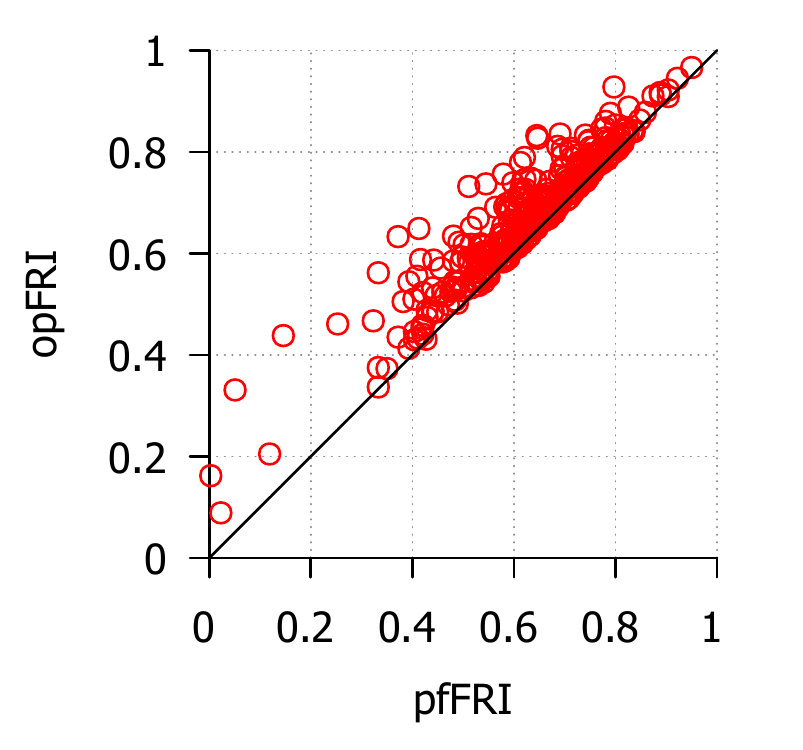}
\includegraphics[width=0.4\textwidth]{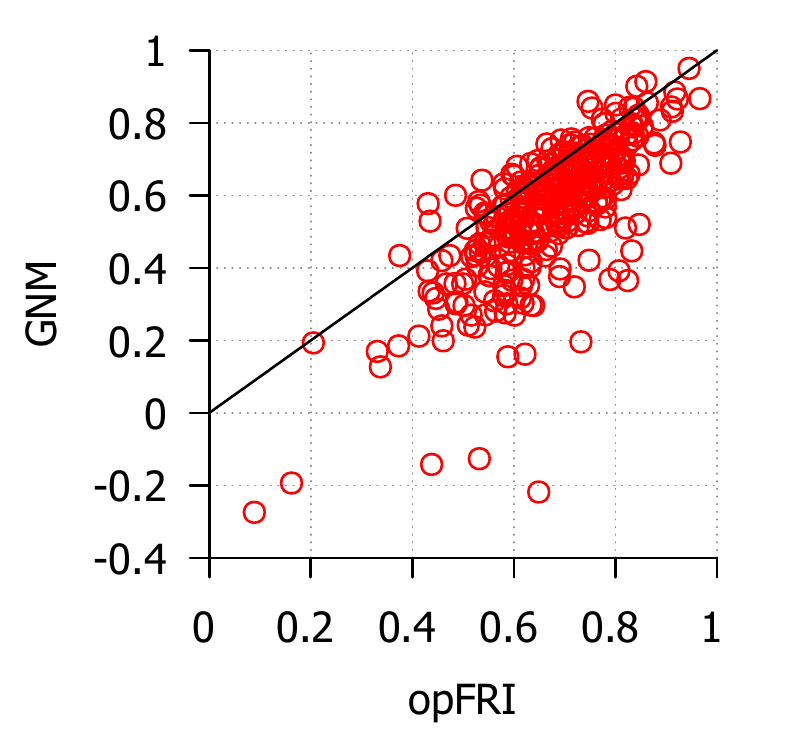}
\end{tabular}
\begin{tabular}{c}
\includegraphics[width=0.4\textwidth]{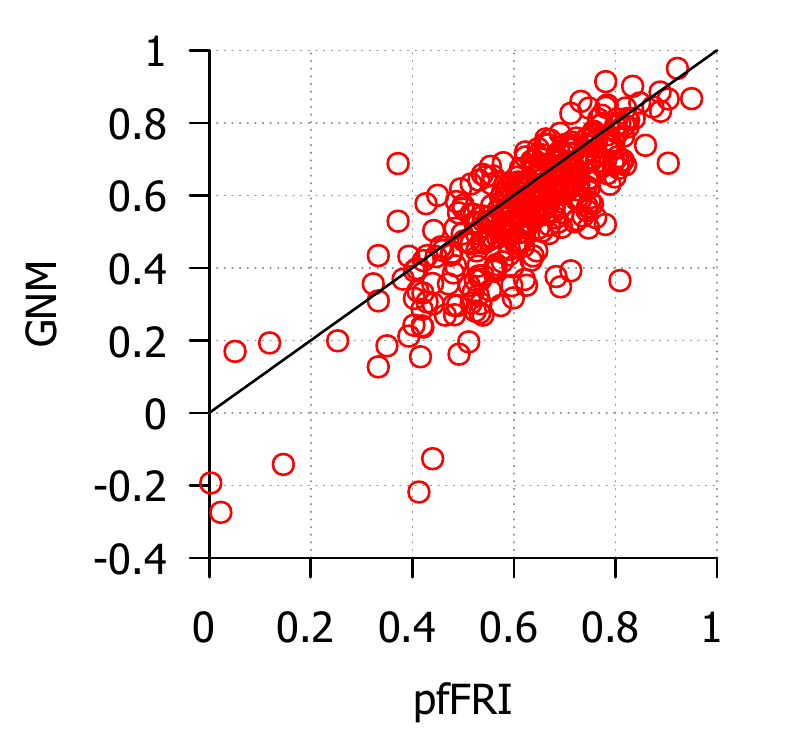}
\includegraphics[width=0.4\textwidth]{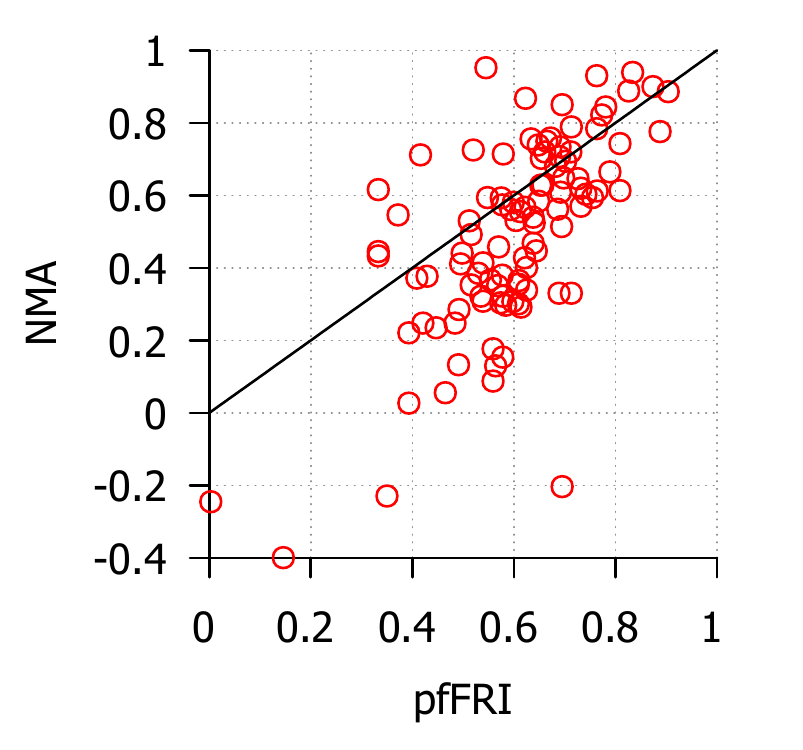}
\end{tabular}
\end{center}
\caption{ Comparison of correlation coefficients from B-factor prediction using  GNM, coarse-grained (C$_\alpha$) NMA and FRI methods.
Top left:     pfFRI vs opFRI for 365 proteins;
Top right:    opFRI vs GNM for 365 proteins;
Bottom left:  pfFRI vs GNM for 365 proteins;
Bottom right: pfFRI vs NMA for three sets of proteins used by Park et al.   \cite{JKPark:2013}.
The correlation coefficients for NMA are adopted from Park et al.   \cite{JKPark:2013} for three sets of proteins.
For optimal FRI,  parameter  $\upsilon$ is optimized for a range from 0.1 to 10.0.
For the parameter free version of the FRI (pfFRI), we set  $\upsilon =3$ and $\eta=3$\AA.
The line $y=x$ is included to aid in comparing scores. }
\label{accuracycomp}
\end{figure}

In order to compare the FRI and GNM, we re-analyzed the structures from Park et al. \cite{JKPark:2013} with the GNM method with a cutoff value of 7 \AA, the same value used by the authors. It was found that some correlation coefficients were artificially low for GNM due to multiple coordinates for some C$_\alpha$ atoms in some PDB data and missing C$\alpha$ atoms in others. To ensure a fair comparison between the FRI and GNM we re-analyzed the structures using GNM after processing the PDB files to fix these issues. We removed all but the highest occupancy coordinates for each atom and used every C$\alpha$ atom from the original PDB files to run the GNM B-factor prediction code and calculate corrected correlation coefficients. In Tables \ref{small_table}, \ref{medium_table} and \ref{large_table}, optimal and parameter free FRI is compared to the GNM data reported by Park et al. \cite{JKPark:2013} The newly calculated correlation coefficient is shown only if there is a significant improvement using our processed PDB files. On the other hand, Table \ref{long_table} lists all correlation coefficients for GNM from our own tests using our processed PDB files. These correlation coefficients are typically the same as those reported by Park et al. \cite{JKPark:2013} although some have changed. The use of our processed PDB files leads to a slight increase in the average scores for the GNM in our analysis.

\newcolumntype{L}[1]{>{\raggedright\let\newline\\\arraybackslash\hspace{0pt}}m{#1}}
\newcolumntype{C}[1]{>{\centering\let\newline\\\arraybackslash\hspace{0pt}}m{#1}}
\newcolumntype{R}[1]{>{\raggedleft\let\newline\\\arraybackslash\hspace{0pt}}m{#1}}

\begin{table}[htbp]
  \centering
			\caption{ Average correlation coefficients for C$_\alpha$ B-factor prediction with FRI, GNM and NMA for three structure sets from Park et al. \cite{JKPark:2013} and a superset of 365 structures.   }
 \begin{tabular}{lC{2.0cm}C{2.0cm}C{2.0cm}C{2.0cm}}
    \toprule
PDB set & opFRI & pfFRI & GNM &  NMA \\
    \midrule
Small & 0.667 & 0.594 & 0.541 & 0.480 \\
Medium & 0.664 & 0.605 & 0.550 & 0.482 \\
Large & 0.636 & 0.591 & 0.529 & 0.494 \\
Superset & 0.673 & 0.626 & 0.565 & NA \\
    \bottomrule
    \end{tabular}%
  \label{avgtable}%
\end{table}

\newcolumntype{a}{>{\columncolor{Gray}}c}
\newcolumntype{b}{>{\columncolor{white}}c}

To directly compare the FRI with GNM and NMA, we calculated the correlation coefficient of C$_\alpha$ B-factor predictions for the three structure sets taken from Park et al. \cite{JKPark:2013}  To further compare the FRI and GNM, we also calculated the accuracy of these two methods on a superset of 365 structures. Two versions of the FRI are used for these tests. The first, optimal FRI (opFRI), searches a wide range of parameters for the highest scoring parameter and the second, parameter free FRI (pfFRI), uses $\upsilon=3$ and $\eta=3$\AA~ in all cases. The correlation coefficients for three  sets proposed by Park et al. are reported in Tables \ref{small_table}, \ref{medium_table} and \ref{large_table} for FRI, GNM and NMA. The results of the B-factor predictions for the superset are shown in Fig. \ref{accuracycomp}. 
 {Using the top left chart as an example,  both axises are correlation coefficients. For each circle, its $x$-coordinate is its correlation coefficient for  pfFRI, while its $y$-coordinate is its correlation coefficient for  opFRI. Since all circles are located above the diagonal line, opFRI always  outperform pfFRI.}   
The average correlation scores for optimal FRI, parameter free FRI, GNM and NMA for each set of structures are listed in Table \ref{avgtable}. As shown in  Table \ref{avgtable} and Fig. \ref{accuracycomp}, opFRI outperforms pfFRI in many cases although the majority of structures have little difference in their score for each method. Both optimal and parameter free FRI methods outperform GNM and NMA for most structures.  B-factor prediction with the FRI is most accurate for smaller structures (<70 residues). All three methods tend to perform worse as the structures get larger except in the case of NMA where the medium-sized structures scored slightly lower than the large-sized structures. This behavior is expected because as proteins get larger their structures become more complex and may include structural co-factors and more amino acid side chain interactions that contribute to the protein's stability. The coarse-grained C$_\alpha$ representation used in these methods is unable to capture these kinds of details. The average increase in correlation coefficients when using the FRI over GNM on the superset of 365 proteins is 0.096 for opFRI and 0.059 for pfFRI. Additionally, opFRI and pfFRI are more accurate on average than GNM and NMA for all three sets of structures used by Park et al. \cite{JKPark:2013}  From these results we conclude that both FRI and pfFRI are more accurate on average than either GNM or NMA.

\subsubsection{fFRI  vs GNM }\label{sec:B-factor2}

\begin{table}[htbp]
  \centering
\caption{ Average correlation coefficients (CC) of B-factor prediction for a set of 365 proteins using fFRI ($R=12$).  The improvements of the fFRI over the GNM
prediction (0.565) are given in parentheses. }
\begin{tabular}{clcc}
\toprule
		 Exponential parameters      & Avg. CC &  Lorentz parameters   & Avg. CC \\
			\midrule
$\kappa$=0.5, $\eta$=0.5 & 0.615 (8.8\%) & $\upsilon$=2.5, $\eta$=2.0 & 0.622  (10.1\%) \\
$\kappa$=1.0, $\eta$=3.0 & 0.623 (10.3\%) & $\upsilon$=3.0, $\eta$=3.0 & 0.626  (10.8\%) \\
$\kappa$=1.5, $\eta$=6.0 & 0.619 (9.6\%) & $\upsilon$=3.5, $\eta$=4.0 & 0.623 (10.3\%) \\
\bottomrule
\end{tabular}
\label{ACC}
\end{table}

 Table  \ref{ACC} lists  the average correlation coefficients of B-factor prediction for 365 proteins using fFRI schemes at a given truncation ($R=12$\AA).  It is seen that the proposed fFRI schemes implemented in either exponential ($\eta=3 $\AA~   and  $\kappa=1$)  or Lorentz  ($\eta=3$\AA~  and  $\upsilon=3$) are at least 10\% more accurate than the GNM.

\subsection{Efficiency comparison for FRI, fFRI and GNM }\label{sec:Efficiency}

This section concerns the computational efficiency of the FRI method. The efficiency of the FRI  and the fFRI  is compared with that of the GNM in this section.

\begin{figure}[ht!]
\begin{center}
\begin{tabular}{cc}
\includegraphics[width=0.45\textwidth]{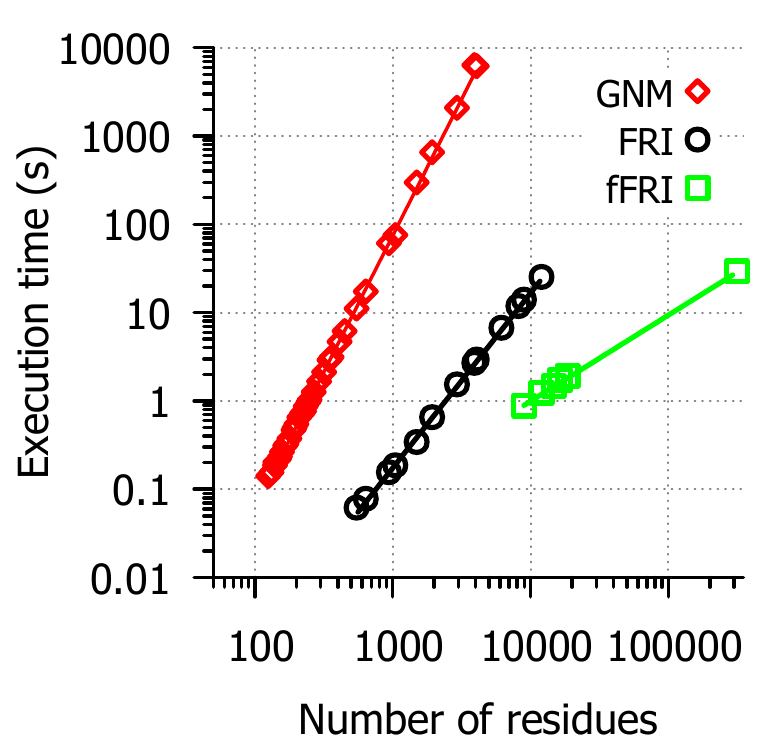}
\end{tabular}
\end{center}
\caption{Efficiency comparison between FRI algorithms and GNM.
Execution time in second (s)  vs. number of residues for FRI (circle), fFRI (square) and GNM (diamond). A set of 44 C$_\alpha$ only PDB files was used to evaluate the computational complexity of GNM,  FRI and fFRI. Available correlation coefficient values are listed in Table \ref{eff_table}.
}
\label{efcomp}
\end{figure}

Computational efficiency in the B-factor prediction becomes important for large proteins and for repeated predictions in molecular dynamics simulation and flexible docking analysis. High efficiency in the rigidity analysis is also a  requirement for CEWAR dynamics, where atomic rigidity functions are to be evaluated during the time evolution.  The previously described set of 44 proteins as listed in Table  \ref{eff_table} are used to test the computational complexity of the FRI, fFRI and GNM algorithms. The method used to obtain the structure of the HIV virus  capsid, which has more than 313,000 amino acid residues, does not provide experimental B-factors. To ensure a fair test, we have added some random noise to the predicted B-factors. The resulting B-factors  of the HIV structure are employed in our efficiency test as if they were real experimental data.
	
\begin{figure} 
\begin{center}
\begin{tabular}{ccc}
\includegraphics[width=0.333\textwidth]{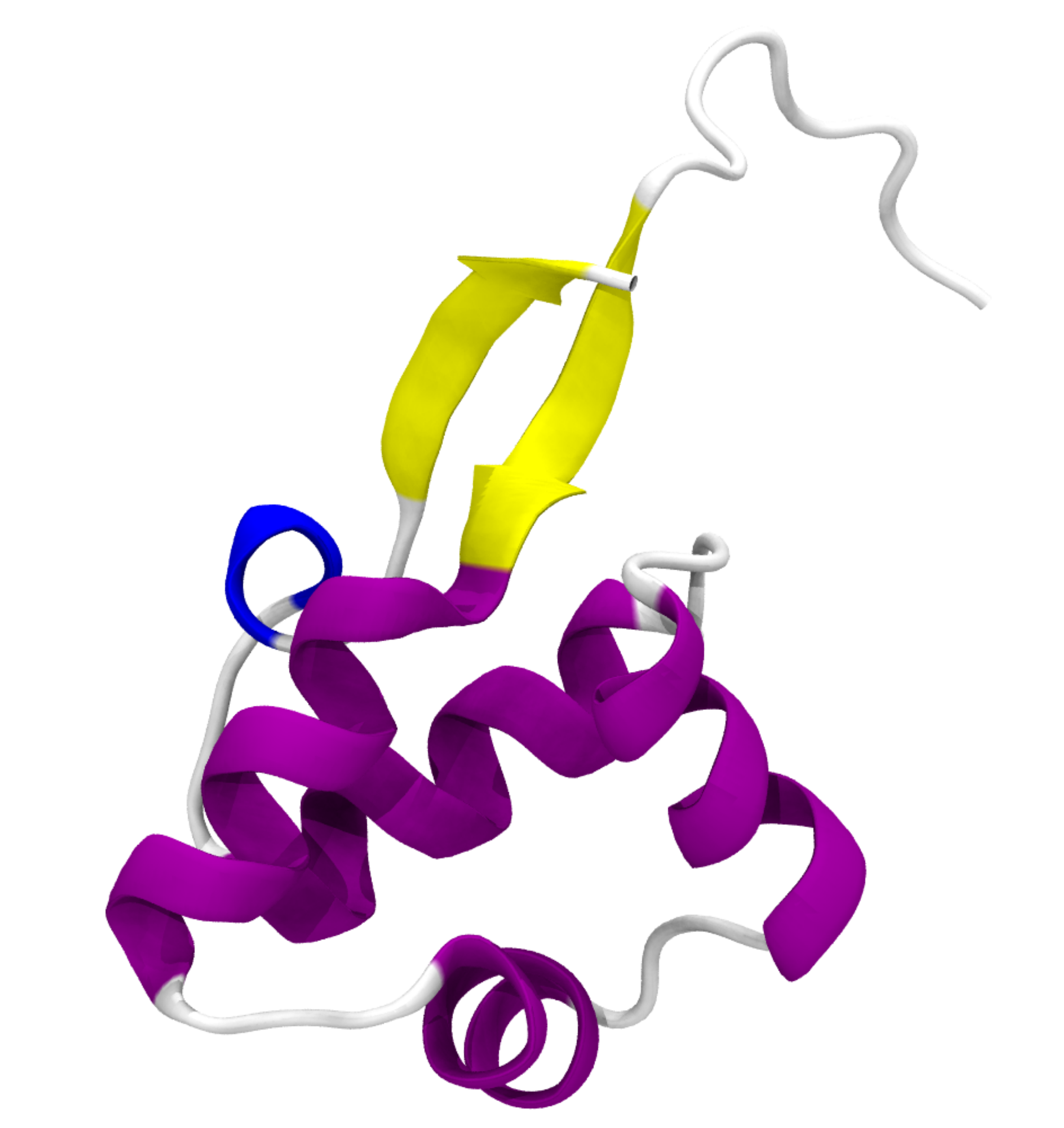}
\includegraphics[width=0.333\textwidth]{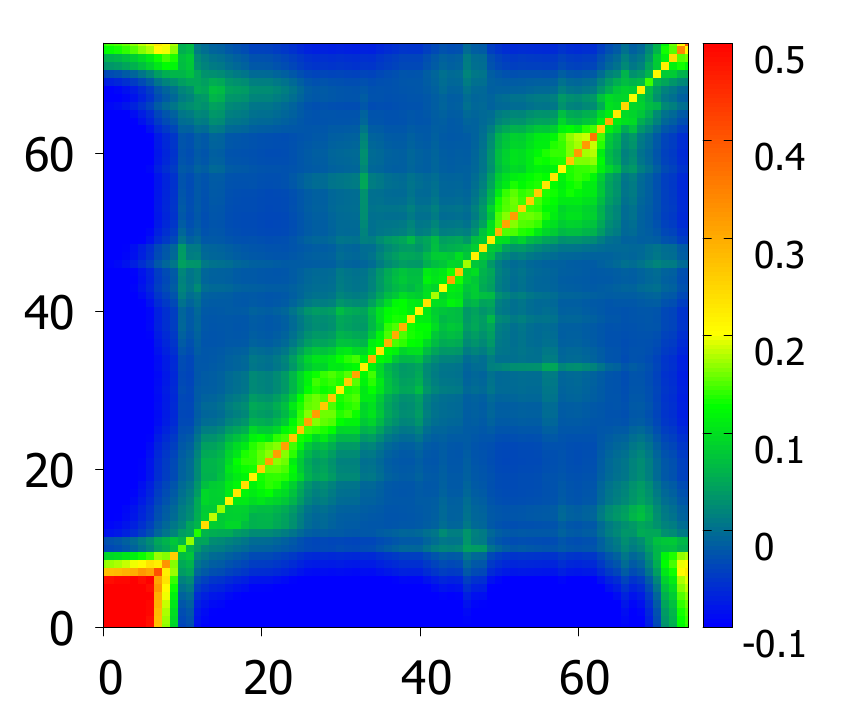}
\includegraphics[width=0.333\textwidth]{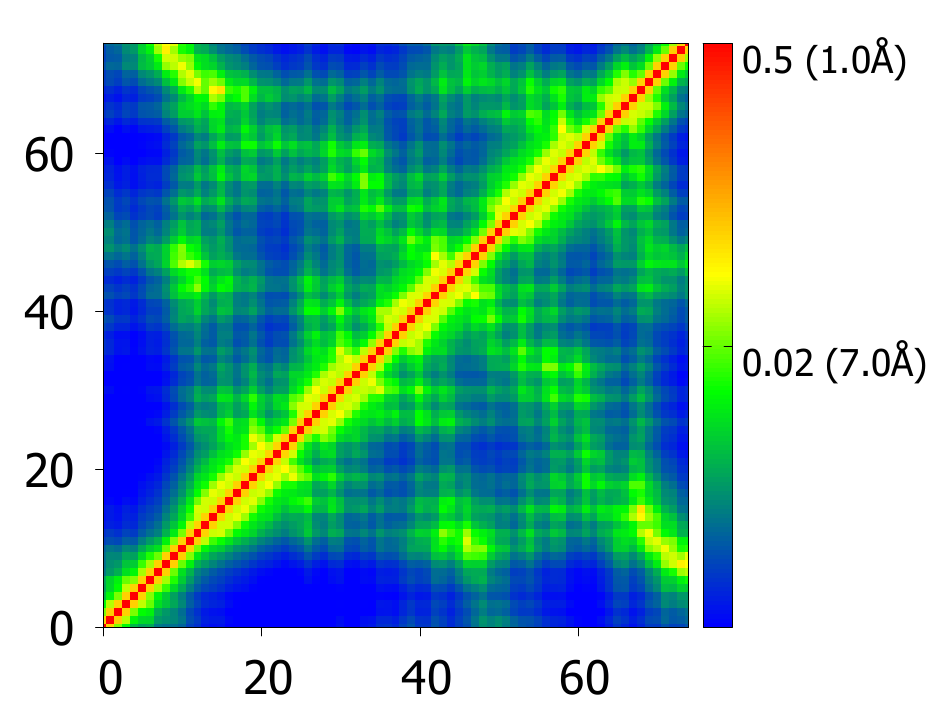}

\end{tabular}
\begin{tabular}{ccc}
\includegraphics[width=0.333\textwidth]{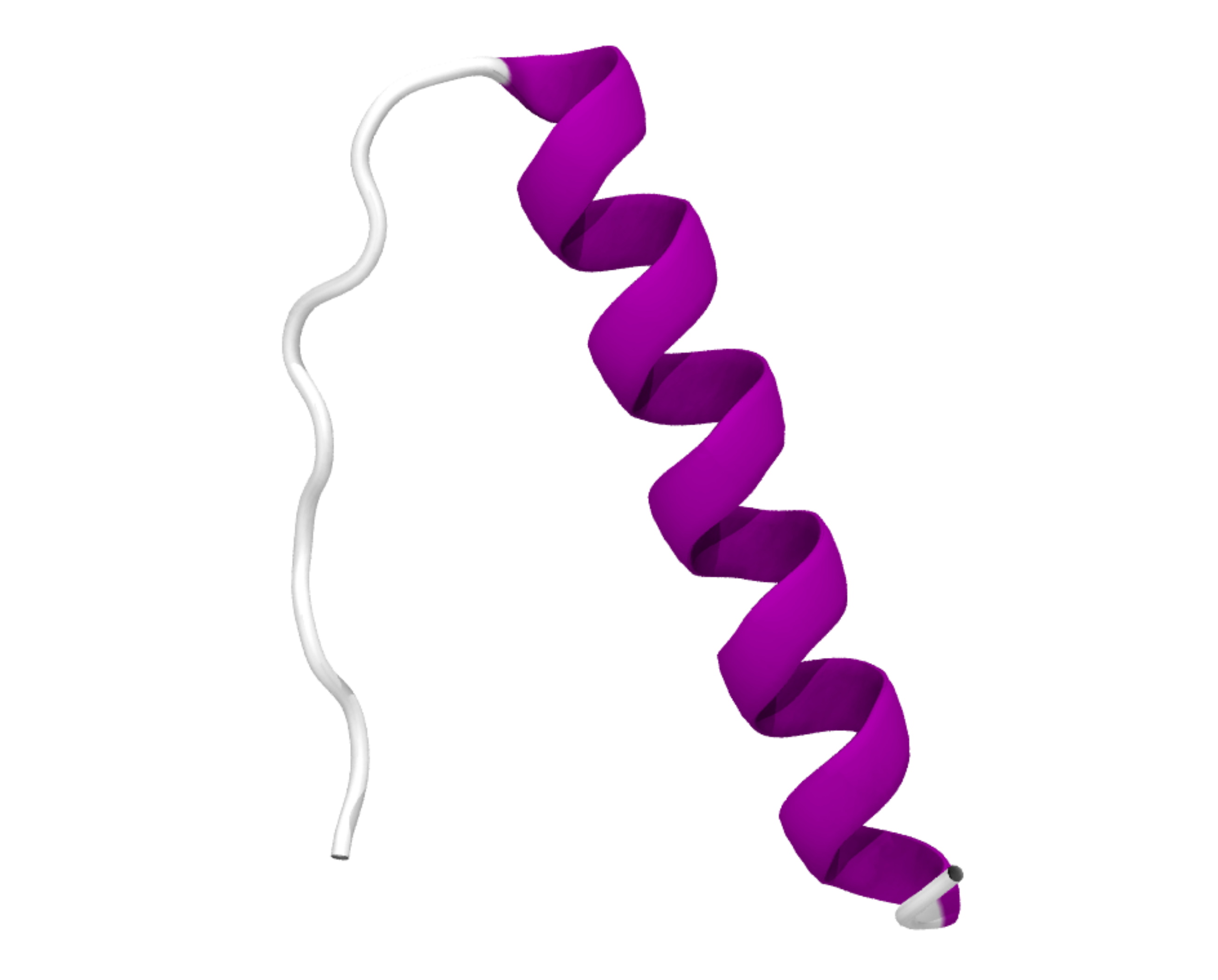}
\includegraphics[width=0.333\textwidth]{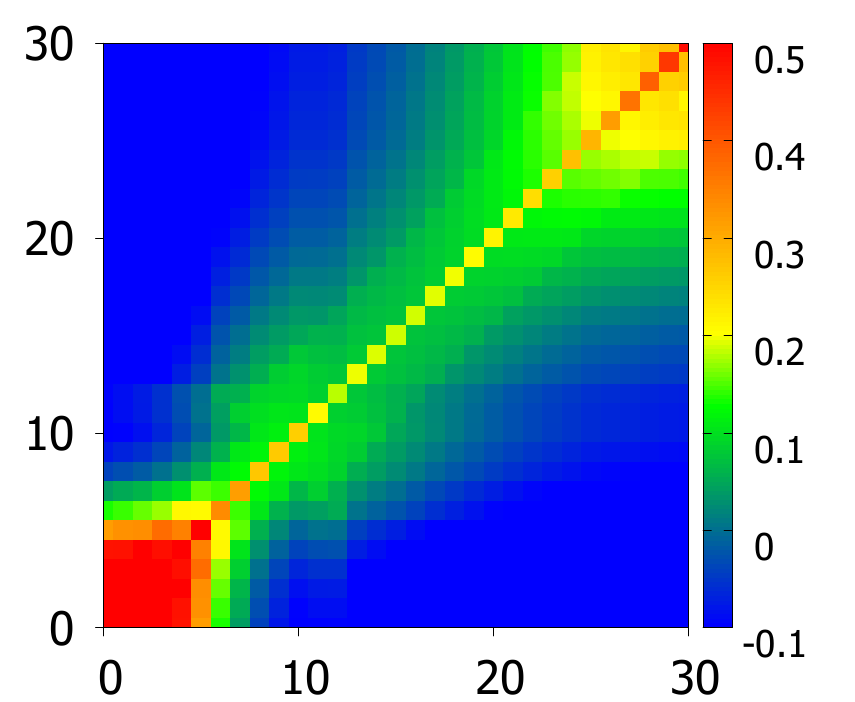}
\includegraphics[width=0.333\textwidth]{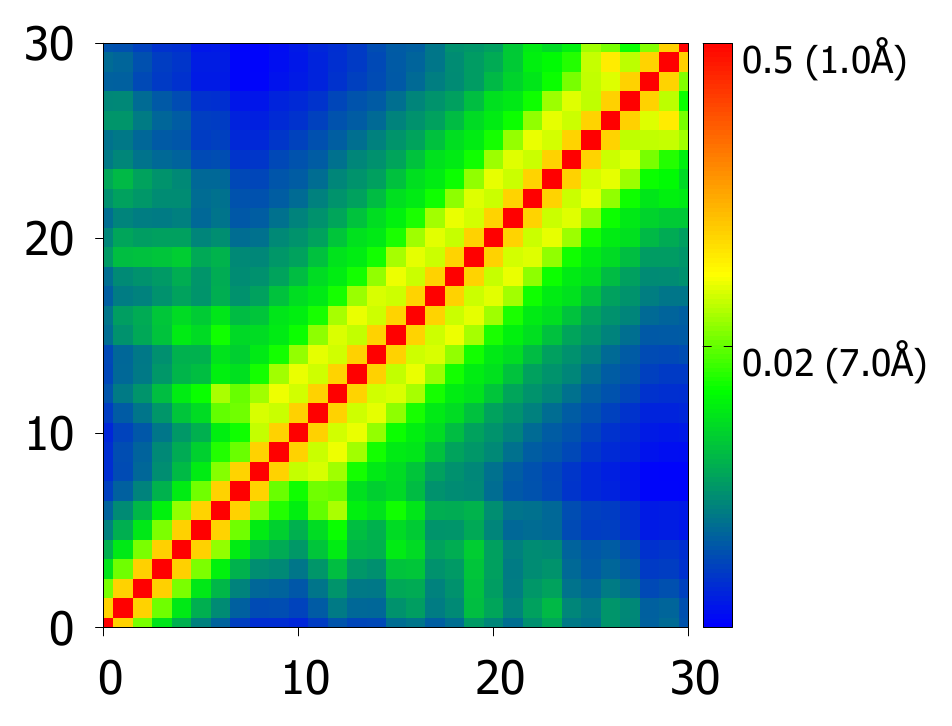}

\end{tabular}
\begin{tabular}{ccc}
\includegraphics[width=0.333\textwidth]{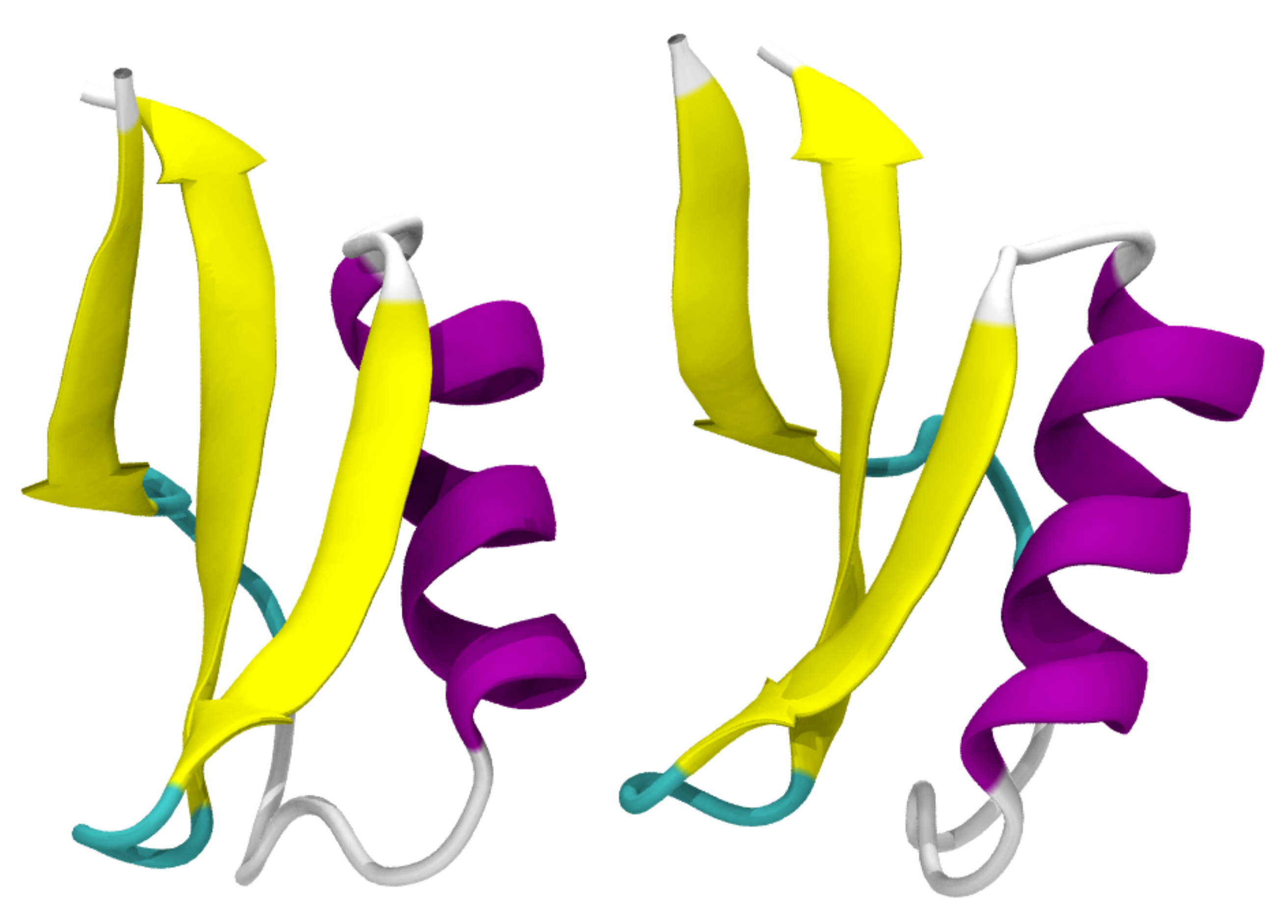}
\includegraphics[width=0.333\textwidth]{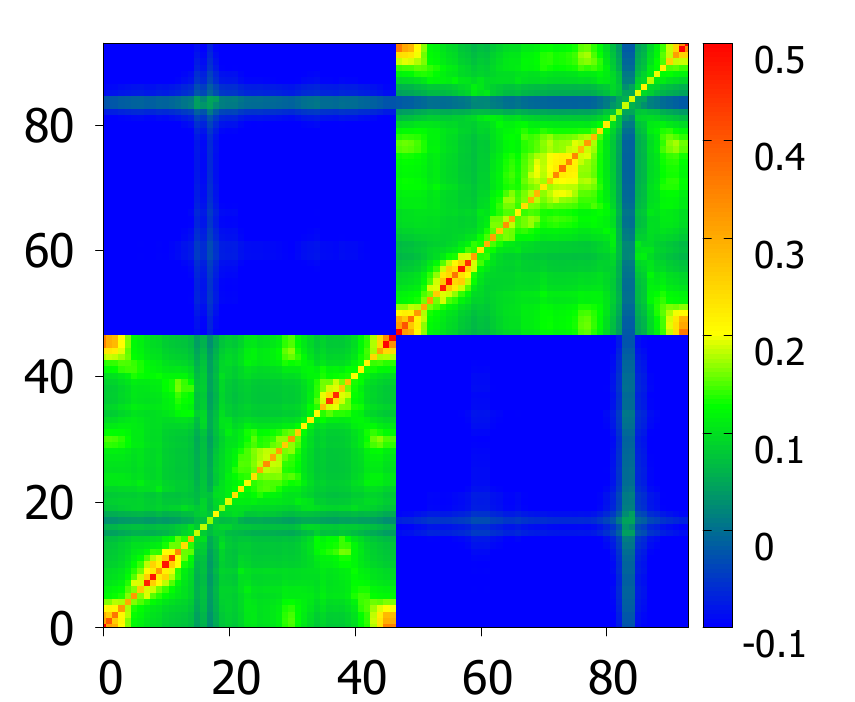}
\includegraphics[width=0.333\textwidth]{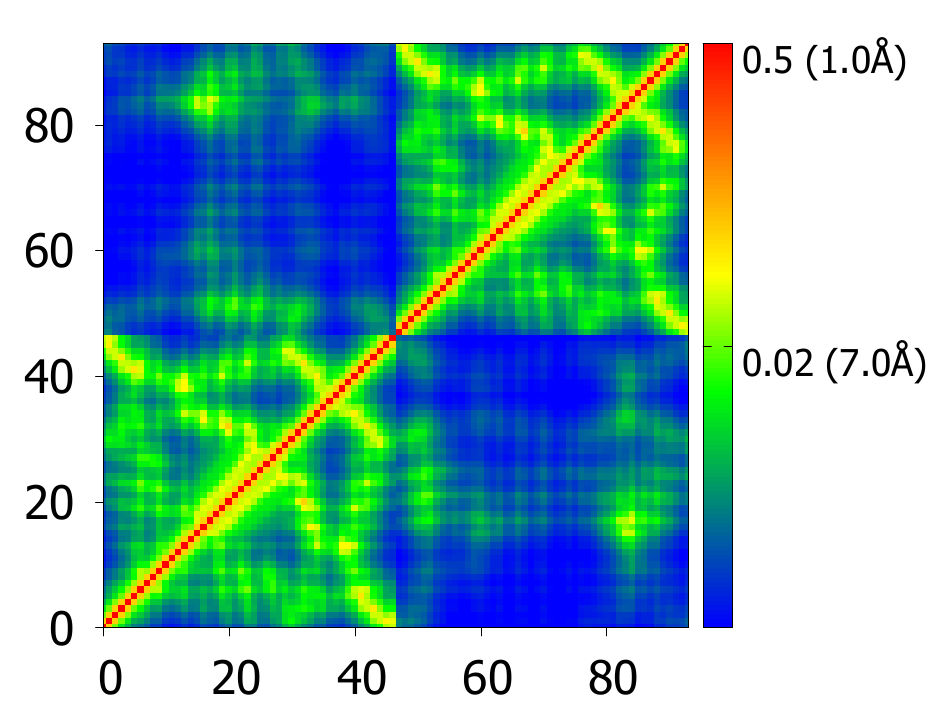}

\end{tabular}
\end{center}
\caption{ Comparison of correlation maps generated by GNM (middle) and FRI (right) for three proteins which are displayed in a secondary structure representation (left).  From top to bottom, PDB IDs for the structures displayed are 3TYS, 1AIE and 3PSM. The three-dimensional representation of each protein is generated in VMD\cite{VMD} and colored by secondary structure. The correlation maps in the middle column are computed using the GNM method and the  correlation maps from the FRI method are shown in the right column.
}
\label{matrixcomp}
\end{figure}

Table \ref{eff_table} and Figure \ref{efcomp} are the running times for each method in our FORTRAN implementations of GNM and FRI.  {Tests were conducted using a single core of an AMD Phenom II X6 1100T processor and include the entire GNM and FRI algorithm leaving out only the time it takes to load PDB files}. As expected,  GNM has a computational  complexity close to ${\cal O}(N^3)$ due to the matrix decomposition, while the FRI is approximately of ${\cal O}(N^2)$, mainly because of  the computation of correlation functions. As for the fFRI, its computational  complexity  is of
${\cal O}(N)$ due to the nature of its sparse matrix. The lines of best fit for CPU time ($t$) are: $t=(4\e{-8})*N^{3.09}$ for GNM, $t=(2\e{-7})*N^{1.98}$ for FRI and $t=(1\e{-7})*N^{0.975}$ for fFRI.  Some of the 44 structures used for efficiency testing were excluded from the final analysis of the FRI and fFRI methods because they required so little time to run that it was not possible to get an accurate measure of execution time.  A few of the largest structures were only tested with FRI and  fFRI methods because they require much more CPU time to run with GNM and the efficiency data are already sufficient to show that GNM scales at approximately  ${\cal O}(N^3)$.  For a protein of seven thousand amino acid residues, it takes close to ten thousand seconds for GNM and only a few seconds for  the  FRI to predict the B-factors in our test. The fFRI is significantly faster than other methods. It takes less than 30 seconds for the fFRI to predict the B-factors of the HIV virus  capsid with 313236 residues.

\subsection{Visualization of correlation maps }\label{sec:C-matrix}

The correlation functions used to generate  FRI correlation maps are based on monotonically decreasing radial basis functions. NMA, GNM and related tools, on the other hand, use either a Kirchhoff (i.e., contact matrix) or a cross correlation map  for connectivity between atoms. A Kirchhoff matrix is similar to our correlation map except the values are set to -1 for pairs of atoms within a cutoff distance and set to 0 for pairs of atoms outside the cutoff distance. The downside of using the Kirchhoff matrix is that, by definition, it treats all bonds within a certain cutoff distance the same and neglects all interactions outside that distance.
{We know that the closest atoms, such as covalently bonded atoms, will contribute more to the rigidity of a particular atom. With a rapidly decaying distance function, such as those used in our correlation functions, the impact of covalently bonded atoms is emphasized over the interactions that are just slightly more distant. This relationship is important because while structural features at a distance can play a significant role in stability and flexibility of a molecule, local structure has a much greater impact. Additionally, the cross correlation map in GNM and other methods does not reflect the atomic distance relations in a direct manner. }
In contrast, given a correlation matrix and a function from the FRI method it is easy to reconstruct the position of  each C$_\alpha$ atom in our coarse-grained model. The FRI correlation maps on the right side of Fig. \ref{matrixcomp} display distance values along side the correlation values to reflect this property of the FRI. These maps were calculated using the Lorentz function with $\upsilon$=2.5. The second case (PDB: 1AIE), a single $\alpha$-helix, is a good example of how the distance based correlation map reflects secondary structure information. The width of the band of high correlation is four amino acids, approximately the number of amino acids in one turn of an $\alpha$-helix. For atoms within one turn of the helix, correlation values to nearby atoms follow a predictable pattern based on their distance. The cross-correlation matrix for 1AIE from GNM shows a similar overall pattern but without the same kind of atomic  detail from the distance based correlation functions.

The other structures in Fig. \ref{matrixcomp} are from larger proteins and they show how the FRI and GNM represent complex arrangements of secondary structures. The FRI correlation maps clearly indicate where secondary structures are in the protein and what other residues they interact closely with. In these correlation maps, secondary structures are typically shown as small bands of relatively high correlation (yellow to red) while interactions between them appear in green and they often appear as regularly spaced green spots. These spots have space between them because they involve interactions with one face of an $\alpha$-helix or similar fold and so the residues on the far side have a lower correlation. The cross-correlation matrix from GNM also gives some indication of where secondary structures are, $\alpha$-helix is a square of green and beta sheets are lines, however these shapes are less defined on the atomic scale. Similarly the interactions between secondary structures are harder to pinpoint atomically as these interactions appear as a green smear in the matrices while in the correlation maps of the FRI method there are discrete spots with individual correlation values for atom to atom interactions. In the last example of Fig. \ref{matrixcomp} we can also see in the top left and bottom right corners how each map displays the interaction between two images of a structure from a single X-ray crystallography experiment.

\begin{figure} 
\begin{center}
\begin{tabular}{cc}
\includegraphics[width=0.4\textwidth]{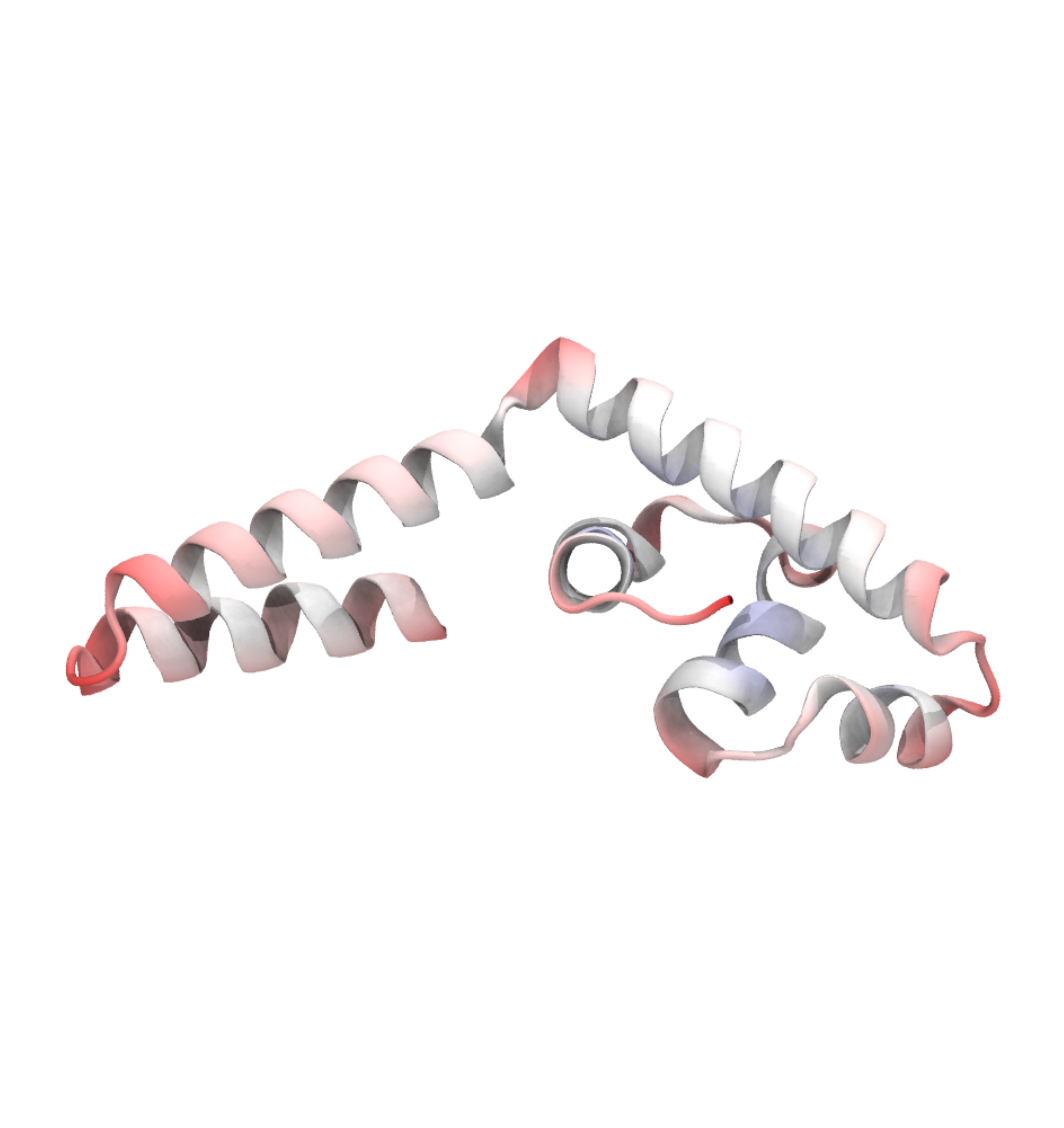}
\includegraphics[width=0.4\textwidth]{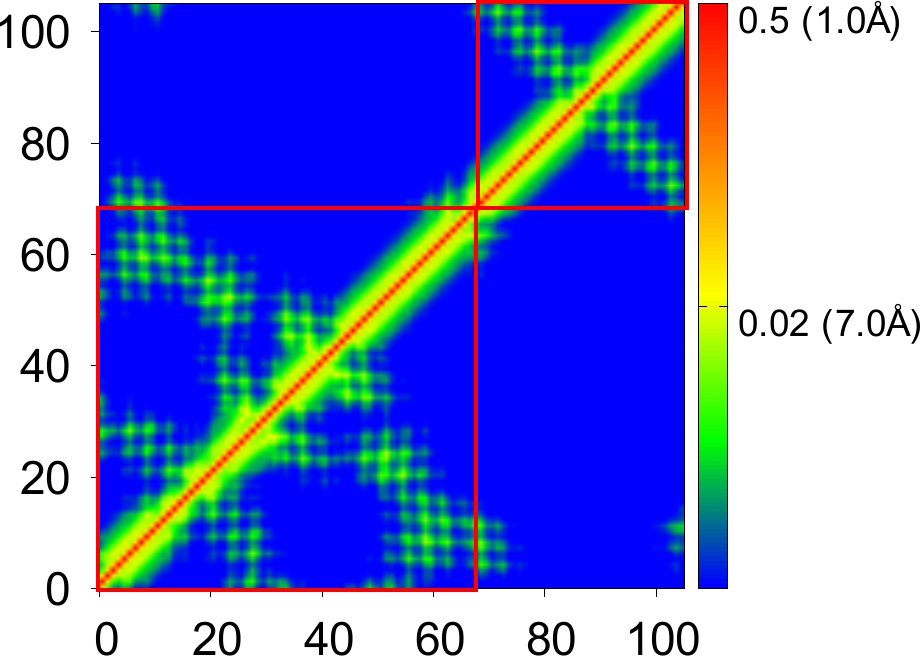}
\end{tabular}
\begin{tabular}{cc}
\includegraphics[width=0.4\textwidth]{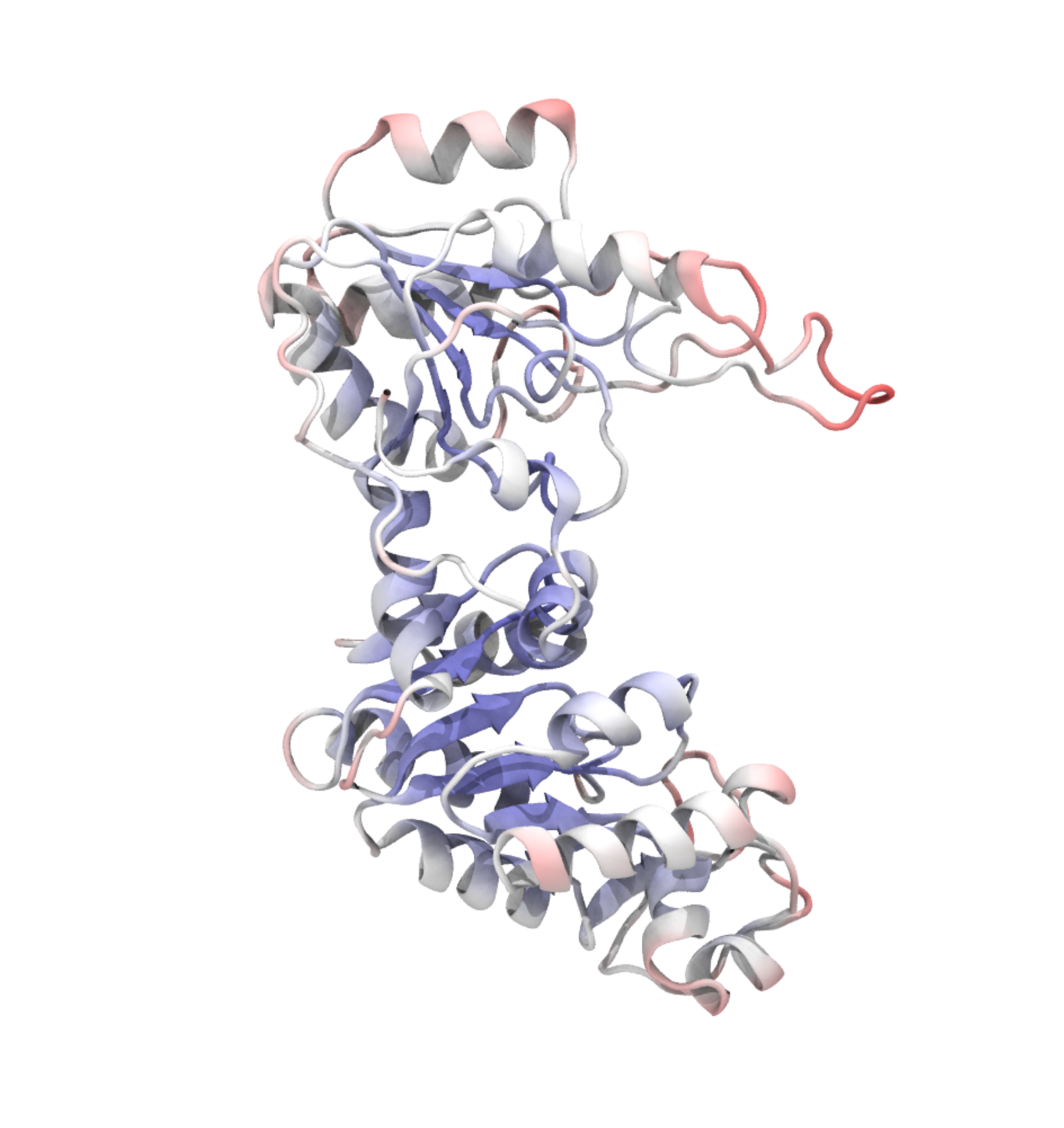}
\includegraphics[width=0.4\textwidth]{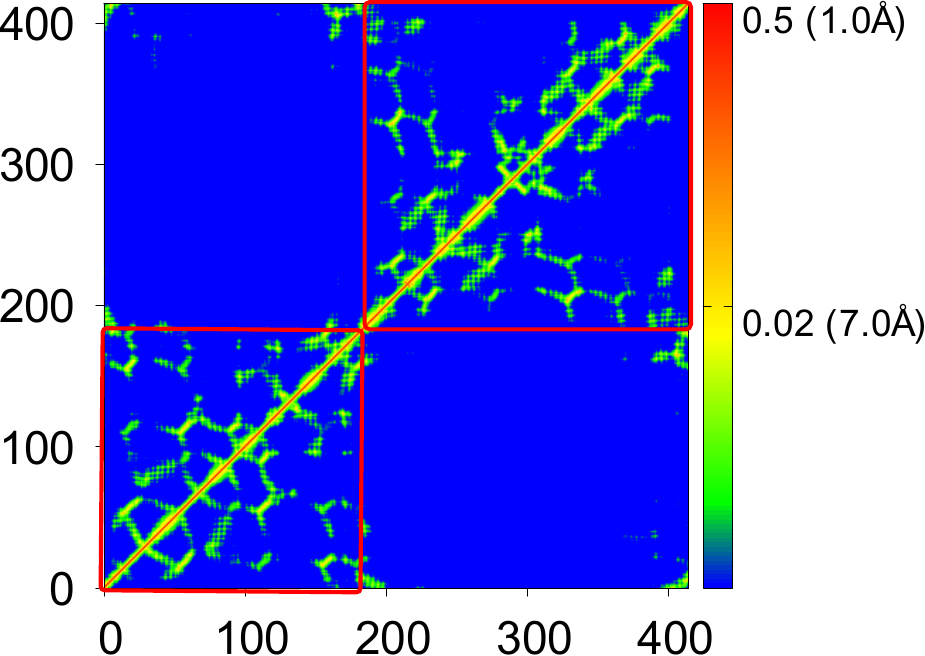}
\end{tabular}
\begin{tabular}{cc}
\includegraphics[width=0.4\textwidth]{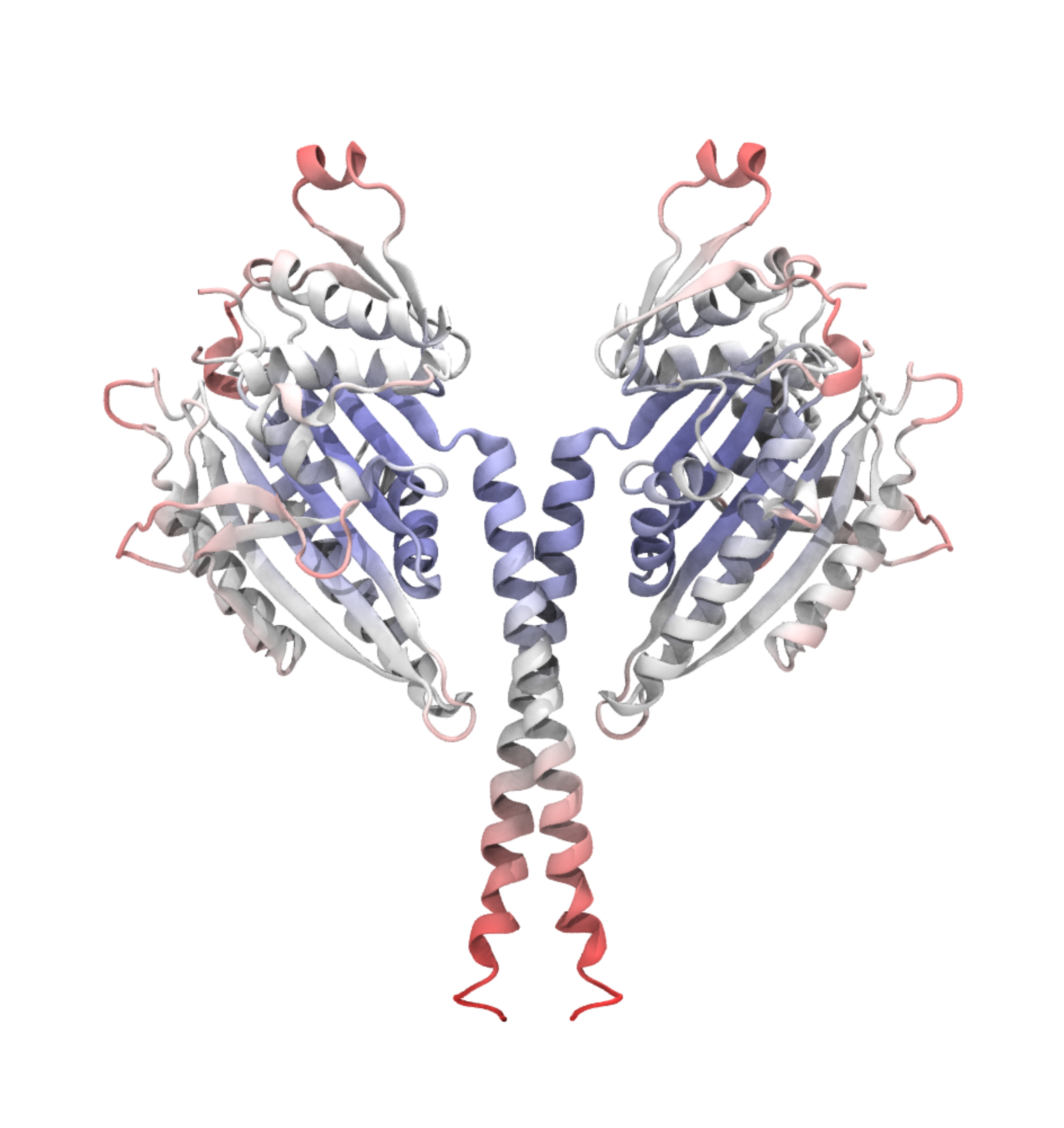}
\includegraphics[width=0.4\textwidth]{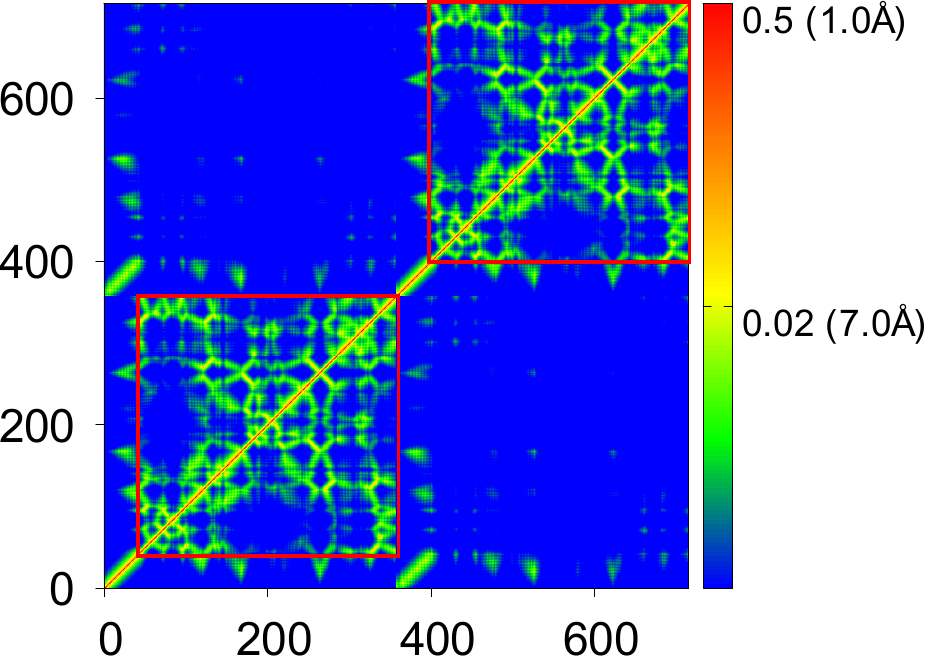}
\end{tabular}
\end{center}
\caption{ Secondary structure representations and correlation maps for PDB structures 1S7O, 3PGK and 2NCD. Structures colored by values of the correlation function of the FRI using the exponential function with $\kappa$=0.4 and $\eta$=0.8\AA, which are the optimal values for 2NCD kinesin with a correlation coefficient of 0.671. All three-dimensional images are rendered using VMD \cite{VMD}.}
\label{domfig2}
\end{figure}

\begin{figure} 
\begin{center}
\begin{tabular}{cc}
\includegraphics[width=0.33\textwidth]{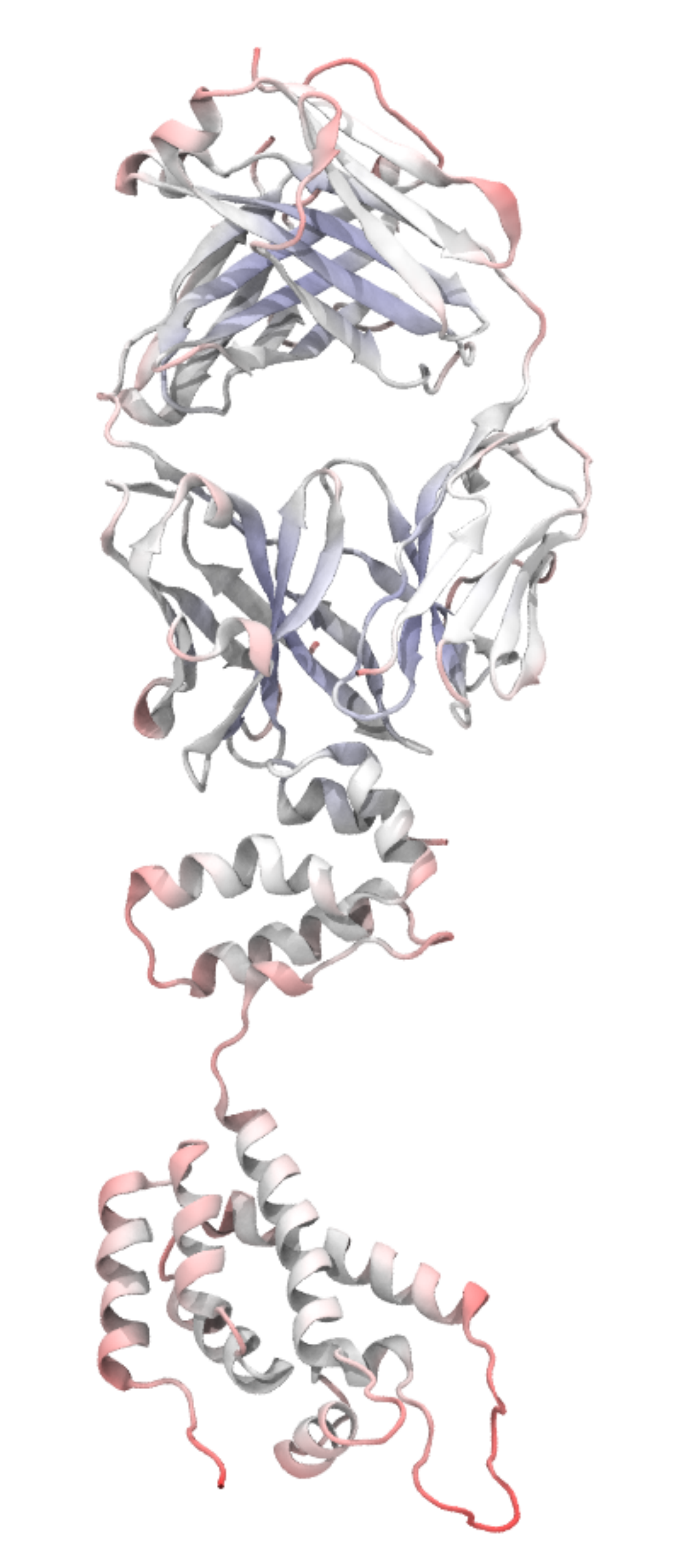}
\includegraphics[width=0.33\textwidth]{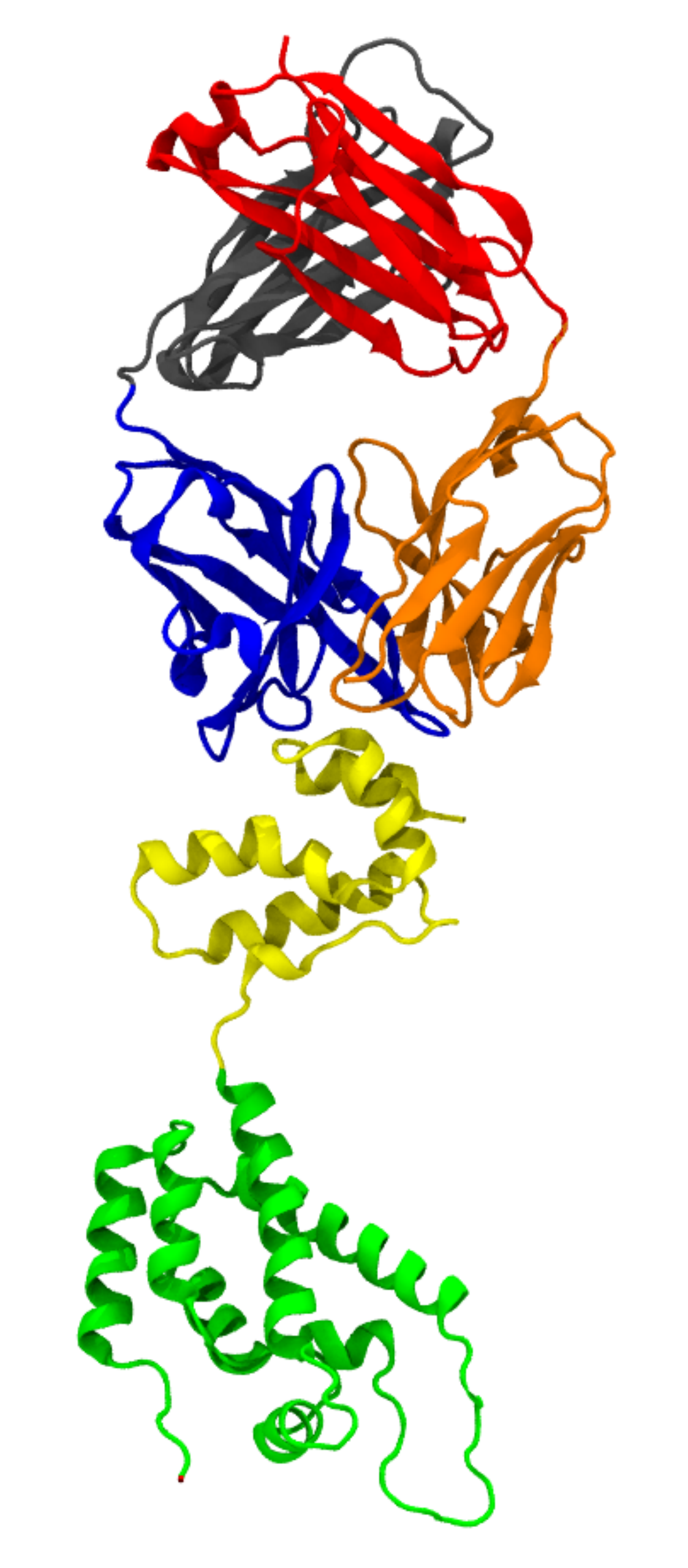}
\includegraphics[width=0.33\textwidth]{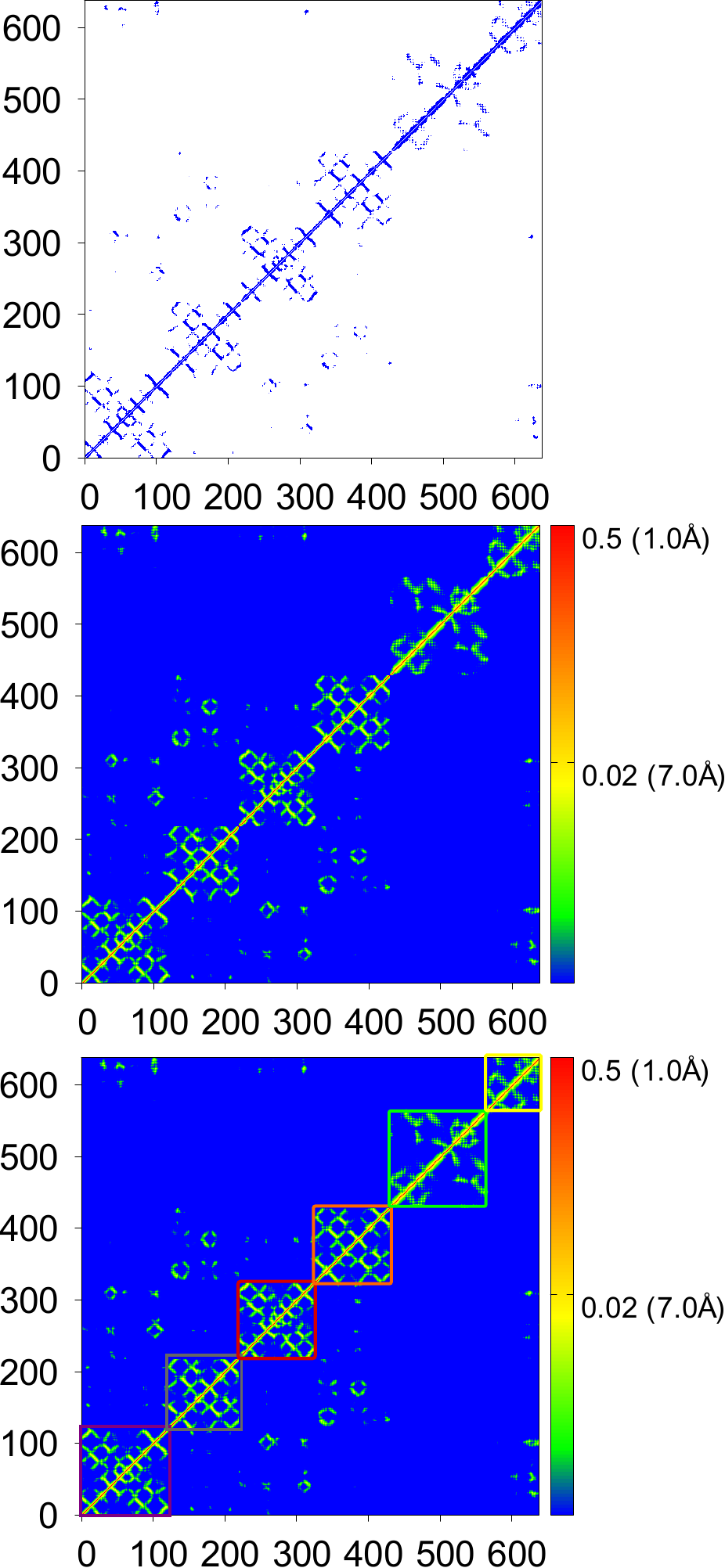}
\end{tabular}
\end{center}
\caption{ HIV capsid protein secondary structures (left and middle) and contact matrix from GNM (top right) and correlation matrix from FRI (middle right and bottom right). Secondary structure representations are colored by flexibility from FRI (left) and by domain (middle). Domain separations are highlighted in color coded boxes in the FRI correlation matrix at bottom right. All three-dimensional images are rendered using VMD \cite{VMD}.}
\label{domfig1}
\end{figure}

\section{ Protein domain analysis with FRI and aFRI }\label{Sec:Domains}

\subsection{ FRI for protein identification }\label{Sec:DomainsFRI}
 
Structural domains, as opposed to functional domains, are the basic unit of protein structures. Typically these portions of a protein fold independently and are stable on their own. The way various domains are assembled in a protein dictates its shape and function and is therefore interesting to biologists. One simple way to identify these domains and their interactions is by looking at the correlation matrix. The FRI correlation matrix or correlation map allows us to identify protein structural domains because the values of a correlation function directly reflect the pairwise distances between C$\alpha$-C$\alpha$ atoms. Therefore FRI correlation maps are also distance maps, which  have been shown to be an effective tool for identifying protein domains as early as 1974 \cite{Rossman:1974}. This method involves generating a C$\alpha$-C$\alpha$ pairwise distance map and identifying dense, triangular areas of contacts near the diagonal. The correlation matrix generated for FRI is similar to a basic C$\alpha$-C$\alpha$ distance map with the benefit of a rapidly decaying distance function. This results in more pronounced separation between domains in the map, potentially making it easier to identify distinct domains. Figure \ref{domfig2} demonstrates how domains are identified on a C$\alpha$-C$\alpha$ distance map or a FRI correlation matrix for three proteins of increasing size and complexity.  Each protein is divided into two structural domains outlined with red boxes for clarity.

In more complex cases, we see larger domains made up of multiple subdomains. Figure \ref{domfig1} shows the HIV capsid protein (PDB ID: 1E6J) consisting of six distinct structural domains. Alongside the protein's secondary structure are the contact matrix used in GNM and the correlation maps from the FRI. By adjusting the distance cutoff in the GNM and correlation function parameters in the FRI we arrive at similar results. The six compact structural subdomains can be identified visually from these matrices by finding areas with high correlation or large numbers of contacts in a triangular shape near the diagonal. Some interactions between subdomains can also be seen in the matrices. Subdomains with a significant level of correlation or contacts between them can be considered as single larger domain. In the example of Figure \ref{domfig1}, there are contacts in the matrices between five subdomains. The last subunit (green color) shows little or no correlation or contacts to the other subdomains and is therefor a separate domain and expected to move fully independently. However, predicting domain motions with the FRI requires an alternative formulation, the anisotropic FRI.

\begin{figure}[ht!]
\begin{center}
\begin{tabular}{c}\vspace*{1cm} 
\includegraphics[width=0.8\textwidth]{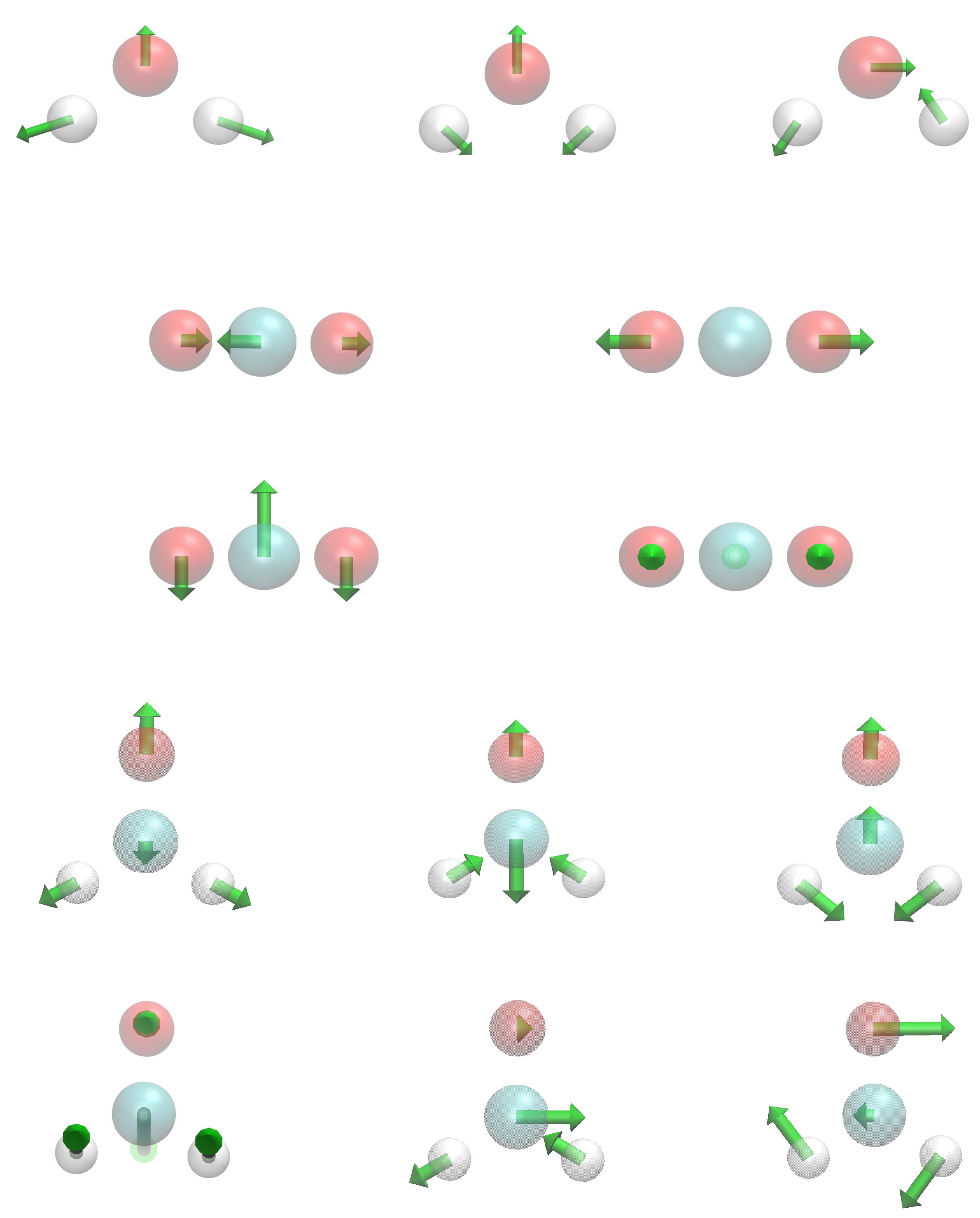} 
\end{tabular}
\end{center}
\caption{Validation of the completely global aFRI for the vibrational analysis of  three small molecules ($\upsilon=2$). 
First row: three vibrational modes for H$_2$O ($\eta=1$); 
Second and third rows: four vibrational modes for CO$_2$ ($\eta=1$); 
Last two rows: six vibrational modes for CH$_2$O ($\eta=2$).  
All three-dimensional images are rendered using VMD \cite{VMD}.}
\label{smallmolecues}
\end{figure}

\begin{figure}[ht!]
\begin{center}
\begin{tabular}{ccc}
\includegraphics[width=0.22\textwidth]{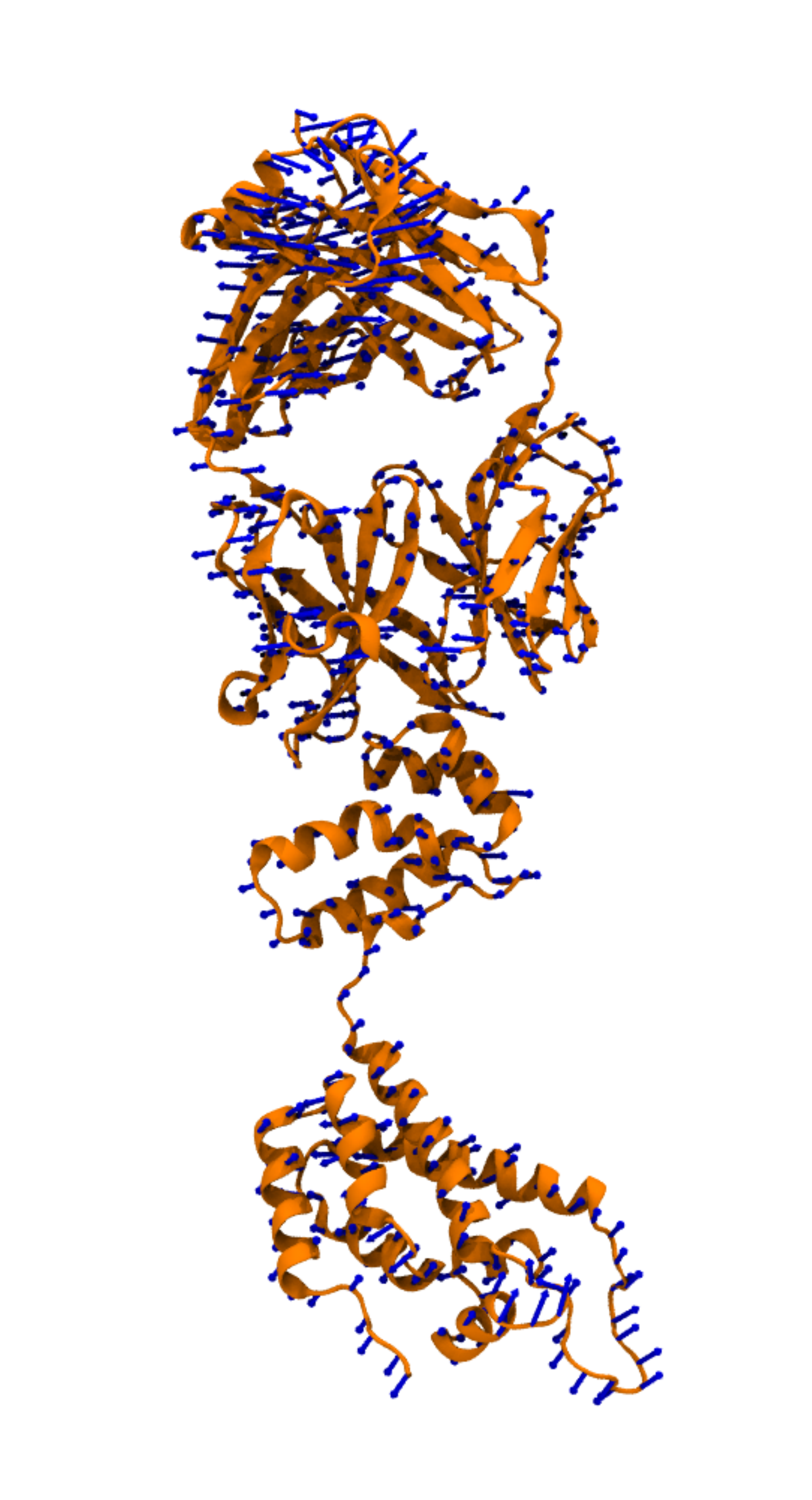}
\includegraphics[width=0.22\textwidth]{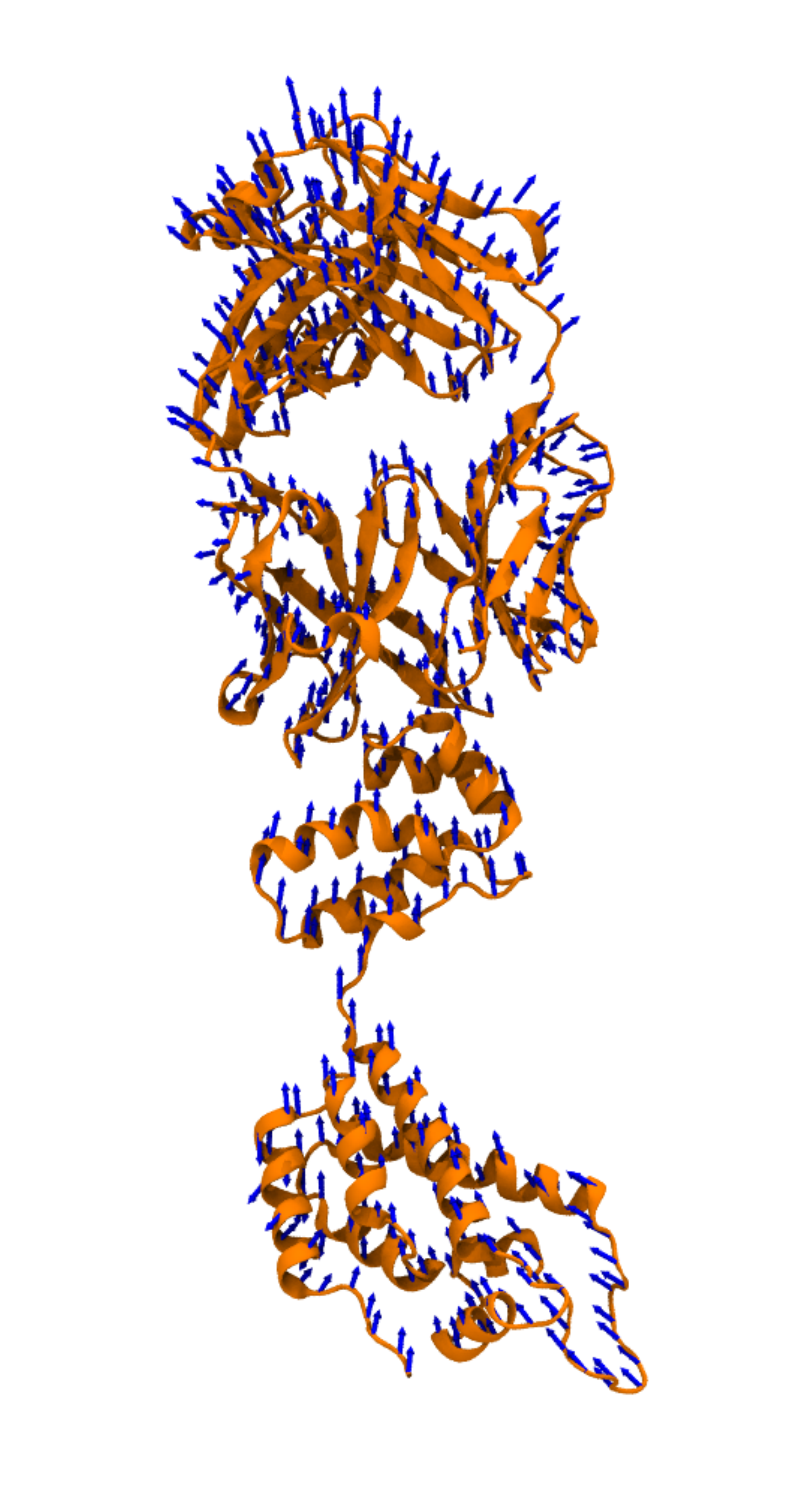}
\includegraphics[width=0.22\textwidth]{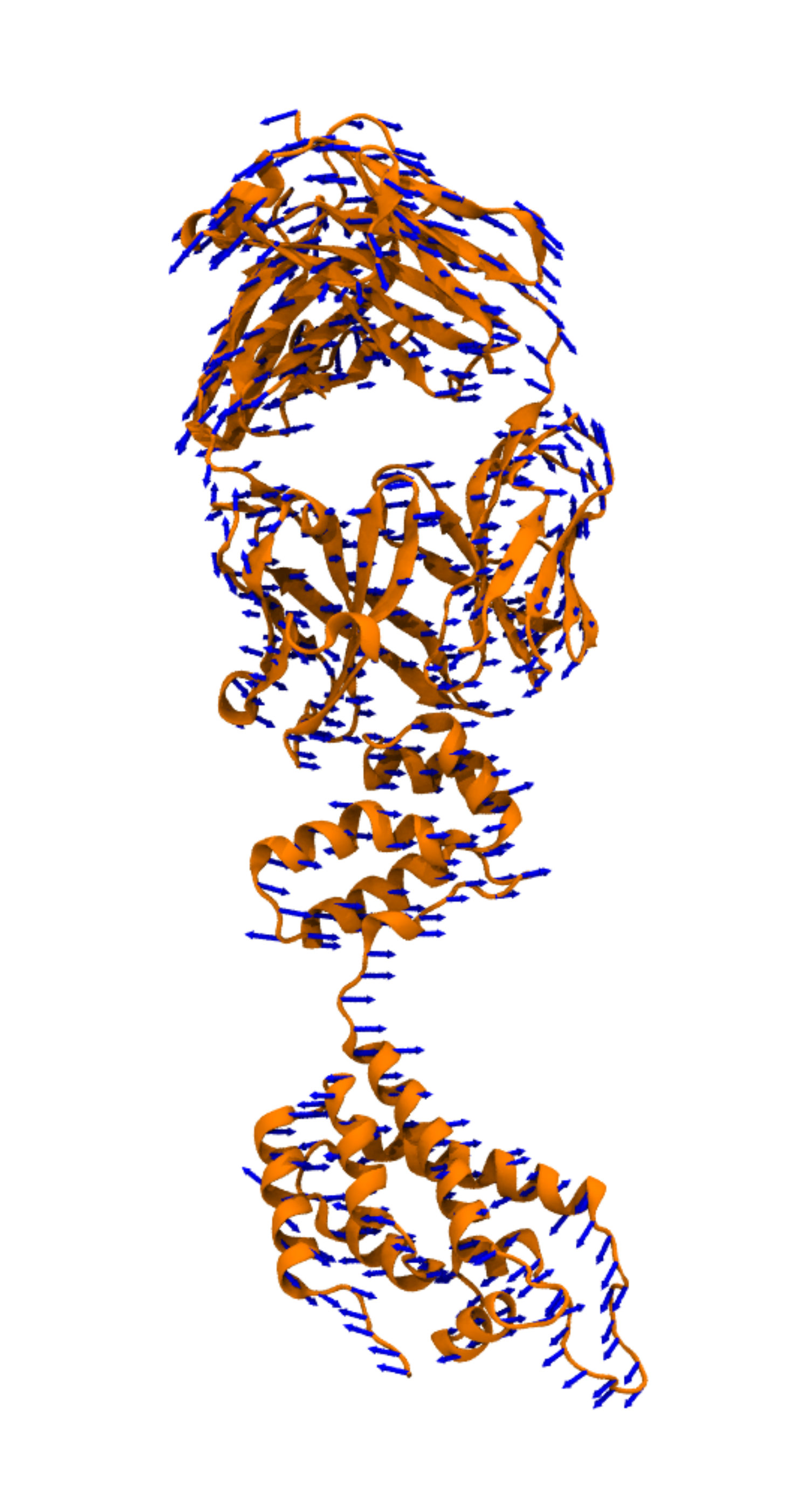}
\end{tabular}
\begin{tabular}{ccc}
\includegraphics[width=0.22\textwidth]{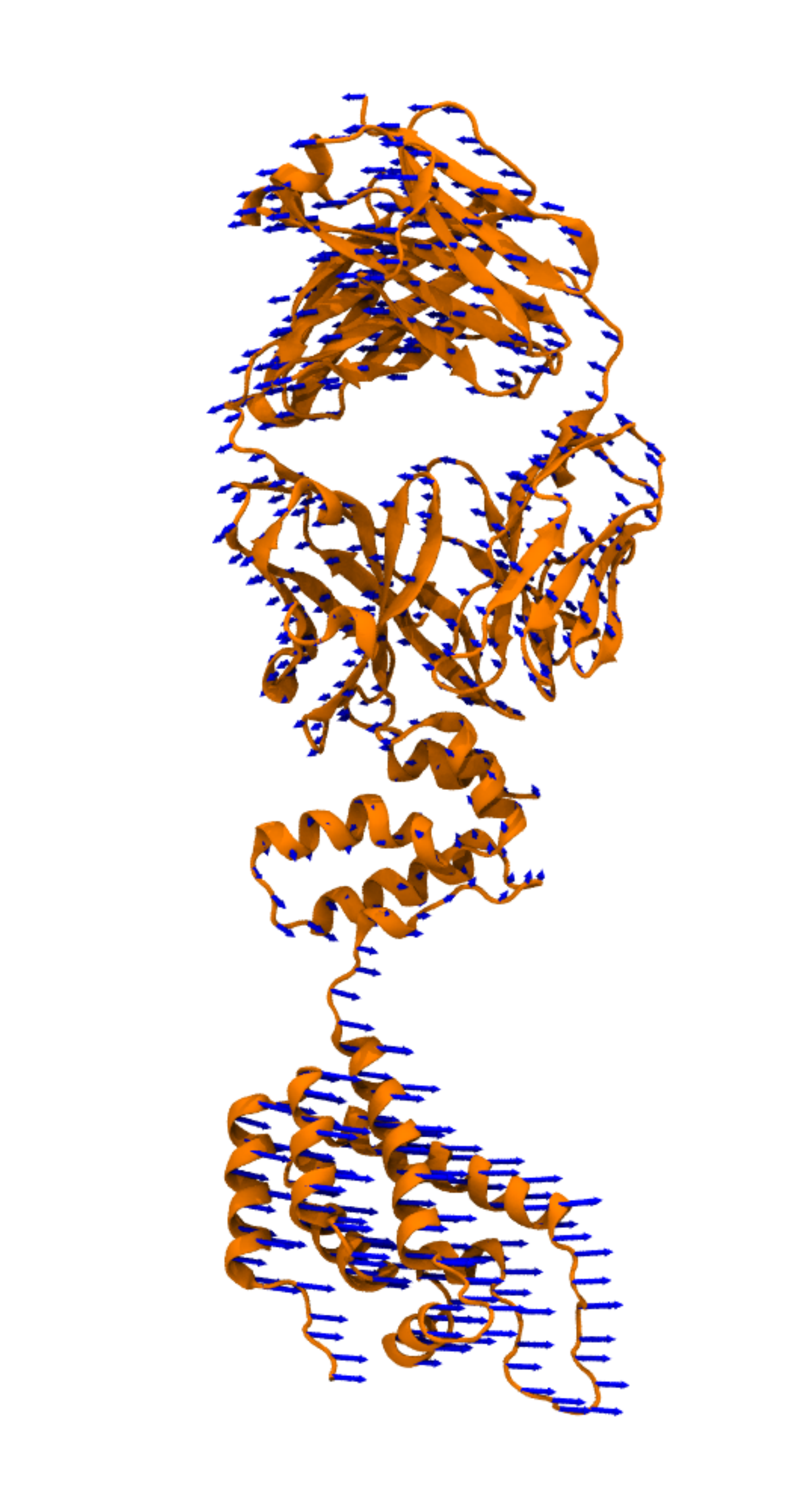}
\includegraphics[width=0.22\textwidth]{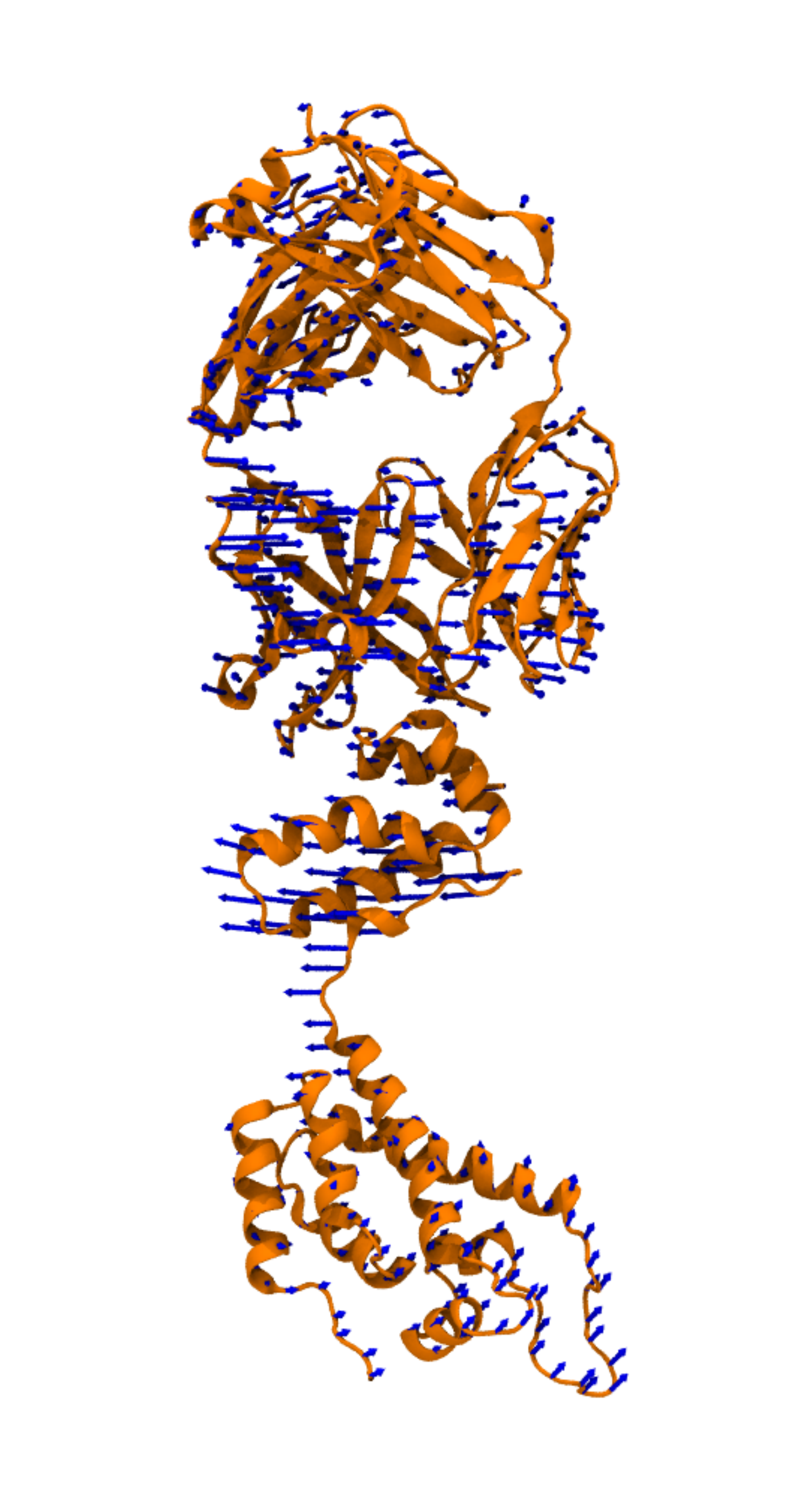}
\includegraphics[width=0.22\textwidth]{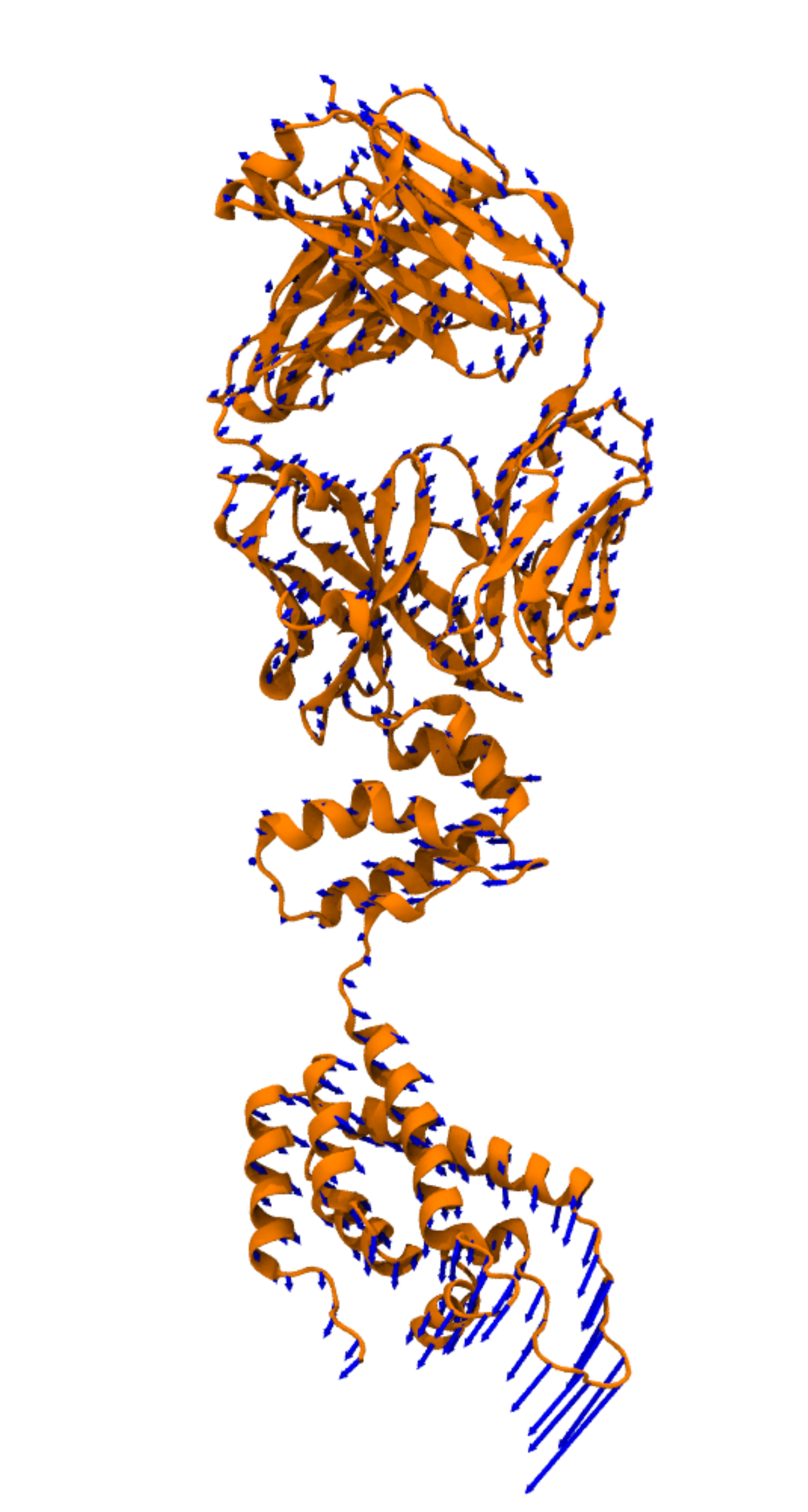}
\end{tabular}
\begin{tabular}{ccc}
\includegraphics[width=0.24\textwidth]{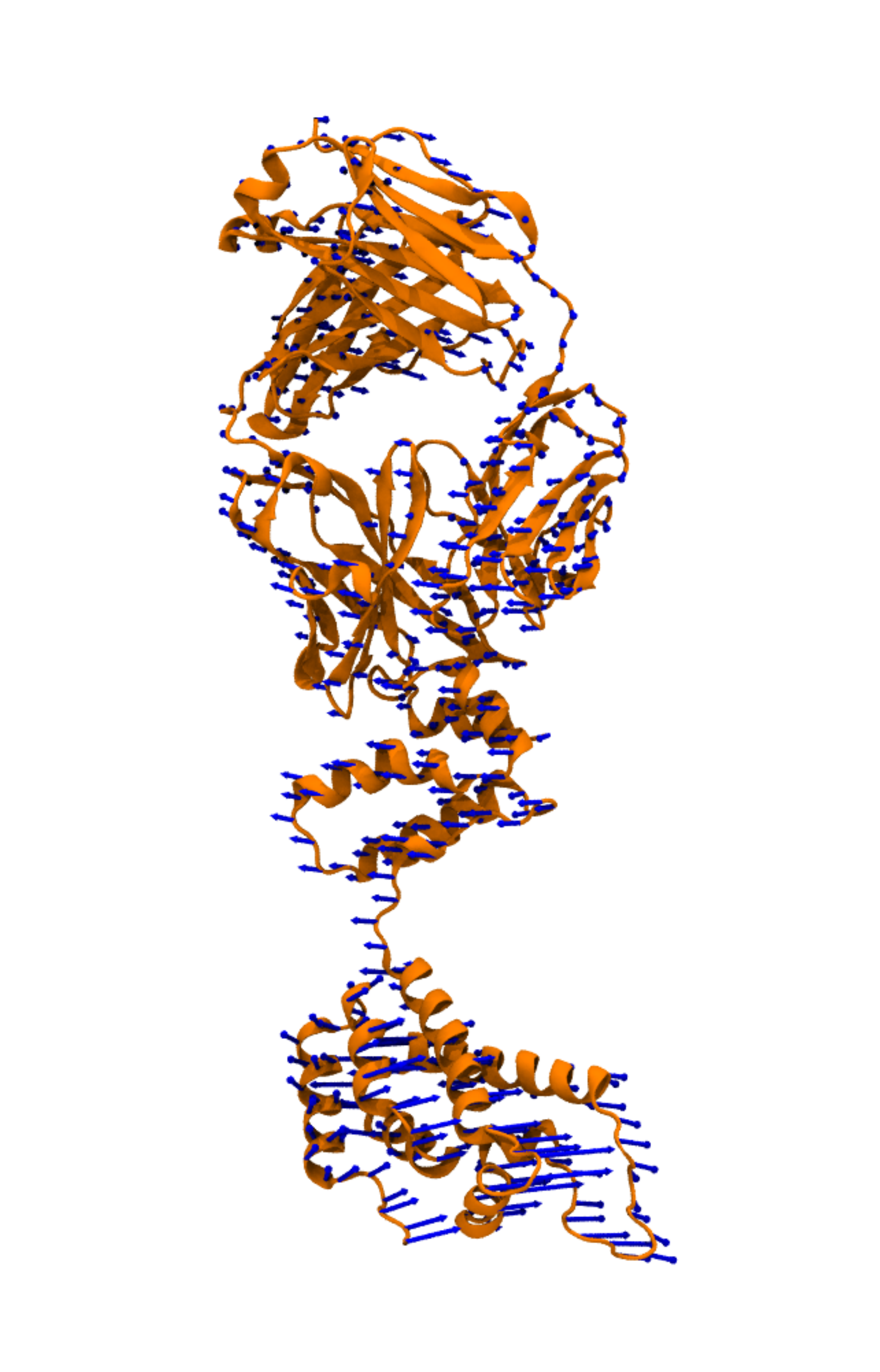}
\includegraphics[width=0.24\textwidth]{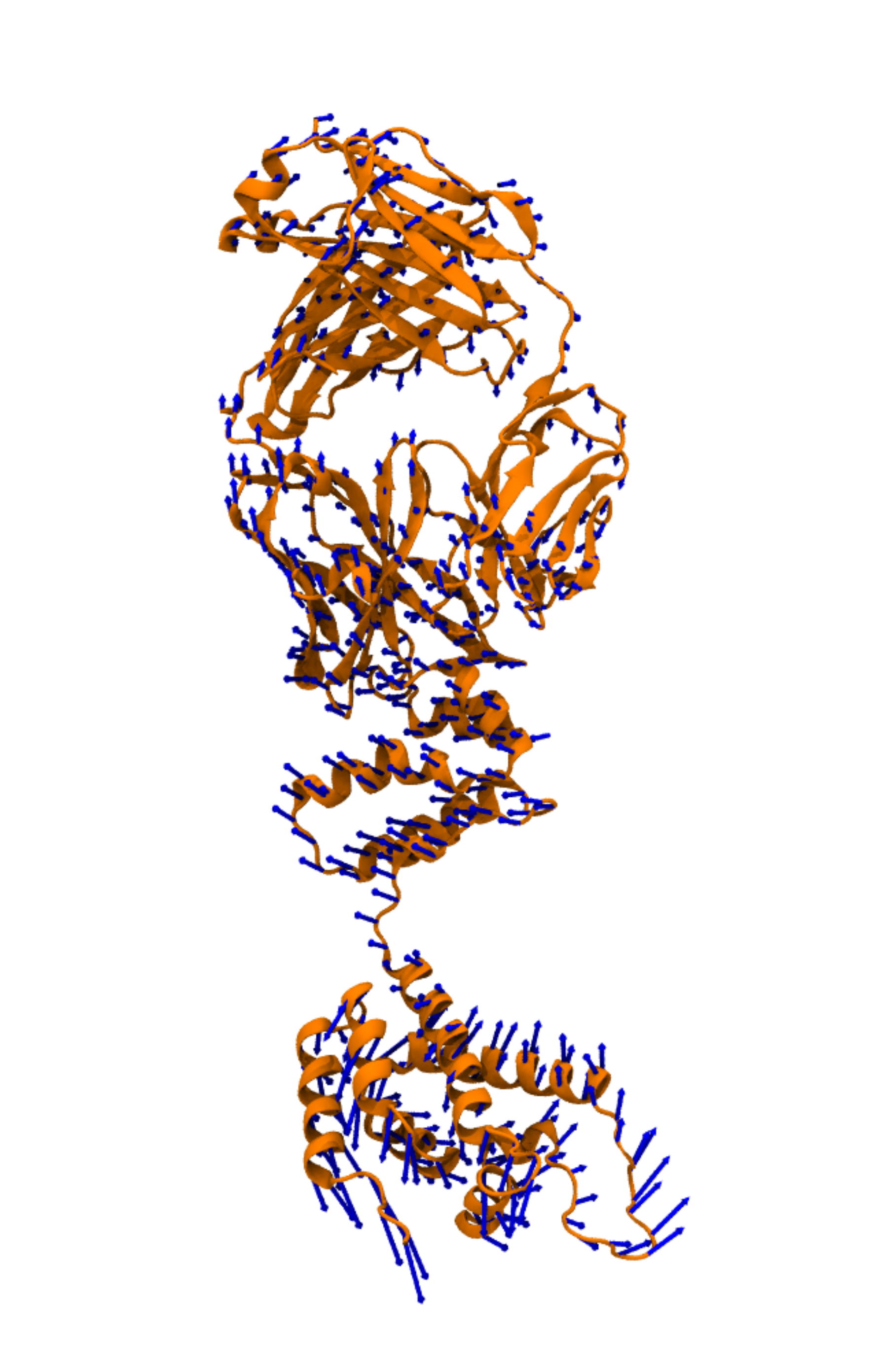}
\includegraphics[width=0.24\textwidth]{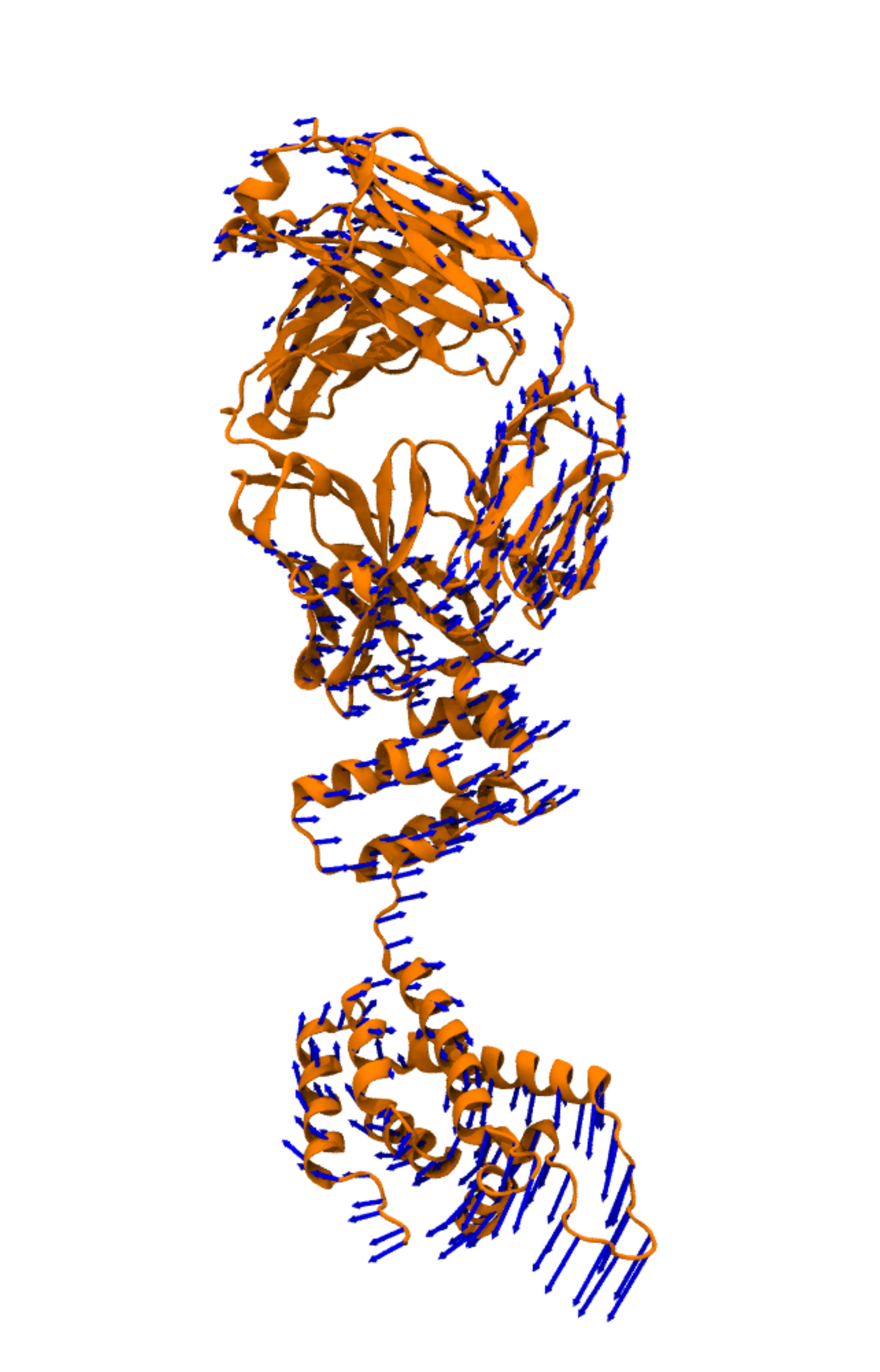}
\end{tabular}
\end{center}
\caption{ Comparison of modes for HIV capsid protein (PDB ID: 1E6J). 
The top row is generated by using the completely local aFRI   with $\upsilon=2$ and $\eta=50$. 
The middle row is generated by using the completely global aFRI with $\upsilon=2$ and $\eta=50$. 
The bottom row is generated by using  the ANM with ProDy v1.5 \cite{Bakan:2011} using default settings. All three-dimensional images are rendered using VMD \cite{VMD}.
}
\label{modes}
\end{figure}

\begin{figure}[ht!]
\begin{center}
\begin{tabular}{ccc}
\includegraphics[width=0.25\textwidth]{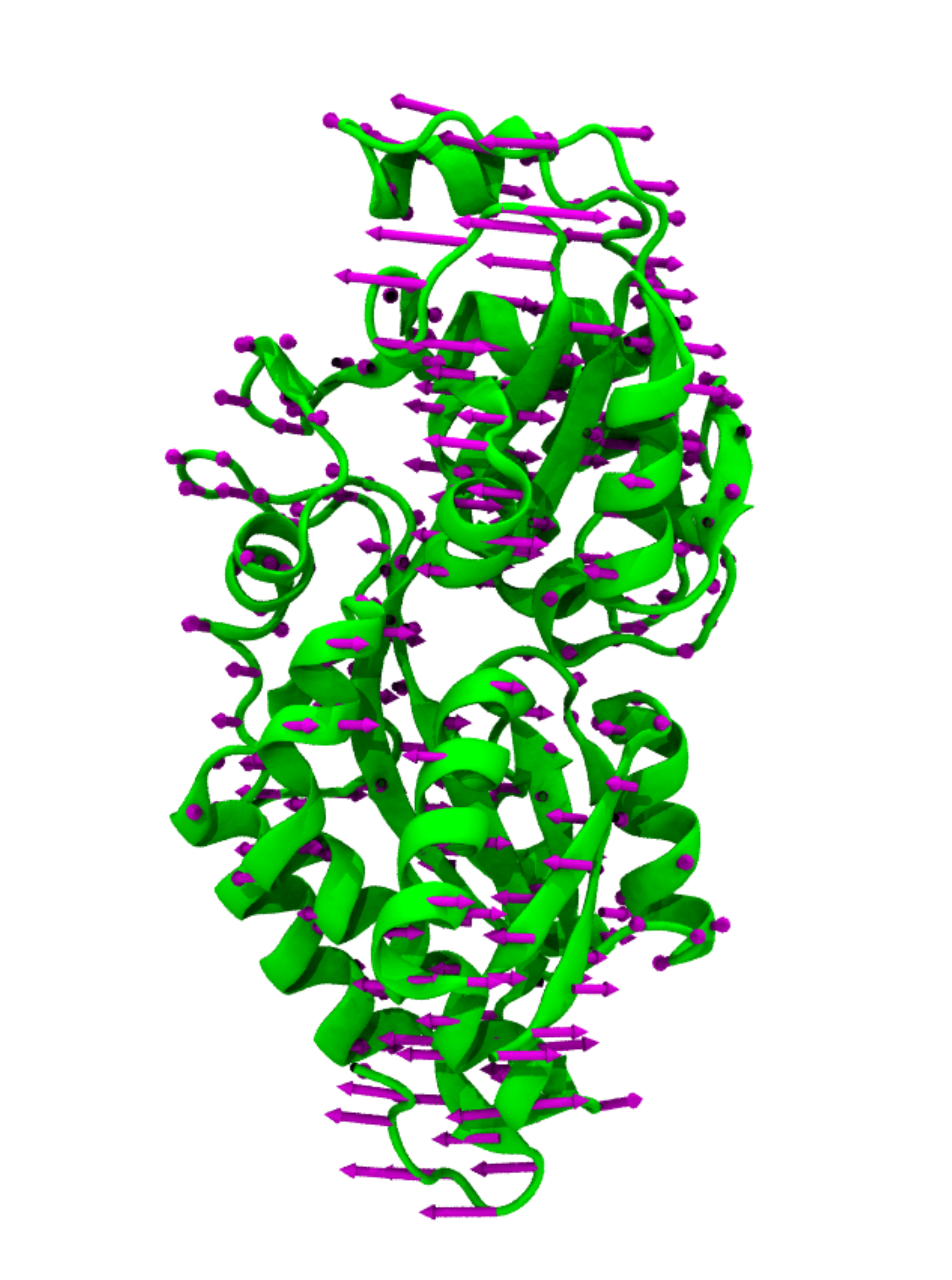}
\includegraphics[width=0.25\textwidth]{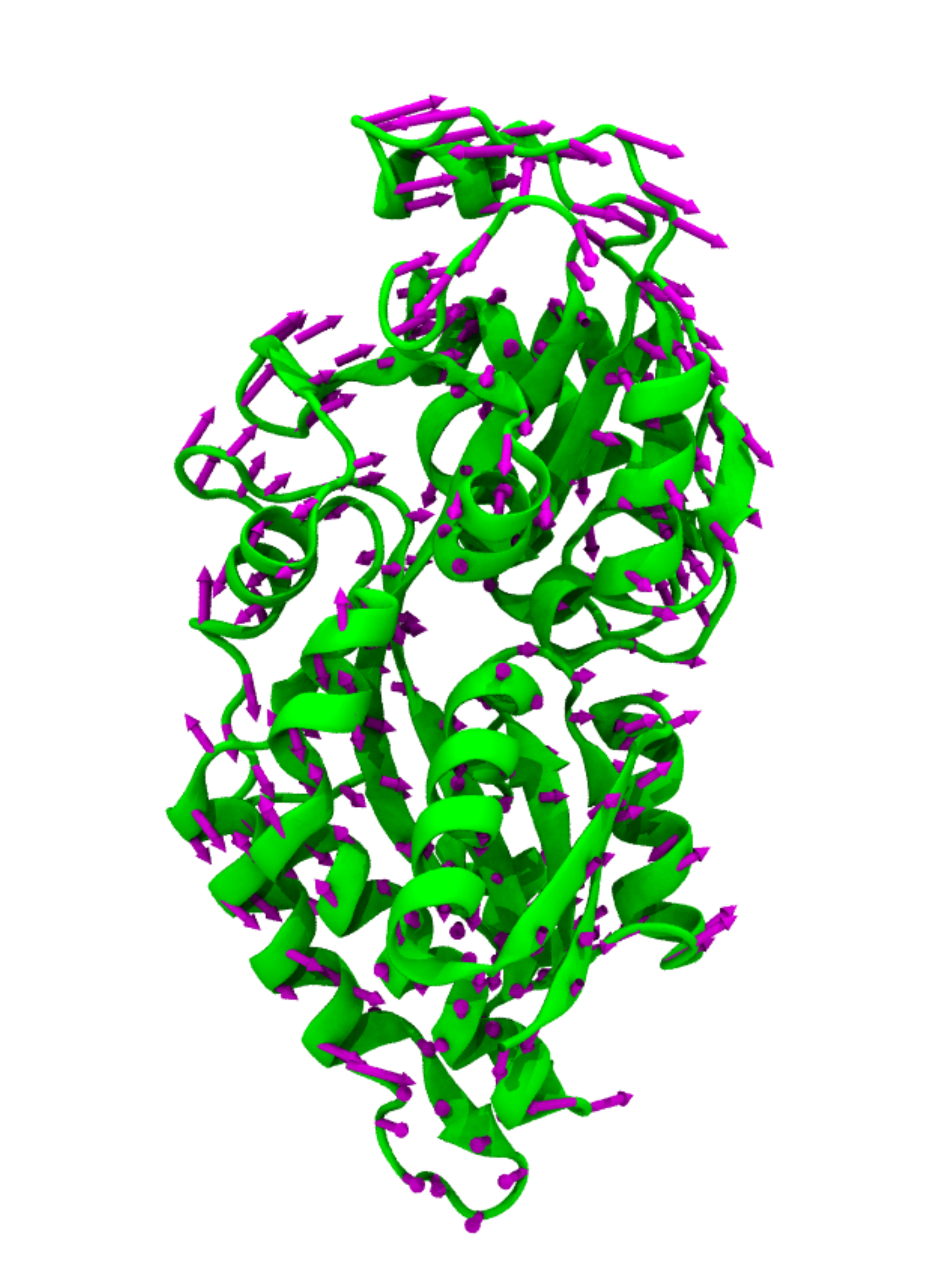}
\includegraphics[width=0.25\textwidth]{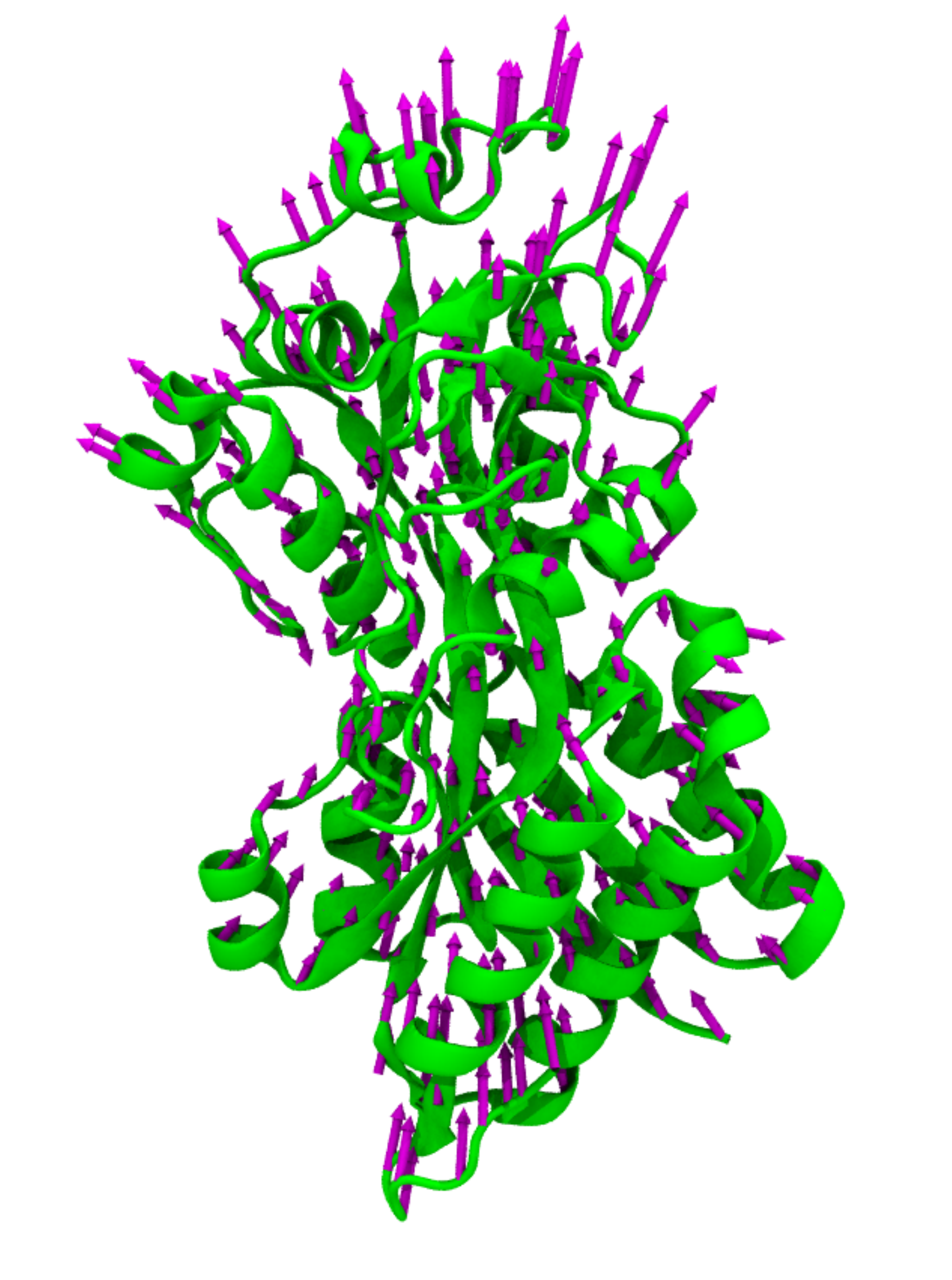}
\end{tabular}
\begin{tabular}{ccc}
\includegraphics[width=0.25\textwidth]{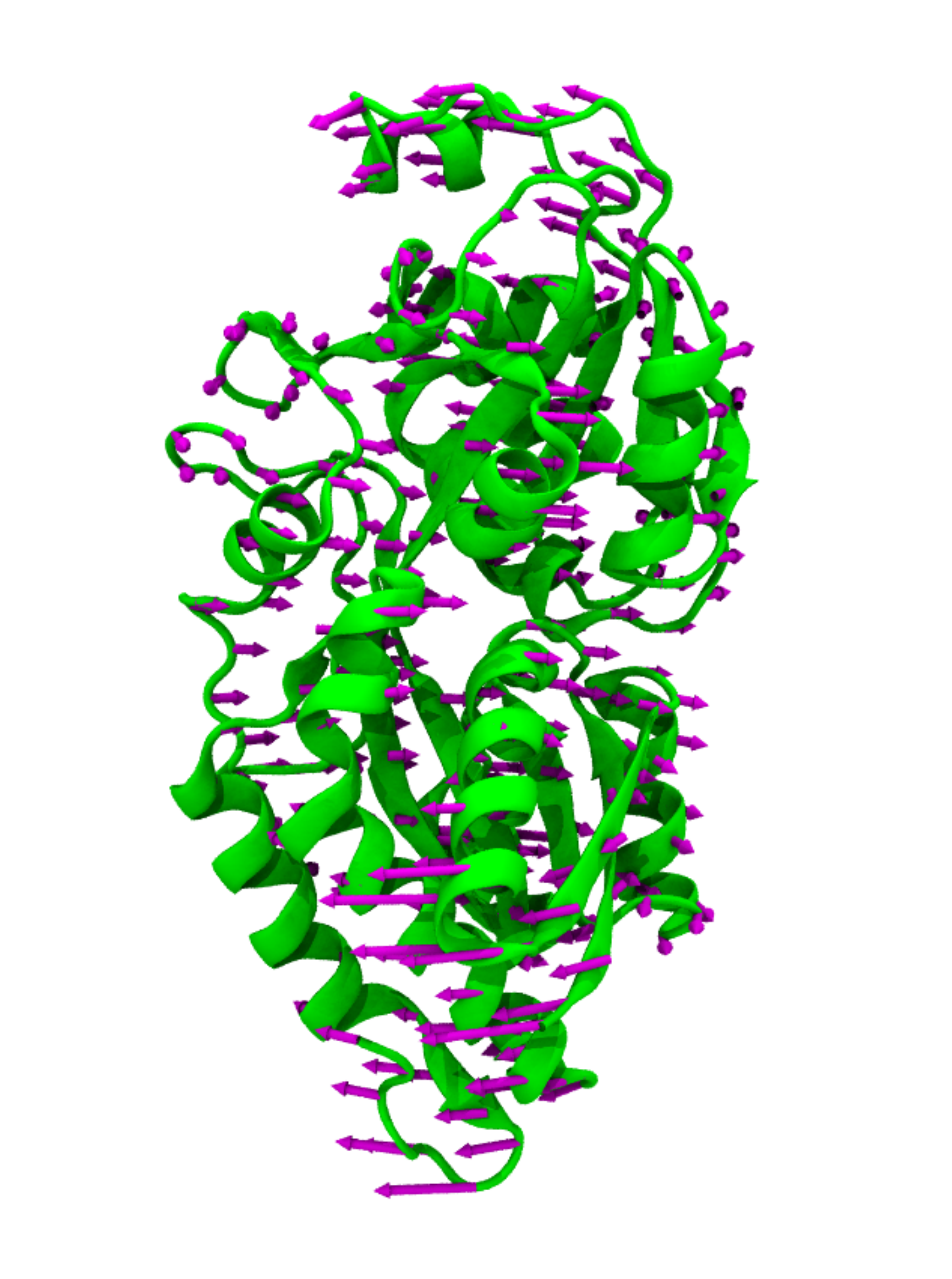}
\includegraphics[width=0.25\textwidth]{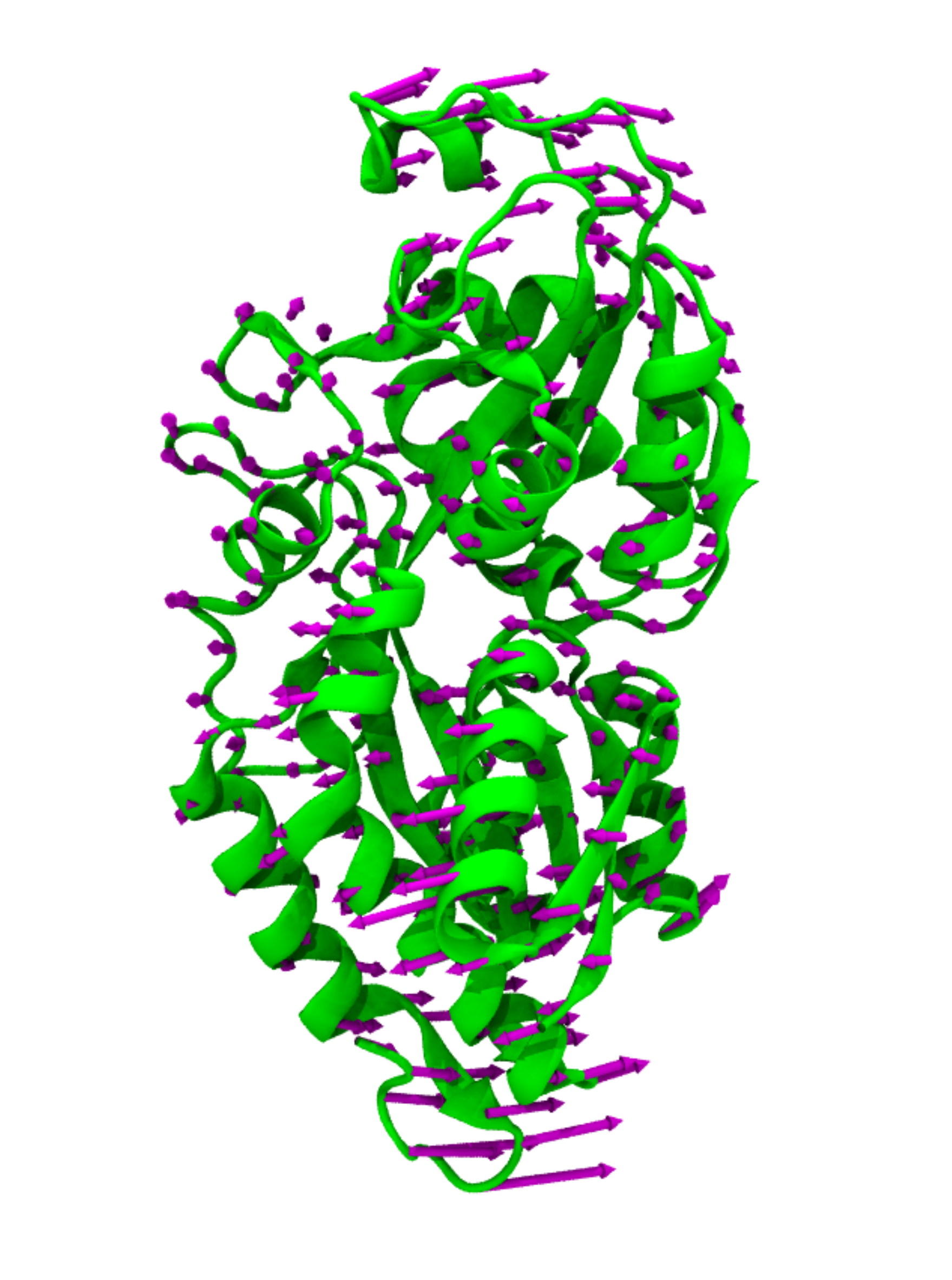}
\includegraphics[width=0.25\textwidth]{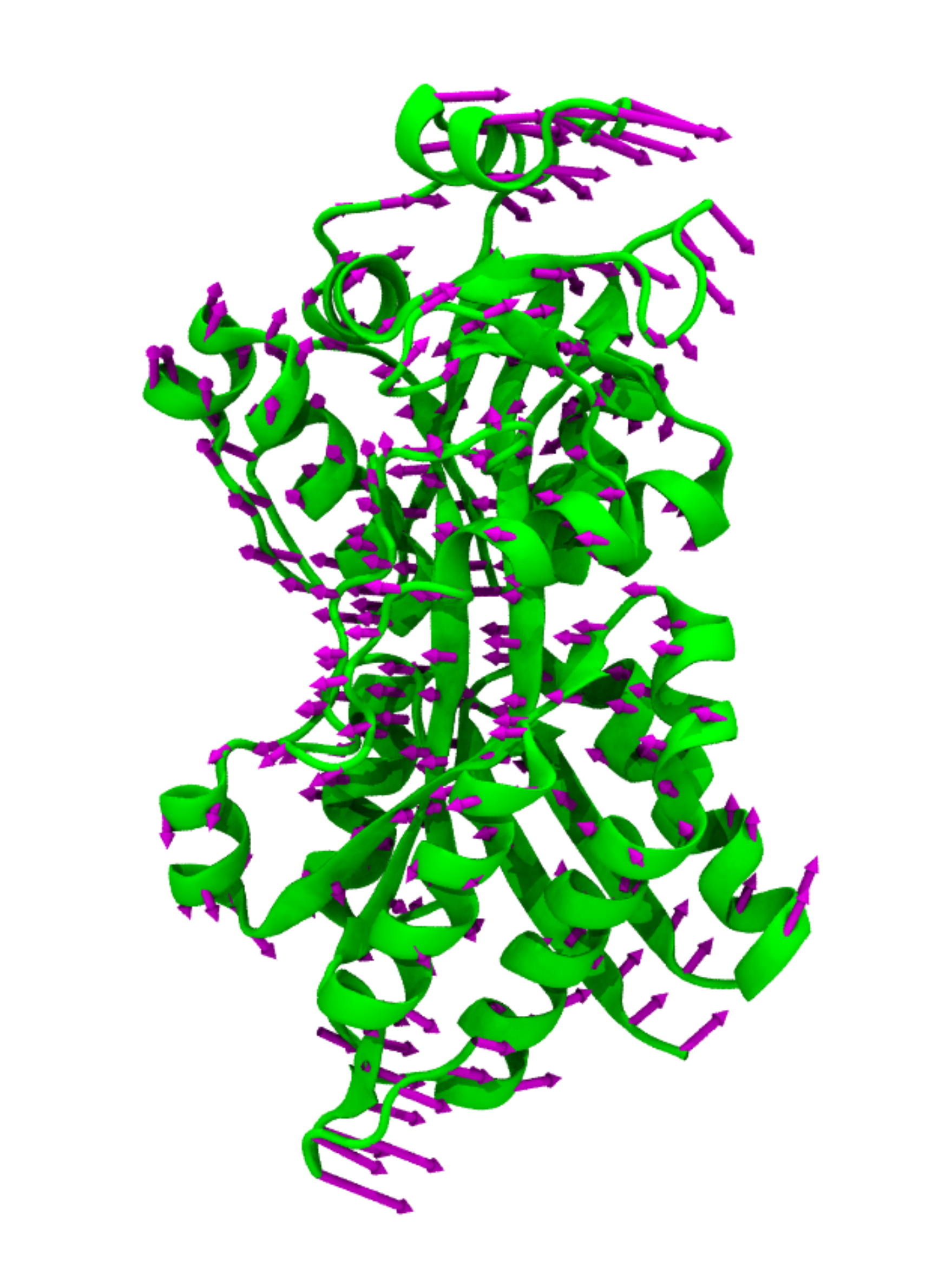}
\end{tabular}
\begin{tabular}{ccc}
\includegraphics[width=0.25\textwidth]{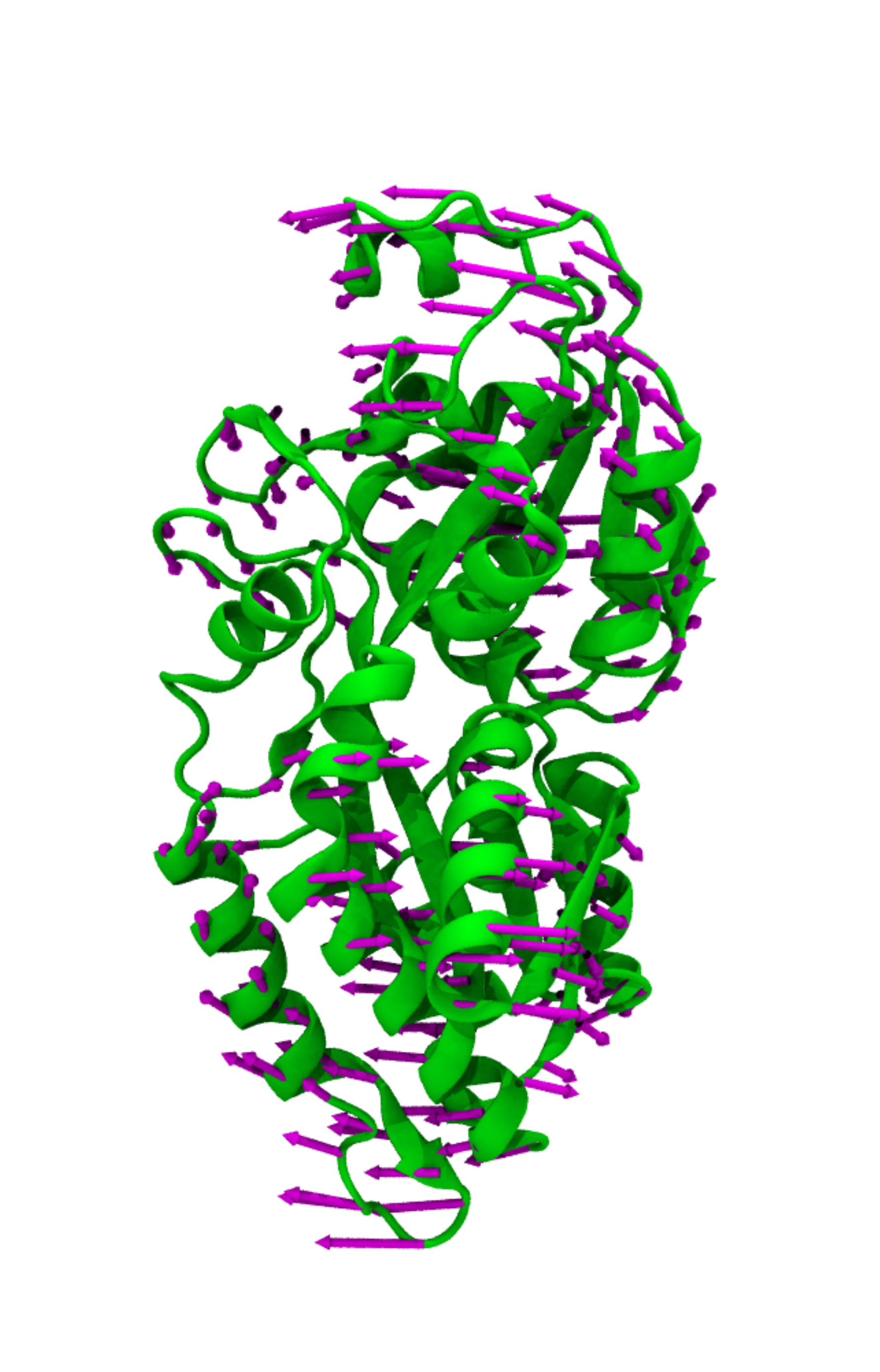}
\includegraphics[width=0.25\textwidth]{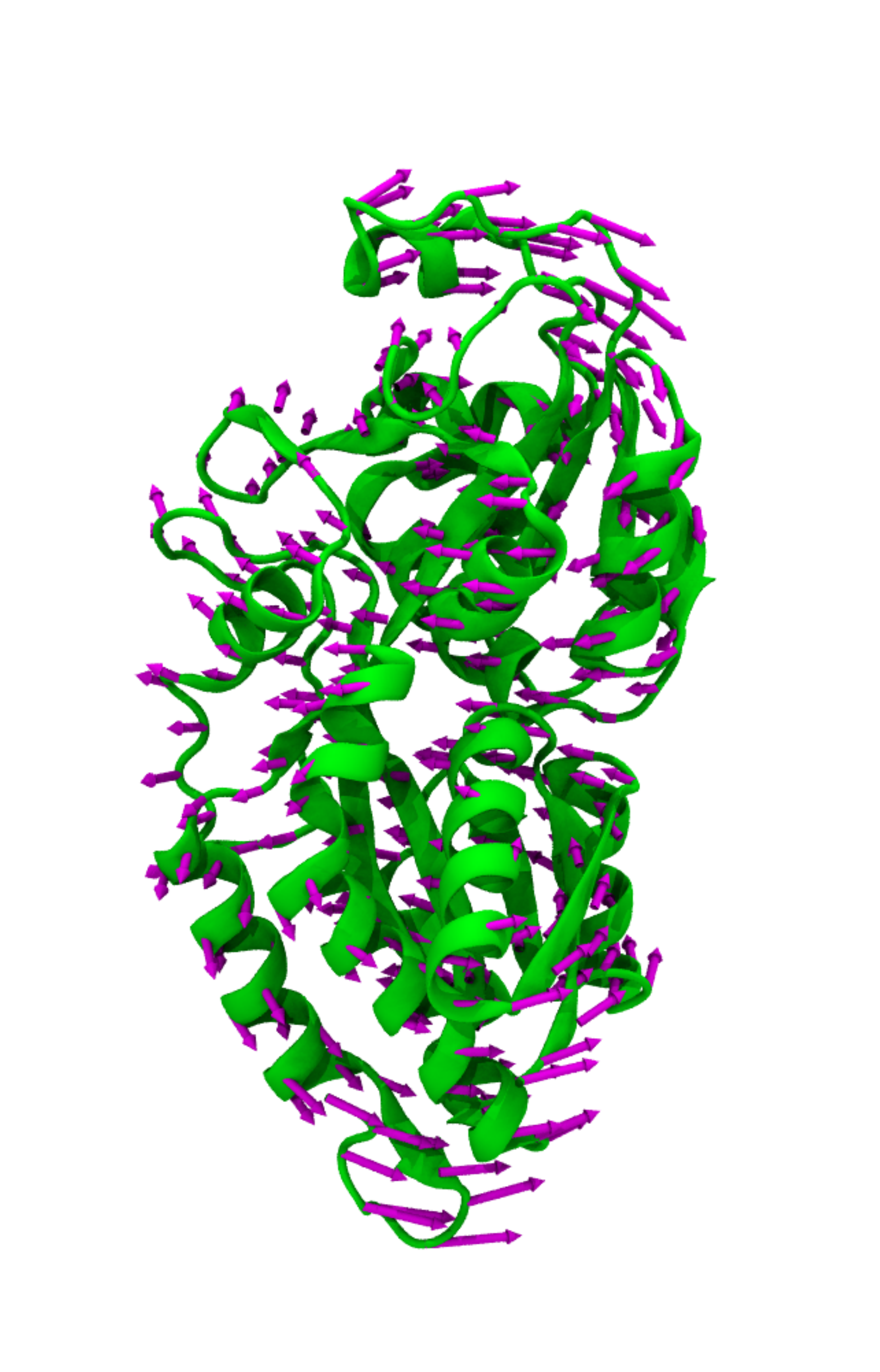}
\includegraphics[width=0.25\textwidth]{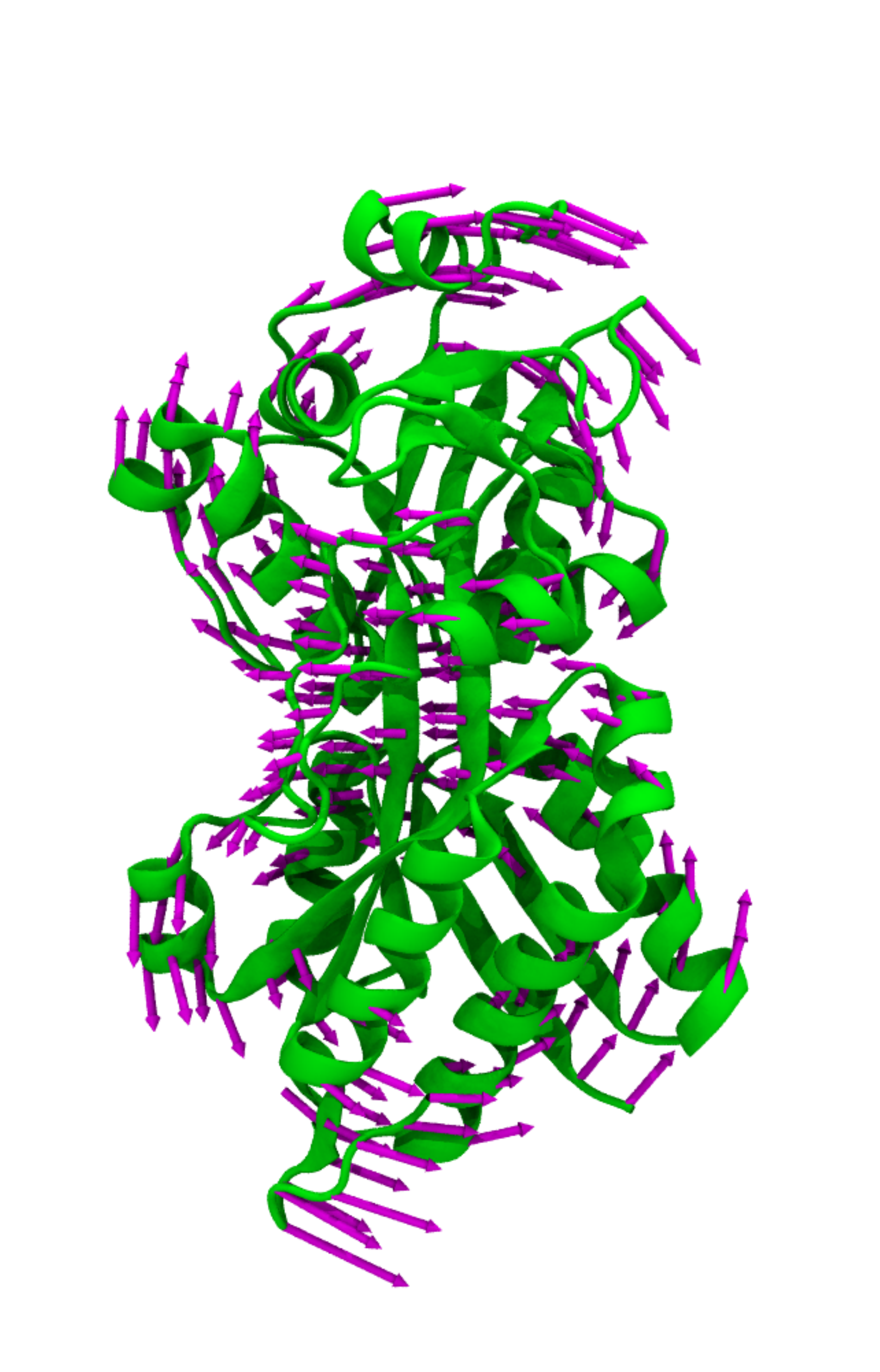}
\end{tabular}
\end{center}
\caption{ Comparison of modes for phosphate active transport protein (PDB ID: 2ABH). 
The top row is generated by using the completely local aFRI with $\upsilon=2$ and $\eta=60$. 
The middle row is generated by using the completely global aFRI with   $\upsilon=2$ and $\eta=35$. 
The bottom row is generated by using  the ANM with ProDy v1.5 \cite{Bakan:2011} using default settings. All three-dimensional images are rendered using VMD \cite{VMD}.
 }
\label{modes2}
\end{figure}

\subsection{ aFRI for protein identification }\label{Sec:DomainsaFRI}
 
Anisotropic FRI, like ANM, predicts the amplitudes and directions of atomic fluctuations. To test the accuracy of this new method we can compare aFRI fluctuation predictions to experimental B-factors as we have done with FRI and GNM.  {Both anisotropic rigidity based flexibility $(f_i^{\rm AR})$ and anisotropic flexibility based flexibility ($f_i^{\rm AF}$) are examined for their  B-factor predictions and their average correlation coefficients are 0.602 ($\upsilon=2$ and $\eta=9$) and 0.572 ($\upsilon=2$ and $\eta=18$) for the superset of 365 structures.} This means that aFRI is  more accurate than GNM (average correlation coefficient 0.565 for the superset). However, aFRI is slightly less accurate than pfFRI (average correlation coefficient 0.626 for the superset), which is similar to the fact that ANM is not as accurate as GNM.

  {
A major utility of the proposed   aFRI theory is the prediction  of protein motions by using the anisotropic flexibility. Since aFRI is adaptive, its cluster can be as large as the whole molecule and as small as a single particle. The completely global aFRI   has  a Hessian matrix of $3N\times3N$ elements and produces $3N$ eigenmodes. Depending on symmetry, 5 or 6 modes are due to the translational and rotational motions. Therefore, the remaining $3N-5$ or $3N-6$ vibrational modes can be obtained. To validate the proposed aFRI theory for normal mode  analysis, we have computed the vibrational modes for a few simple molecules, namely,  H$_2$O,  CO$_2$ and CH$_2$O, whose   vibrational modes are well-known. Our results are displayed in Figure  \ref{smallmolecues}. In addition to these vibrational modes,  appropriate translational and rotational modes are also observed from our calculations, but are omitted in our presentation. Therefore, the proposed aFRI works well for the analysis of small molecular translation, rotation, and vibration. 
}

  {
Having established our aFRI for small molecular vibrational analysis, we are interested in examining its behavior for macromolecules.  Anisotropic normal mode analysis of large biomolecules can be very expensive because the computational complexity of the global matrix scales as    ${\cal O}((3N)^3)$. As such,  the adaptive cluster analysis option provided by the aFRI algorithm can be useful.  In this work, we explore two extreme aFRI options, i.e., the completely global cluster and completely local clusters, for protein vibrational analysis.   Similar to the ANM, the completely global aFRI algorithm has a Hessian matrix of $3N\times3N$ elements and produces $3N$ eigenmodes. The motions predicted by these eigenmodes are typically very similar to those produced by ANM. In contrast, the completely localized aFRI has only a total of $N$  $3\times3$ Hessian matrices and gives rise to $3N$ eigenmodes for $N$ particles. We assemble these $3N$ eigenmodes into 3 modes for the molecule and weight the amplitude of each eigenmode by the corresponding B-factor for the particle.  Due to the non-local correlation built in the aFRI matrices, these three modes of motion obtained by the completely local aFRI algorithm are often  similar to certain low-order   modes  calculated by ANM for a protein.   The first three modes of motion for the HIV capsid protein  are shown in Figure \ref{modes} for two aFRI algorithms and ANM. It is seen that three eigenmodes obtained from  the completely global aFRI    resemble those calculated by the ANM.  Although three modes produced by the completely local aFRI algorithm show different motions, it is amazing to note that there is much collective motion in these modes.   
 }

 {
Figure \ref{modes2} depicts three  modes for  phosphate active transport receptor protein generated by using  two aFRI algorithms and ANM. Once again, we see a good similarity between eigenmodes calculated by using the   completely global aFRI algorithm and those computed by using the ANM. However, the modes   generated with the completely localized aFRI  demonstrate somewhat different motions.} In each method, the relative motion of two domains can be clearly identified.   The domain relative motions in the eigenmodes  of the   completely global aFRI and the ANM exhibit a better  synergistic effect in general. Whereas, modes from the completely local aFRI  are slightly less collective. Since there is no standard answer to domain fluctuations, it is difficult to say which one is right or wrong. An  interesting observation is that although  aFRI matrices can be completely local, they have built in non-local correlation and thus are able to simulate highly collective protein  motions.

\section{Concluding remarks}\label{sec:Conclusion}

The fundamental challenges that hinder the current quantitative understanding of biomolecular systems are their tremendous complexity and excessively large number of degrees of freedom. A  multiscale approach, the continuum elasticity with atomic rigidity (CEWAR), provides a new method for the reduction of the number of degrees of freedom in biomolecular systems \cite{KLXia:2013d}.  The performance of the CEWAR method relies on the accurate and efficient evaluation of a continuous atomic rigidity function. The flexibility-rigidity index (FRI) is proposed as a potential algorithm for such an evaluation. The underlying assumption of the FRI is that protein interactions uniquely determine the protein structure which, in turn,  determines the protein functions, such as stability and flexibility. Therefore, one just needs the structural information to predict protein B-factors without reconstructing the protein interaction Hamiltonian.  In particular, we assume that biomolecular flexibility and rigidity are local structural properties. Therefore, the flexibility at an atom is completely determined by its local environment, namely, local geometry and local topological connectivity.  We treat the (local) flexibility as an inverse of the (local) rigidity. As a consequence, we do not need  to solve the (global) eigenvalue problem of the Hamiltonian.  The first step of the FRI method is to measure protein topological connectivity from the distance geometry via smooth and monotonically decreasing radial basis functions. The atomic rigidity index is then associated with the total connectivity or interaction strength at each residue. Consequently, the atomic flexibility index, which is the inverse function of the atomic rigidity index, is associated with protein B-factors.

Protein flexibility is an intrinsic property that strongly correlates with protein functions. The analysis of protein flexibility is a crucial task in computational biophysics. Many established methods,  {including   normal mode analysis (NMA)  \cite{Go:1983,Tasumi:1982,Brooks:1983,Levitt:1985} and  elastic network model (ENM) \cite{Tirion:1996} such as   Gaussian network model (GNM)   \cite{Flory:1976, Bahar:1997,Bahar:1998},} have been developed in the past. These approaches typically depend on  the Hamiltonian mechanics of elastic interactions and matrix decomposition of the interaction Hamiltonian. The present work examines the performance of the proposed FRI for protein flexibility analysis in a comparison with some cutting edge methods, specifically, GNM and NMA.

We calibrate  the accuracy, reliability and computational efficiency of the FRI method by using GNM and NMA on three sets of proteins, i.e,   relatively small-sized, medium sized and large sized structures employed by Park et al. \cite{JKPark:2013} in a recent study. Additional calibration with the GNM is carried out on  extended set. As a result, a total of  365 proteins is studied in the present comparative study. As an internal validation, the FRI method is realized by using three families of correlation functions. Correlation  functions of generalized exponential type and   Lorentz type are found to  deliver better results. In particular,   correlation  functions of Lorentz type are simple and can be made parameter free, which is desirable for general use.  Although GNM and NMA may outperform the proposed FRI method on certain proteins in terms of the accuracy of the B-factor prediction, the FRI method is able to improve on the average correlation coefficient of GNM and NMA on all three sets of proteins.  Additionally, the FRI is found to significantly outperform the GNM on the extended superset of 365 structures as well.

A possible reason for the FRI to outperform the existing methods is that GNM and  NMA are essentially global methods in a sense that they rely on the solution of the global eigenvalue problem to predict local  atomic properties, e.g., B-factors.  In contrast, the FRI is a local method and utilizes the local geometric information to predict local atomic properties. In parallel, there are   (global) band theory of solids and (local) atomic orbital model of solids. The former is good for describing many global physical properties, such as electrical conductivity and thermal lattice motions in terms of excitations,  while the latter is more powerful for explaining localized chemical reactivity and catalysis of solids. 

The GNM is known for its superb computational  efficiency \cite{LWYang:2008}. The matrix diagonalization is of ${\cal O}(N^3)$ in computational complexity, where $N$ is the number of residues. The  computational complexity of our original FRI is of ${\cal O}(N^2)$. In the present work, we propose a fast FRI (fFRI) algorithm, which further reduces the  computational complexity of FRI to ${\cal O}(N)$. Both FRI and fFRI do not involve the time consuming matrix decomposition. As a result, it takes less than 30 seconds for the fFRI to predict the B Factors of an HIV virus structure with more than three hundred thousands of residues, which otherwise requires many years for the GNM to compute.  Additionally, both the exponential based  parameter-free fFRI and the Lorentz based  parameter-free fFRI  are about 10\% more accurate than the GNM in the B-factor prediction of  365 proteins.

Anisotropic motions between protein domains are known to correlate with protein functions. To describe protein anisotropic fluctuations, we introduce anisotropic FRI (aFRI) algorithms.  {We  introduce an adaptive aFRI method which partitions the molecule into many clusters  with variable sizes. We specifically examine two extreme cases, i.e., a one-cluster partition and $N$-cluster partition, which result in a completely global $3N\times3N$ Hessian matrix and $N$ completely localized $3\times3$ Hessian matrices, respectively.    The  computational complexity of aFRI varies from  ${\cal O}(N^3)$ to ${\cal O}(N)$.} Although aFRI Hessian matrices can be completely local, they still contain much non-location correlation. As such, all of three protein modes predicted by the completely local aFRI  exhibit  highly collective global motions. The eigenmodes  obtained from 
the completely global aFRI closely resemble those of the anisotropic network model (ANM) \cite{Atilgan:2001,Bakan:2011}. However, mode constructed from the completely local aFRI show different collective motion patterns. Since there is no analytical solution for collective motions, it is not possible to judge whose collective motions are more correct.  In general, the   eigenmodes of ANM and the completely global aFRI exhibit a slightly better synergistic effect than modes generated by using the completely local aFRI.

The proposed FRI has a few visual applications. First, the correlation maps of the FRI are capable of revealing both short- and long-distance interactions or connectivities. Since correlation map elements are directly related to the original distances by a known radial basis function, the distances can be labeled on the map as well. Additionally, the predicted B-factors can be plotted as the radii of residues to visualize the amplitude of thermal fluctuations. This plot becomes even more interesting when atomic spheres are colored with the electrostatics \cite{KLXia:2013d}. The close correlation between flexibility and large electrostatic potentials can be unveiled, which  sheds light on protein intrinsic structural properties.  Moreover, the predicted B-factors can be plotted with  secondary structures to have an overall picture of structural flexibility. Finally, as  continuous functions, the atomic rigidity function and atomic flexibility function can be projected onto  protein molecular surfaces or other surface representations to analyze flexibility.

Another application of FRI and aFRI is the analysis of protein domains. Existing methods, such as GNM and ANM, are well known to do well for domain analysis.  The present FRI provides a clear correlation map for domain identifications. It is found that  aFRI gives rise to highly collective domain motion patterns, although  not all parts of a domain move uniformly in our aFRI representations.

\vspace{1cm}
\section*{Acknowledgments}

 This work was supported in part by NSF grants   IIS-1302285 and DMS-1160352,   NIH grant R01GM-090208 and MSU Center for Mathematical Molecular Biosciences Initiative.   KO thanks Dr Minxin Chen for useful discussion about the  data structure of cell lists. The authors acknowledge the Mathematical Biosciences Institute for hosting valuable workshops.

\vspace{1cm}
%

\newpage
\appendix
\centerline{\bf Efficiency or accuracy results for five sets of proteins}

\begin{table}[htbp]
  \centering
\caption{ Correlation coefficients for B-factor prediction obtained by optimal FRI (opFRI), parameter free FRI (pfFRI) and Gaussian normal mode (GNM) for small-size structures. \textdagger GNM and NMA values are taken from the coarse-grained (C$\alpha$) GNM and NMA results reported in Park et al. \cite{JKPark:2013} except where starred (*). Starred values indicate correlation coefficients, from our own test of GNM, that have significantly increased compared to the values reported by Park et al.\cite{JKPark:2013}  See Section \ref{sec:B-factor} for details regarding the calculation of new GNM scores.  }
\begin{tabular}{C{1.2cm}C{1.2cm}R{1.4cm}R{1.4cm}R{1.4cm}R{1.4cm}R{1.4cm}R{1.4cm}R{1.4cm}R{1.4cm}}
\toprule
PDB ID & $N$ & opFRI & pfFRI   & GNM \textdagger  & NMA \textdagger   \\
\midrule
1AIE  & 31    & 0.588 & 0.416 & 0.155 & 0.712 \\
1AKG  & 16    & 0.373 & 0.350 & 0.185 & -0.229 \\
1BX7  & 51    & 0.726 & 0.623 & 0.706 & 0.868 \\
1ETL  & 12    & 0.710 & 0.609 & 0.628 & 0.355 \\
1ETM  & 12    & 0.544 & 0.393 & 0.432 & 0.027 \\
1ETN  & 12    & 0.089 & 0.023 & -0.274 & -0.537 \\
1FF4  & 65    & 0.718 & 0.613 & 0.674 & 0.555 \\
1GK7  & 39    & 0.845 & 0.773 & 0.821 & 0.822 \\
1GVD  & 52    & 0.781 & 0.732 & 0.591 & 0.570 \\
1HJE  & 13    & 0.811 & 0.686 & 0.616 & 0.562 \\
1KYC  & 15    & 0.796 & 0.763 & 0.754 & 0.784 \\
1NOT  & 13    & 0.746 & 0.622 & 0.523 & 0.567 \\
1O06  & 20    & 0.910 & 0.874 & 0.844 & 0.900 \\
1OB4  & 16    & 0.776 & 0.763 & 0.750* & 0.930 \\
1OB7  & 16    & 0.737 & 0.545 & 0.652* & 0.952 \\
1P9I  & 29    & 0.754 & 0.742 & 0.625 & 0.603 \\
1PEF  & 18    & 0.888 & 0.826 & 0.808 & 0.888 \\
1PEN  & 16    & 0.516 & 0.465 & 0.270 & 0.056 \\
1Q9B  & 43    & 0.746 & 0.726 & 0.656 & 0.646 \\
1RJU  & 36    & 0.517 & 0.447 & 0.431 & 0.235 \\
1U06  & 55    & 0.474 & 0.429 & 0.434 & 0.377 \\
1UOY  & 64    & 0.713 & 0.653 & 0.671 & 0.628 \\
1USE  & 40    & 0.438 & 0.146 & -0.142 & -0.399 \\
1VRZ  & 21    & 0.792 & 0.695 & 0.677* & -0.203 \\
1XY2  & 8     & 0.619 & 0.570 & 0.562 & 0.458 \\
1YJO  & 6     & 0.375 & 0.333 & 0.434 & 0.445 \\
1YZM  & 46    & 0.842 & 0.834 & 0.901 & 0.939 \\
2DSX  & 52    & 0.337 & 0.333 & 0.127 & 0.433 \\
2JKU  & 35    & 0.805 & 0.695 & 0.656 & 0.850 \\
2NLS  & 36    & 0.605 & 0.559 & 0.530 & 0.088 \\
2OL9  & 6     & 0.909 & 0.904 & 0.689 & 0.886 \\
2OLX  & 4     & 0.917 & 0.888 & 0.885 & 0.776 \\
6RXN  & 45    & 0.614 & 0.574 & 0.594 & 0.304 \\

    \bottomrule
    \end{tabular}%
  \label{small_table}%
\end{table}%

\begin{table}[htbp]
  \centering
\caption{  Correlation coefficients for B-factor prediction obtained by optimal FRI (opFRI), parameter free FRI (pfFRI) and Gaussian normal mode (GNM) for medium-size structures. \textdagger GNM and NMA values are taken from the coarse-grained (C$\alpha$) GNM and NMA results reported in Park et al. \cite{JKPark:2013} except where starred (*). Starred values indicate correlation coefficients, from our own test of GNM, that have significantly increased compared to the values reported by Park et al. \cite{JKPark:2013}  See Section \ref{sec:B-factor} for details regarding the calculation of new GNM scores. }
\begin{tabular}{C{1.2cm}C{1.2cm}R{1.4cm}R{1.4cm}R{1.4cm}R{1.4cm}R{1.4cm}R{1.4cm}R{1.4cm}R{1.4cm}}
\toprule
PDB ID & $N$ & opFRI & pfFRI   & GNM \textdagger  & NMA \textdagger   \\
\midrule
1ABA  & 87    & 0.727 & 0.698  & 0.613 & 0.057\\
1CYO  & 88    & 0.751 & 0.702  & 0.741 & 0.774 \\
1FK5  & 93    & 0.590 & 0.568  & 0.485 & 0.362 \\
1GXU  & 88    & 0.748 & 0.634  & 0.421 & 0.581\\
1I71  & 83    & 0.549 & 0.516  & 0.549 & 0.380\\
1LR7  & 73    & 0.679 & 0.657  & 0.620 & 0.795\\
1N7E  & 95    & 0.651 & 0.609  & 0.497 & 0.385\\
1NNX  & 93    & 0.795 & 0.789  & 0.631 & 0.517\\
1NOA  & 113   & 0.622 & 0.604  & 0.615 & 0.485\\
1OPD  & 85    & 0.555 & 0.409  & 0.398 & 0.796\\
1QAU  & 112   & 0.678 & 0.672  & 0.620 & 0.533\\
1R7J  & 90    & 0.789 & 0.621  & 0.368 & 0.078\\
1UHA  & 83    & 0.726 & 0.665  & 0.638* & 0.308\\
1ULR  & 87    & 0.639 & 0.594  & 0.495 & 0.223\\
1USM  & 77    & 0.832 & 0.809  & 0.798 & 0.780\\
1V05  & 96    & 0.629 & 0.599  & 0.632 & 0.389\\
1W2L  & 97    & 0.691 & 0.564  & 0.397 & 0.432\\
1X3O  & 80    & 0.600 & 0.559  & 0.654 & 0.453\\
1Z21  & 96    & 0.662 & 0.638  & 0.433 & 0.289\\
1ZVA  & 75    & 0.756 & 0.579  & 0.690 & 0.579\\
2BF9  & 36    & 0.606 & 0.554  & 0.680* & 0.521\\
2BRF  & 100   & 0.795 & 0.764  & 0.710 & 0.535\\
2CE0  & 99    & 0.706 & 0.598  & 0.529 & 0.628\\
2E3H  & 81    & 0.692 & 0.682  & 0.605 & 0.632\\
2EAQ  & 89    & 0.753 & 0.690  & 0.695 & 0.688\\
2EHS  & 75    & 0.720 & 0.713  & 0.747 & 0.565\\
2FQ3  & 85    & 0.719 & 0.692  & 0.348 & 0.508\\
2IP6  & 87    & 0.654 & 0.578  & 0.572 & 0.826\\
2MCM  & 113   & 0.789 & 0.713  & 0.639 & 0.643\\
2NUH  & 104   & 0.835 & 0.691  & 0.771 & 0.685\\
2PKT  & 93    & 0.162 & 0.003  & -0.193* &-0.165 \\
2PLT  & 99    & 0.508 & 0.484  & 0.509* & 0.187\\
2QJL  & 99    & 0.594 & 0.584  & 0.594 & 0.497\\
2RB8  & 93    & 0.727 & 0.614  & 0.517 & 0.485\\
3BZQ  & 99    & 0.532 & 0.516  & 0.466 & 0.351\\
5CYT  & 103   & 0.441 & 0.421  & 0.331 & 0.102\\
    \bottomrule
    \end{tabular}%
  \label{medium_table}%
\end{table}%

\begin{table}[htbp]
  \centering
\caption{ Correlation coefficients for B-factor prediction obtained by optimal FRI (opFRI), parameter free FRI (pfFRI) and Gaussian normal mode (GNM) for large-size structures. \textdagger GNM and NMA values are taken from the coarse-grained (C$\alpha$) GNM and NMA results reported in Park et al. \cite{JKPark:2013} except where starred (*). Starred values indicate correlation coefficients, from our own test of GNM, that have significantly increased compared to the values reported by Park et al. \cite{JKPark:2013}. See Section \ref{sec:B-factor} for details regarding the calculation of new GNM scores.  }
\begin{tabular}{C{1.2cm}C{1.2cm}R{1.4cm}R{1.4cm}R{1.4cm}R{1.4cm}R{1.4cm}R{1.4cm}R{1.4cm}R{1.4cm}}
\toprule
PDB ID & $N$ & opFRI & pfFRI   & GNM \textdagger  & NMA \textdagger   \\
\midrule
1AHO  & 64    & 0.698 & 0.625 & 0.562 & 0.339 \\
1ATG  & 231   & 0.613 & 0.578 & 0.497 & 0.154 \\
1BYI  & 224   & 0.543 & 0.491 & 0.552 & 0.133 \\
1CCR  & 111   & 0.580 & 0.512 & 0.351 & 0.530 \\
1E5K  & 188   & 0.746 & 0.732 & 0.859 & 0.620 \\
1EW4  & 106   & 0.650 & 0.644 & 0.547 & 0.447 \\
1IFR  & 113   & 0.697 & 0.689 & 0.637 & 0.330 \\
1NKO  & 122   & 0.619 & 0.535 & 0.368 & 0.322 \\
1NLS  & 238   & 0.669 & 0.530 & 0.523* & 0.385 \\
1O08  & 221   & 0.562 & 0.333 & 0.309 & 0.616 \\
1PMY  & 123   & 0.671 & 0.654 & 0.685 & 0.702 \\
1PZ4  & 114   & 0.828 & 0.781 & 0.843 & 0.844 \\
1QTO  & 122   & 0.543 & 0.520 & 0.334 & 0.725 \\
1RRO  & 112   & 0.435 & 0.372 & 0.529 & 0.546 \\
1UKU  & 102   & 0.665 & 0.661 & 0.742 & 0.720 \\
1V70  & 105   & 0.622 & 0.492 & 0.162 & 0.285 \\
1WBE  & 204   & 0.591 & 0.577 & 0.549 & 0.574 \\
1WHI  & 122   & 0.601 & 0.539 & 0.270 & 0.414 \\
1WPA  & 107   & 0.634 & 0.577 & 0.417 & 0.380 \\
2AGK  & 233   & 0.705 & 0.694 & 0.512 & 0.514 \\
2C71  & 205   & 0.658 & 0.649 & 0.560 & 0.584 \\
2CG7  & 90    & 0.551 & 0.539 & 0.379 & 0.308 \\
2CWS  & 227   & 0.647 & 0.640 & 0.696 & 0.524 \\
2HQK  & 213   & 0.824 & 0.809 & 0.365 & 0.743 \\
2HYK  & 238   & 0.585 & 0.575 & 0.510 & 0.593 \\
2I24  & 113   & 0.593 & 0.498 & 0.494 & 0.441 \\
2IMF  & 203   & 0.652 & 0.625 & 0.514 & 0.401 \\
2PPN  & 107   & 0.677 & 0.638 & 0.668 & 0.468 \\
2R16  & 176   & 0.582 & 0.495 & 0.618* & 0.411 \\
2V9V  & 135   & 0.555 & 0.548 & 0.528 & 0.594 \\
2VIM  & 104   & 0.413 & 0.393 & 0.212 & 0.221 \\
2VPA  & 204   & 0.763 & 0.755 & 0.576 & 0.594 \\
2VYO  & 210   & 0.675 & 0.648 & 0.729 & 0.739 \\
3SEB  & 238   & 0.801 & 0.712 & 0.826 & 0.720 \\
3VUB  & 101   & 0.625 & 0.610 & 0.607 & 0.365 \\

 \bottomrule
    \end{tabular}%
  \label{large_table}%
\end{table}%

\begin{table}[htbp]
  \centering
	\renewcommand\thetable{7}
\caption{Correlation coefficients for B-factor prediction obtained by optimal FRI (opFRI), parameter free FRI (pfFRI) and Gaussian normal mode (GNM) for a set of 365 proteins. GNM scores reported here are the result of our tests with a processed set of PDB files as described in Section \ref{sec:B-factor}}
\begin{tabular}{C{1.2cm}C{1.2cm}C{1.2cm}C{1.2cm}C{1.2cm}C{1.2cm}C{1.2cm}C{1.2cm}C{1.2cm}C{1.2cm}}

\toprule
PDB ID & $N$  & opFRI & pfFRI & GNM        & PDB ID & $N$   & opFRI & pfFRI & GNM \\
\midrule

1ABA  & 87    & 0.727 & 0.698 & 0.613 & 1PEF  & 18    & 0.888 & 0.826 & 0.808 \\
1AGN  & 1492  & 0.331 & 0.051 & 0.170 & 1PEN  & 16    & 0.516 & 0.465 & 0.270 \\
1AHO  & 64    & 0.698 & 0.625 & 0.562 & 1PMY  & 123   & 0.671 & 0.654 & 0.685 \\
1AIE  & 31    & 0.588 & 0.416 & 0.155 & 1PZ4  & 114   & 0.828 & 0.781 & 0.843 \\
1AKG  & 16    & 0.373 & 0.350 & 0.185 & 1Q9B  & 43    & 0.746 & 0.726 & 0.656 \\
1ATG  & 231   & 0.613 & 0.578 & 0.497 & 1QAU  & 112   & 0.678 & 0.672 & 0.620 \\
1BGF  & 124   & 0.603 & 0.539 & 0.543 & 1QKI  & 3912  & 0.809 & 0.751 & 0.645 \\
1BX7  & 51    & 0.726 & 0.623 & 0.706 & 1QTO  & 122   & 0.543 & 0.520 & 0.334 \\
1BYI  & 224   & 0.543 & 0.491 & 0.552 & 1R29  & 122   & 0.650 & 0.631 & 0.556 \\
1CCR  & 111   & 0.580 & 0.512 & 0.351 & 1R7J  & 90    & 0.789 & 0.621 & 0.368 \\
1CYO  & 88    & 0.751 & 0.702 & 0.741 & 1RJU  & 36    & 0.517 & 0.447 & 0.431 \\
1DF4  & 57    & 0.912 & 0.889 & 0.832 & 1RRO  & 112   & 0.435 & 0.372 & 0.529 \\
1E5K  & 188   & 0.746 & 0.732 & 0.859 & 1SAU  & 114   & 0.742 & 0.671 & 0.596 \\
1ES5  & 260   & 0.653 & 0.638 & 0.677 & 1TGR  & 104   & 0.720 & 0.711 & 0.714 \\
1ETL  & 12    & 0.710 & 0.609 & 0.628 & 1TZV  & 141   & 0.837 & 0.820 & 0.841 \\
1ETM  & 12    & 0.544 & 0.393 & 0.432 & 1U06  & 55    & 0.474 & 0.429 & 0.434 \\
1ETN  & 12    & 0.089 & 0.023 & -0.274 & 1U7I  & 267   & 0.778 & 0.762 & 0.691 \\
1EW4  & 106   & 0.650 & 0.644 & 0.547 & 1U9C  & 221   & 0.600 & 0.577 & 0.522 \\
1F8R  & 1932  & 0.878 & 0.859 & 0.738 & 1UHA  & 83    & 0.726 & 0.665 & 0.638 \\
1FF4  & 65    & 0.718 & 0.613 & 0.674 & 1UKU  & 102   & 0.665 & 0.661 & 0.742 \\
1FK5  & 93    & 0.590 & 0.568 & 0.485 & 1ULR  & 87    & 0.639 & 0.594 & 0.495 \\
1GCO  & 1044  & 0.766 & 0.693 & 0.646 & 1UOY  & 64    & 0.713 & 0.653 & 0.671 \\
1GK7  & 39    & 0.845 & 0.773 & 0.821 & 1USE  & 40    & 0.438 & 0.146 & -0.142 \\
1GVD  & 52    & 0.781 & 0.732 & 0.591 & 1USM  & 77    & 0.832 & 0.809 & 0.798 \\
1GXU  & 88    & 0.748 & 0.634 & 0.421 & 1UTG  & 70    & 0.691 & 0.610 & 0.538 \\
1H6V  & 2927  & 0.488 & 0.429 & 0.306 & 1V05  & 96    & 0.629 & 0.599 & 0.632 \\
1HJE  & 13    & 0.811 & 0.686 & 0.616 & 1V70  & 105   & 0.622 & 0.492 & 0.162 \\
1I71  & 83    & 0.549 & 0.516 & 0.549 & 1VRZ  & 21    & 0.792 & 0.695 & 0.677 \\
1IDP  & 441   & 0.735 & 0.715 & 0.690 & 1W2L  & 97    & 0.691 & 0.564 & 0.397 \\
1IFR  & 113   & 0.697 & 0.689 & 0.637 & 1WBE  & 204   & 0.591 & 0.577 & 0.549 \\
1K8U  & 89    & 0.553 & 0.531 & 0.378 & 1WHI  & 122   & 0.601 & 0.539 & 0.270 \\
1KMM  & 1499  & 0.749 & 0.744 & 0.558 & 1WLY  & 322   & 0.695 & 0.679 & 0.666 \\
1KNG  & 144   & 0.547 & 0.536 & 0.512 & 1WPA  & 107   & 0.634 & 0.577 & 0.417 \\
1KR4  & 110   & 0.635 & 0.612 & 0.466 & 1X3O  & 80    & 0.600 & 0.559 & 0.654 \\
1KYC  & 15    & 0.796 & 0.763 & 0.754 & 1XY1  & 18    & 0.832 & 0.645 & 0.447 \\
1LR7  & 73    & 0.679 & 0.657 & 0.620 & 1XY2  & 8     & 0.619 & 0.570 & 0.562 \\
1MF7  & 194   & 0.687 & 0.681 & 0.700 & 1Y6X  & 87    & 0.596 & 0.524 & 0.366 \\
1N7E  & 95    & 0.651 & 0.609 & 0.497 & 1YJO  & 6     & 0.375 & 0.333 & 0.434 \\
1NKD  & 59    & 0.750 & 0.703 & 0.631 & 1YZM  & 46    & 0.842 & 0.834 & 0.901 \\
1NKO  & 122   & 0.619 & 0.535 & 0.368 & 1Z21  & 96    & 0.662 & 0.638 & 0.433 \\
1NLS  & 238   & 0.669 & 0.530 & 0.523 & 1ZCE  & 146   & 0.808 & 0.757 & 0.770 \\
1NNX  & 93    & 0.795 & 0.789 & 0.631 & 1ZVA  & 75    & 0.756 & 0.579 & 0.690 \\
1NOA  & 113   & 0.622 & 0.604 & 0.615 & 2A50  & 457   & 0.564 & 0.524 & 0.281 \\
1NOT  & 13    & 0.746 & 0.622 & 0.523 & 2AGK  & 233   & 0.705 & 0.694 & 0.512 \\
1O06  & 20    & 0.910 & 0.874 & 0.844 & 2AH1  & 939   & 0.684 & 0.593 & 0.521 \\
1O08  & 221   & 0.562 & 0.333 & 0.309 & 2B0A  & 186   & 0.639 & 0.603 & 0.467 \\
1OB4  & 16    & 0.776 & 0.763 & 0.750 & 2BCM  & 413   & 0.555 & 0.551 & 0.477 \\
1OB7  & 16    & 0.737 & 0.545 & 0.652 & 2BF9  & 36    & 0.606 & 0.554 & 0.680 \\
1OPD  & 85    & 0.555 & 0.409 & 0.398 & 2BRF  & 100   & 0.795 & 0.764 & 0.710 \\
1P9I  & 29    & 0.754 & 0.742 & 0.625 & 2C71  & 205   & 0.658 & 0.649 & 0.560 \\

    \bottomrule
    \end{tabular}%
  \label{long_table}%
\end{table}%

\begin{table}[htbp]
  \centering
		\renewcommand\thetable{7 (cont.)}
\caption{Correlation coefficients for B-factor prediction obtained by optimal FRI (opFRI), parameter free FRI (pfFRI) and Gaussian normal mode (GNM) for a set of 365 proteins. GNM scores reported here are the result of our tests with a processed set of PDB files as described in Section \ref{sec:B-factor}}
\begin{tabular}{C{1.2cm}C{1.2cm}C{1.2cm}C{1.2cm}C{1.2cm}C{1.2cm}C{1.2cm}C{1.2cm}C{1.2cm}C{1.2cm}}

\toprule
PDB ID & $N$   & opFRI & pfFRI & GNM        & PDB ID & $N$   & opFRI & pfFRI & GNM \\
\midrule
2CE0  & 99    & 0.706 & 0.598 & 0.529 & 2OLX  & 4     & 0.917 & 0.888 & 0.885 \\
2CG7  & 90    & 0.551 & 0.539 & 0.379 & 2PKT  & 93    & 0.162 & 0.003 & -0.193 \\
2COV  & 534   & 0.846 & 0.823 & 0.812 & 2PLT  & 99    & 0.508 & 0.484 & 0.509 \\
2CWS  & 227   & 0.647 & 0.640 & 0.696 & 2PMR  & 76    & 0.693 & 0.682 & 0.619 \\
2D5W  & 1214  & 0.689 & 0.682 & 0.681 & 2POF  & 440   & 0.682 & 0.651 & 0.589 \\
2DKO  & 253   & 0.816 & 0.812 & 0.690 & 2PPN  & 107   & 0.677 & 0.638 & 0.668 \\
2DPL  & 565   & 0.596 & 0.538 & 0.658 & 2PSF  & 608   & 0.526 & 0.500 & 0.565 \\
2DSX  & 52    & 0.337 & 0.333 & 0.127 & 2PTH  & 193   & 0.822 & 0.784 & 0.767 \\
2E10  & 439   & 0.798 & 0.796 & 0.692 & 2Q4N  & 153   & 0.711 & 0.667 & 0.740 \\
2E3H  & 81    & 0.692 & 0.682 & 0.605 & 2Q52  & 412   & 0.756 & 0.748 & 0.621 \\
2EAQ  & 89    & 0.753 & 0.690 & 0.695 & 2QJL  & 99    & 0.594 & 0.584 & 0.594 \\
2EHP  & 248   & 0.804 & 0.804 & 0.773 & 2R16  & 176   & 0.582 & 0.495 & 0.618 \\
2EHS  & 75    & 0.720 & 0.713 & 0.747 & 2R6Q  & 138   & 0.603 & 0.540 & 0.529 \\
2ERW  & 53    & 0.461 & 0.253 & 0.199 & 2RB8  & 93    & 0.727 & 0.614 & 0.517 \\
2ETX  & 389   & 0.580 & 0.556 & 0.632 & 2RE2  & 238   & 0.652 & 0.613 & 0.673 \\
2FB6  & 116   & 0.791 & 0.786 & 0.740 & 2RFR  & 154   & 0.693 & 0.671 & 0.753 \\
2FG1  & 157   & 0.620 & 0.617 & 0.584 & 2V9V  & 135   & 0.555 & 0.548 & 0.528 \\
2FN9  & 560   & 0.607 & 0.595 & 0.611 & 2VE8  & 515   & 0.744 & 0.643 & 0.616 \\
2FQ3  & 85    & 0.719 & 0.692 & 0.348 & 2VH7  & 94    & 0.775 & 0.726 & 0.596 \\
2G69  & 99    & 0.622 & 0.590 & 0.436 & 2VIM  & 104   & 0.413 & 0.393 & 0.212 \\
2G7O  & 68    & 0.785 & 0.784 & 0.660 & 2VPA  & 204   & 0.763 & 0.755 & 0.576 \\
2G7S  & 190   & 0.670 & 0.644 & 0.649 & 2VQ4  & 106   & 0.680 & 0.679 & 0.555 \\
2GKG  & 122   & 0.688 & 0.646 & 0.711 & 2VY8  & 149   & 0.770 & 0.724 & 0.533 \\
2GOM  & 121   & 0.586 & 0.584 & 0.491 & 2VYO  & 210   & 0.675 & 0.648 & 0.729 \\
2GXG  & 140   & 0.847 & 0.780 & 0.520 & 2W1V  & 548   & 0.680 & 0.680 & 0.571 \\
2GZQ  & 191   & 0.505 & 0.382 & 0.369 & 2W2A  & 350   & 0.706 & 0.638 & 0.589 \\
2HQK  & 213   & 0.824 & 0.809 & 0.365 & 2W6A  & 117   & 0.823 & 0.748 & 0.647 \\
2HYK  & 238   & 0.585 & 0.575 & 0.510 & 2WJ5  & 96    & 0.484 & 0.440 & 0.357 \\
2I24  & 113   & 0.593 & 0.498 & 0.494 & 2WUJ  & 100   & 0.739 & 0.598 & 0.598 \\
2I49  & 398   & 0.714 & 0.683 & 0.601 & 2WW7  & 150   & 0.499 & 0.471 & 0.356 \\
2IBL  & 108   & 0.629 & 0.625 & 0.352 & 2WWE  & 111   & 0.692 & 0.582 & 0.628 \\
2IGD  & 61    & 0.585 & 0.481 & 0.386 & 2X1Q  & 240   & 0.534 & 0.478 & 0.443 \\
2IMF  & 203   & 0.652 & 0.625 & 0.514 & 2X25  & 168   & 0.632 & 0.598 & 0.403 \\
2IP6  & 87    & 0.654 & 0.578 & 0.572 & 2X3M  & 166   & 0.744 & 0.717 & 0.655 \\
2IVY  & 88    & 0.544 & 0.483 & 0.271 & 2X5Y  & 171   & 0.718 & 0.705 & 0.694 \\
2J32  & 244   & 0.863 & 0.848 & 0.855 & 2X9Z  & 262   & 0.583 & 0.578 & 0.574 \\
2J9W  & 200   & 0.716 & 0.705 & 0.662 & 2XHF  & 310   & 0.606 & 0.591 & 0.569 \\
2JKU  & 35    & 0.805 & 0.695 & 0.656 & 2Y0T  & 101   & 0.778 & 0.774 & 0.798 \\
2JLI  & 100   & 0.779 & 0.613 & 0.622 & 2Y72  & 170   & 0.780 & 0.754 & 0.766 \\
2JLJ  & 115   & 0.741 & 0.720 & 0.527 & 2Y7L  & 319   & 0.928 & 0.797 & 0.747 \\
2MCM  & 113   & 0.789 & 0.713 & 0.639 & 2Y9F  & 149   & 0.771 & 0.762 & 0.664 \\
2NLS  & 36    & 0.605 & 0.559 & 0.530 & 2YLB  & 400   & 0.807 & 0.807 & 0.675 \\
2NR7  & 194   & 0.803 & 0.785 & 0.727 & 2YNY  & 315   & 0.813 & 0.804 & 0.706 \\
2NUH  & 104   & 0.835 & 0.691 & 0.771 & 2ZCM  & 357   & 0.458 & 0.422 & 0.420 \\
2O6X  & 306   & 0.814 & 0.799 & 0.651 & 2ZU1  & 360   & 0.689 & 0.672 & 0.653 \\
2OA2  & 132   & 0.571 & 0.456 & 0.458 & 3A0M  & 148   & 0.807 & 0.712 & 0.392 \\
2OCT  & 192   & 0.567 & 0.550 & 0.540 & 3A7L  & 128   & 0.713 & 0.663 & 0.756 \\
2OHW  & 256   & 0.614 & 0.539 & 0.475 & 3AMC  & 614   & 0.675 & 0.669 & 0.581 \\
2OKT  & 342   & 0.433 & 0.411 & 0.336 & 3AUB  & 116   & 0.614 & 0.608 & 0.637 \\
2OL9  & 6     & 0.909 & 0.904 & 0.689 & 3B5O  & 230   & 0.644 & 0.629 & 0.601 \\

    \bottomrule
    \end{tabular}%
  \label{long_table2}%
\end{table}%

\begin{table}[htbp]
  \centering
		\renewcommand\thetable{7 (cont.)}
\caption{Correlation coefficients for B-factor prediction obtained by optimal FRI (opFRI), parameter free FRI (pfFRI) and Gaussian normal mode (GNM) for a set of 365 proteins. GNM scores reported here are the result of our tests with a processed set of PDB files as described in Section \ref{sec:B-factor}}
\begin{tabular}{C{1.2cm}C{1.2cm}C{1.2cm}C{1.2cm}C{1.2cm}C{1.2cm}C{1.2cm}C{1.2cm}C{1.2cm}C{1.2cm}}

\toprule
PDB ID & $N$   & opFRI & pfFRI & GNM        & PDB ID & $N$   &  opFRI & pfFRI & GNM \\
\midrule

3BA1  & 312   & 0.661 & 0.624 & 0.621 & 3MD4  & 12    & 0.860 & 0.781 & 0.914 \\
3BED  & 261   & 0.845 & 0.820 & 0.684 & 3MD5  & 12    & 0.649 & 0.413 & -0.218 \\
3BQX  & 139   & 0.634 & 0.481 & 0.297 & 3MEA  & 166   & 0.669 & 0.669 & 0.600 \\
3BZQ  & 99    & 0.532 & 0.516 & 0.466 & 3MGN  & 348   & 0.205 & 0.119 & 0.193 \\
3BZZ  & 100   & 0.485 & 0.450 & 0.600 & 3MRE  & 383   & 0.661 & 0.641 & 0.567 \\
3DRF  & 547   & 0.559 & 0.549 & 0.488 & 3N11  & 325   & 0.614 & 0.583 & 0.517 \\
3DWV  & 325   & 0.707 & 0.661 & 0.547 & 3NE0  & 208   & 0.706 & 0.645 & 0.659 \\
3E5T  & 228   & 0.502 & 0.489 & 0.296 & 3NGG  & 94    & 0.696 & 0.689 & 0.719 \\
3E7R  & 40    & 0.706 & 0.687 & 0.642 & 3NPV  & 495   & 0.702 & 0.653 & 0.677 \\
3EUR  & 140   & 0.431 & 0.427 & 0.577 & 3NVG  & 6     & 0.721 & 0.617 & 0.597 \\
3F2Z  & 149   & 0.824 & 0.792 & 0.740 & 3NZL  & 73    & 0.627 & 0.583 & 0.506 \\
3F7E  & 254   & 0.812 & 0.803 & 0.811 & 3O0P  & 194   & 0.727 & 0.706 & 0.734 \\
3FCN  & 158   & 0.640 & 0.606 & 0.632 & 3O5P  & 128   & 0.734 & 0.698 & 0.630 \\
3FE7  & 91    & 0.583 & 0.533 & 0.276 & 3OBQ  & 150   & 0.649 & 0.645 & 0.655 \\
3FKE  & 250   & 0.525 & 0.476 & 0.435 & 3OQY  & 234   & 0.698 & 0.686 & 0.637 \\
3FMY  & 66    & 0.701 & 0.655 & 0.556 & 3P6J  & 125   & 0.774 & 0.767 & 0.810 \\
3FOD  & 48    & 0.532 & 0.440 & -0.126 & 3PD7  & 188   & 0.770 & 0.723 & 0.589 \\
3FSO  & 221   & 0.831 & 0.817 & 0.793 & 3PES  & 165   & 0.697 & 0.642 & 0.683 \\
3FTD  & 240   & 0.722 & 0.713 & 0.634 & 3PID  & 387   & 0.537 & 0.531 & 0.642 \\
3FVA  & 6     & 0.835 & 0.825 & 0.789 & 3PIW  & 154   & 0.758 & 0.744 & 0.717 \\
3G1S  & 418   & 0.771 & 0.700 & 0.630 & 3PKV  & 221   & 0.625 & 0.597 & 0.568 \\
3GBW  & 161   & 0.820 & 0.747 & 0.510 & 3PSM  & 94    & 0.876 & 0.790 & 0.745 \\
3GHJ  & 116   & 0.732 & 0.511 & 0.196 & 3PTL  & 289   & 0.543 & 0.541 & 0.468 \\
3HFO  & 197   & 0.691 & 0.670 & 0.518 & 3PVE  & 347   & 0.718 & 0.667 & 0.568 \\
3HHP  & 1234  & 0.720 & 0.716 & 0.683 & 3PZ9  & 357   & 0.709 & 0.709 & 0.678 \\
3HNY  & 156   & 0.793 & 0.723 & 0.758 & 3PZZ  & 12    & 0.945 & 0.922 & 0.950 \\
3HP4  & 183   & 0.534 & 0.500 & 0.573 & 3Q2X  & 6     & 0.922 & 0.904 & 0.866 \\
3HWU  & 144   & 0.754 & 0.748 & 0.841 & 3Q6L  & 131   & 0.622 & 0.577 & 0.605 \\
3HYD  & 7     & 0.966 & 0.950 & 0.867 & 3QDS  & 284   & 0.780 & 0.745 & 0.568 \\
3HZ8  & 192   & 0.617 & 0.502 & 0.475 & 3QPA  & 197   & 0.587 & 0.442 & 0.503 \\
3I2V  & 124   & 0.486 & 0.441 & 0.301 & 3R6D  & 221   & 0.688 & 0.669 & 0.495 \\
3I2Z  & 138   & 0.613 & 0.599 & 0.317 & 3R87  & 132   & 0.452 & 0.419 & 0.286 \\
3I4O  & 135   & 0.735 & 0.714 & 0.738 & 3RQ9  & 162   & 0.510 & 0.403 & 0.242 \\
3I7M  & 134   & 0.667 & 0.635 & 0.695 & 3RY0  & 128   & 0.616 & 0.606 & 0.470 \\
3IHS  & 169   & 0.586 & 0.565 & 0.409 & 3RZY  & 139   & 0.800 & 0.784 & 0.849 \\
3IVV  & 149   & 0.817 & 0.797 & 0.693 & 3S0A  & 119   & 0.562 & 0.524 & 0.526 \\
3K6Y  & 227   & 0.586 & 0.535 & 0.301 & 3SD2  & 86    & 0.523 & 0.421 & 0.237 \\
3KBE  & 140   & 0.705 & 0.704 & 0.611 & 3SEB  & 238   & 0.801 & 0.712 & 0.826 \\
3KGK  & 190   & 0.784 & 0.775 & 0.680 & 3SED  & 124   & 0.709 & 0.658 & 0.712 \\
3KZD  & 85    & 0.647 & 0.611 & 0.475 & 3SO6  & 150   & 0.675 & 0.666 & 0.630 \\
3L41  & 220   & 0.718 & 0.716 & 0.669 & 3SR3  & 637   & 0.619 & 0.611 & 0.624 \\
3LAA  & 169   & 0.827 & 0.647 & 0.659 & 3SUK  & 248   & 0.644 & 0.633 & 0.567 \\
3LAX  & 106   & 0.734 & 0.730 & 0.584 & 3SZH  & 697   & 0.817 & 0.815 & 0.697 \\
3LG3  & 833   & 0.658 & 0.614 & 0.589 & 3T0H  & 208   & 0.808 & 0.775 & 0.694 \\
3LJI  & 272   & 0.612 & 0.608 & 0.551 & 3T3K  & 122   & 0.796 & 0.748 & 0.735 \\
3M3P  & 249   & 0.584 & 0.554 & 0.338 & 3T47  & 141   & 0.592 & 0.527 & 0.447 \\
3M8J  & 178   & 0.730 & 0.728 & 0.628 & 3TDN  & 357   & 0.458 & 0.419 & 0.240 \\
3M9J  & 210   & 0.639 & 0.574 & 0.296 & 3TOW  & 152   & 0.578 & 0.556 & 0.571 \\
3M9Q  & 176   & 0.591 & 0.510 & 0.471 & 3TUA  & 210   & 0.665 & 0.658 & 0.588 \\
3MAB  & 173   & 0.664 & 0.591 & 0.451 & 3TYS  & 75    & 0.853 & 0.800 & 0.791 \\

    \bottomrule
    \end{tabular}%
  \label{long_table3}%
\end{table}%

\begin{table}[htbp]
  \centering
		\renewcommand\thetable{7 (cont.)}
\caption{Correlation coefficients for B-factor prediction obtained by optimal FRI (opFRI), parameter free FRI (pfFRI) and Gaussian normal mode (GNM) for a set of 365 proteins. GNM scores reported here are the result of our tests with a processed set of PDB files as described in Section \ref{sec:B-factor}}
\begin{tabular}{C{1.2cm}C{1.2cm}C{1.2cm}C{1.2cm}C{1.2cm}C{1.2cm}C{1.2cm}C{1.2cm}C{1.2cm}C{1.2cm}}

\toprule
PDB ID & $N$   & opFRI & pfFRI & GNM        & PDB ID & $N$   &  opFRI & pfFRI & GNM \\
\midrule

3U6G  & 248   & 0.635 & 0.632 & 0.526 & 4DT4  & 160   & 0.776 & 0.738 & 0.716 \\
3U97  & 77    & 0.753 & 0.736 & 0.712 & 4EK3  & 287   & 0.680 & 0.680 & 0.674 \\
3UCI  & 72    & 0.589 & 0.526 & 0.495 & 4ERY  & 318   & 0.740 & 0.701 & 0.688 \\
3UR8  & 637   & 0.666 & 0.652 & 0.597 & 4ES1  & 95    & 0.648 & 0.625 & 0.551 \\
3US6  & 148   & 0.698 & 0.586 & 0.553 & 4EUG  & 225   & 0.570 & 0.529 & 0.405 \\
3V1A  & 48    & 0.531 & 0.487 & 0.583 & 4F01  & 448   & 0.633 & 0.372 & 0.688 \\
3V75  & 285   & 0.604 & 0.596 & 0.491 & 4F3J  & 143   & 0.617 & 0.598 & 0.551 \\
3VN0  & 193   & 0.840 & 0.837 & 0.812 & 4FR9  & 141   & 0.671 & 0.655 & 0.501 \\
3VOR  & 182   & 0.602 & 0.557 & 0.484 & 4G14  & 15    & 0.467 & 0.323 & 0.356 \\
3VUB  & 101   & 0.625 & 0.610 & 0.607 & 4G2E  & 151   & 0.760 & 0.755 & 0.758 \\
3VVV  & 108   & 0.833 & 0.741 & 0.753 & 4G5X  & 550   & 0.786 & 0.754 & 0.743 \\
3VZ9  & 163   & 0.785 & 0.749 & 0.695 & 4G6C  & 658   & 0.591 & 0.590 & 0.528 \\
3W4Q  & 773   & 0.737 & 0.725 & 0.649 & 4G7X  & 194   & 0.688 & 0.587 & 0.624 \\
3ZBD  & 213   & 0.651 & 0.516 & 0.632 & 4GA2  & 144   & 0.528 & 0.485 & 0.406 \\
3ZIT  & 152   & 0.430 & 0.404 & 0.392 & 4GMQ  & 92    & 0.678 & 0.628 & 0.550 \\
3ZRX  & 221   & 0.590 & 0.562 & 0.391 & 4GS3  & 90    & 0.544 & 0.522 & 0.547 \\
3ZSL  & 138   & 0.691 & 0.687 & 0.526 & 4H4J  & 236   & 0.810 & 0.806 & 0.689 \\
3ZZP  & 74    & 0.524 & 0.460 & 0.448 & 4H89  & 168   & 0.682 & 0.588 & 0.596 \\
3ZZY  & 226   & 0.746 & 0.709 & 0.728 & 4HDE  & 168   & 0.745 & 0.728 & 0.615 \\
4A02  & 166   & 0.618 & 0.516 & 0.303 & 4HJP  & 281   & 0.703 & 0.649 & 0.510 \\
4ACJ  & 167   & 0.748 & 0.746 & 0.759 & 4HWM  & 117   & 0.638 & 0.622 & 0.499 \\
4AE7  & 186   & 0.724 & 0.717 & 0.717 & 4IL7  & 85    & 0.446 & 0.404 & 0.316 \\
4AM1  & 345   & 0.674 & 0.619 & 0.460 & 4J11  & 357   & 0.620 & 0.562 & 0.401 \\
4ANN  & 176   & 0.551 & 0.536 & 0.470 & 4J5O  & 220   & 0.793 & 0.757 & 0.777 \\
4AVR  & 188   & 0.680 & 0.605 & 0.650 & 4J5Q  & 146   & 0.742 & 0.742 & 0.689 \\
4AXY  & 54    & 0.700 & 0.623 & 0.720 & 4J78  & 305   & 0.658 & 0.648 & 0.608 \\
4B6G  & 558   & 0.765 & 0.756 & 0.669 & 4JG2  & 185   & 0.746 & 0.736 & 0.543 \\
4B9G  & 292   & 0.844 & 0.816 & 0.763 & 4JVU  & 207   & 0.723 & 0.697 & 0.553 \\
4DD5  & 387   & 0.615 & 0.596 & 0.351 & 4JYP  & 534   & 0.688 & 0.682 & 0.538 \\
4DKN  & 423   & 0.781 & 0.761 & 0.539 & 4KEF  & 133   & 0.580 & 0.530 & 0.324 \\
4DND  & 95    & 0.763 & 0.750 & 0.582 & 5CYT  & 103   & 0.441 & 0.421 & 0.331 \\
4DPZ  & 109   & 0.730 & 0.726 & 0.651 & 6RXN  & 45    & 0.614 & 0.574 & 0.594 \\
4DQ7  & 328   & 0.690 & 0.683 & 0.376 &       &       &       &       &  \\

    \bottomrule
    \end{tabular}%
  \label{long_table4}%
\end{table}%

\begin{table}[htbp]
  \centering
			\renewcommand\thetable{8}
\caption{ CPU execution times (in seconds) from efficiency comparison between FRI, fFRI and GNM   }
\begin{tabular}{C{1.3cm}R{1.3cm}R{1.3cm}R{1.3cm}R{1.6cm}}
\toprule
PDB ID & $N$ & FRI & fFRI   & GNM  \\
\midrule
3P6J   & 125   &  *      &  *     & 0.141 \\
3R87   & 132   &  *      &  *      & 0.156 \\
3KBE   & 140   &  *      &  *      & 0.187 \\
1TZV   & 141   &  *      &  *      & 0.203 \\
2VY8   & 149   &  *      &  *      & 0.219 \\
3ZIT   & 152   &  *      &  *      & 0.234 \\
2FG1   & 157   &  *      &  *      & 0.265 \\
2X3M   & 166   &  *      &  *      & 0.312 \\
3LAA   & 169   &  *      &  *      & 0.327 \\
3M8J   & 178   &  *      &  *      & 0.375 \\
2GZQ   & 191   &  *      &  *      & 0.468 \\
4G7X   & 194   &  *      &  *      & 0.499 \\
2J9W   & 200   &  *      &  *      & 0.546 \\
3TUA   & 210   &  *      &  *      & 0.655 \\
1U9C   & 221   &  *      &  *      & 0.733 \\
3ZRX   & 221   &  *      &  *      & 0.718 \\
3K6Y   & 227   &  *      &  *      & 0.765 \\
3OQY   & 234   &  *      &  *      & 0.873 \\
2J32   & 244   &  *      &  *      & 0.967 \\
3M3P   & 249   &  *      &  *      & 1.029 \\
1U7I   & 267   &  *      &  *      & 1.263 \\
4B9G   & 292   &  *      &  *      & 1.669 \\
4ERY   & 318   &  *      &  *      & 2.122 \\
3MGN   & 348   &  *      &  *      & 2.902 \\
2ZU1   & 360   &  *      &  *      & 3.136 \\
2Q52   & 412   &  *      &  *      & 4.696 \\
4F01   & 448   &  *      &  *      & 6.178 \\
3DRF   & 547   & 0.062 &  *      & 11.154 \\
3UR8   & 637   & 0.07  &  *      & 17.409 \\
2AH1   & 939   & 0.156 &  *      & 61.012 \\
1GCO   & 1044  & 0.187 &  *      & 75.801 \\
1AGN   & 1492  & 0.343 &  *      & 298.632 \\
1F8R   & 1932  & 0.655 &  *      & 654.127 \\
1H6V   & 2927  & 1.545 &  *      & 2085.842 \\
1QKI   & 3912  & 2.699 &  *      & 6365.668 \\
3KGV   & 4064  & 2.949 &  *      & 6194.518 \\
1K32   & 6138  & 6.755 &  *      &  *  \\
1JZ0   & 8168  & 11.87 &  *      &  *  \\
4BGR   & 8949  & 14.056 & 0.889 &  *  \\
1VSZ   & 12012 & 25.413 & 1.248 &  *  \\
GroEL  & 14700 & *      & 1.467 &  *  \\
B Gal  & 16336 & *      & 1.716 &  *  \\
1VRI   & 18540 & *      & 1.934 &  *  \\
HIV    & 313236 & *     & 29.344 &  *  \\

 \bottomrule
\end{tabular}%
  \label{eff_table}%
\end{table}%

\end{document}